\documentclass[reprint,superscriptaddress,tightenlines,onecolumn,showpacs,floatfix,longbibliography,footinbib]{revtex4-2}
\pdfoutput=1

\usepackage{latexsym,amsmath,amsfonts,amssymb,bm,empheq}   
\usepackage{nicefrac} 
\usepackage[dvipsnames]{xcolor}
\usepackage{tabularx}   
\usepackage{booktabs}
\usepackage{multirow}
\usepackage{array}
\usepackage{tabularx}
\usepackage{subfigure}  
\usepackage[dvipsnames]{xcolor}
\usepackage{comment}

\usepackage{mathtools}
\DeclarePairedDelimiter\ceil{\lceil}{\rceil}
\DeclarePairedDelimiter\floor{\lfloor}{\rfloor}

\definecolor{nblue}{RGB}{28,130,185}
\definecolor{cgreen}{RGB}{76,153,0}
\definecolor{myorange}{RGB}{245,156,74}

\usepackage{hyperref}
\hypersetup{colorlinks=true, citecolor=magenta, urlcolor=-myorange}

\newcommand{\dif}{\mathrm{d}} 
\newcommand{\calE}{\mathcal{E}} 
\newcommand{\calR}{\mathcal{R}} 
\newcommand{\calA}{\mathcal{A}} 
\newcommand{\calAs}{\mathcal{A}_{\mathrm{s}}} 
\newcommand{\calAt}{\mathcal{A}_{\tau}} 

\newcommand{\calY}{\mathcal{Y}}
\newcommand{\Li}{\mathrm{Li}}
\newcommand{\csch}{\mathrm{csch}}
\newcommand{\sech}{\mathrm{sech}}

\newcommand{\erfc}{\mathrm{erfc}}

\begin{document}

\title{Transient Fluctuation--Induced Forces \\
in Driven Electrolytes after an Electric Field Quench}
%
%
\author{Saeed Mahdisoltani}
\affiliation{Rudolf Peierls Centre for Theoretical Physics, University of Oxford, Oxford OX1 3PU, United Kingdom}
\affiliation{Max Planck Institute for Dynamics and Self-Organization (MPIDS), D-37077 G\"ottingen, Germany}
%
%
\author{Ramin Golestanian}
\email{ramin.golestanian@ds.mpg.de}
\affiliation{Max Planck Institute for Dynamics and Self-Organization (MPIDS), D-37077 G\"ottingen, Germany}
\affiliation{Rudolf Peierls Centre for Theoretical Physics, University of Oxford, Oxford OX1 3PU, United Kingdom}

\date{\today}
\begin{abstract}
Understanding how electrolyte solutions behave out of thermal equilibrium is a long-standing endeavor in many areas of chemistry and biology. 
Although mean-field theories are widely used to model the dynamics of electrolytes, it is also important to characterize the effects of fluctuations in these systems. 
In a previous work, we showed that the dynamics of the ions in a strong electrolyte that is driven by an external electric field can generate long-ranged correlations manifestly different from the equilibrium screened correlations; in the nonequilibrium steady state, these correlations give rise to a novel long-range fluctuation-induced force (FIF). 
Here, we extend these results by considering the dynamics of the strong electrolyte after it is quenched from thermal equilibrium upon the application of a constant electric field. 
We show that the asymptotic long-distance limit of both charge and density correlations is generally diffusive in time. 
These correlations give rise to long-ranged FIFs acting on the neutral confining plates with long-time regimes that are governed by power-law temporal decays  toward the steady-state value of the force amplitude.
These findings show that nonequilibrium fluctuations have nontrivial implications on the dynamics of objects immersed in a driven electrolyte, and they could be useful for exploring new ways of controlling long-distance  forces in charged solutions.
\end{abstract}

\maketitle

\section{Introduction}      \label{sec:intro}

For more than a century, extensive research efforts have focused on understanding the properties of electrolytes and charged solutions~\cite{israelachvili,oosawa1971, verweyoverbeek1948}.
%
Conventional mean-field theories have been remarkably effective in such endeavors  and, in addition, the electrostatic correlation effects have been characterized to a great extent~\cite{levin2002electrostatic}.  
Similar studies of correlation effects and fluctuation phenomena in nonequilibrium electrolytes are, however,  relatively scarce.  
The present manuscript focuses on one such aspect, namely fluctuation-induced forces (FIFs)~\cite{kardar99friction}  in electrolyte  solutions driven out of equilibrium, by considering the transient behavior of the FIF after a quench by an electric field. 
%

Mean-field descriptions of electrolytes are obtained by combining those equations that govern the electrostatic interactions among the charged particles (e.g., the Poisson equation) with the statistical weights that describe the distribution of the charges. 
In thermal equilibrium, where the statistical weights are given by Boltzmann factors, this combination leads to the so-called nonlinear Poisson--Boltzmann (PB) equation~\cite{israelachvili}, while out of equilibrium one needs to consider more general distributions~\cite{onsagerlong1932} which are governed by Fokker--Planck equations~\cite{nanofluidsreview,bazant2009towards}. 
Even though the resulting nonlinear models in principle contain the full mean-field information about the system, it is often difficult to gain insights into the underlying physical mechanisms from them; as such, linear models have been of great importance in the conceptual developments toward understanding electrolytes and charged fluids. 
The celebrated Debye-H\"{u}ckel (DH) theory~\cite{debye1923}, for instance, is obtained from linearizing the PB equation, and it shows how correlations in an electrolyte become short-ranged as a result of the  screening effects of counterions (i.e., opposite charges).  
The simple picture provided by the DH theory serves as a general starting point to understand and construct models for a wide range of charged systems~\cite{israelachvili}.


%
The mean-field theories, however, do not generally take into account the statistical correlations between the ions in the system. It is well-known that the  correlations in electrolytes can give rise to a number of important phenomena, ranging from phase transitions in two-dimensional systems~\cite{kosterlitz}, to charge renormalization in colloids~\cite{alexander1984charge}, and counterion condensation and correlation-induced interactions which are relevant biological processes~\cite{grosberg2002colloquium,wong2010electrostatics,manning1978molecular,zribi2006condensation,golestanian2002conformational}.
%
In these cases, it can still be useful to combine the phenomenology of the correlation effects, which can lead to the breakdown of the mean-field assumptions, with the linearized  DH equations to get mathematically tractable models~\cite{levin2002electrostatic}. 
Although some related phenomena have also been considered in dynamical settings~\cite{golestanian2000dynamics,netz2003electrofriction,boroudjerdi2005statics}, a general  understanding of nonequilibrium electrostatic  correlations is still far from complete and poses a number of outstanding open questions~\cite{nanofluidsreview}. 
This is in part due to the inherent mathematical difficulty of characterizing the distribution functions of interacting ions with Brownian motions out of thermal equilibrium~\cite{onsagerlong1932,wrightelectrolyte}.   
Langevin formulations of the electrolyte dynamics provide a straightforward way of taking into account the stochasticity in the motion of the ions and can be used to obtain the required correlation functions, e.g., to compute the power spectrum of nanopores or the conductivity of strong electrolytes~\cite{zorkot2016power,zorkot2018nanopore,demery2016conductivity}.    
Inspired by some of the recent investigations of electrolytes and ionic liquids~\cite{Perkin2012,gebbie2013dilute,smith2016electrostatic,gebbie2017long,perez2017scaling,perez2019surface,kornyshev,stoneAC}, we use a similar mathematical framework to study how fluctuations in a strong electrolyte in the presence of an external electric field may modify the force that is exerted on boundaries that confine the electrolyte.  
This direction would also be relevant for designing more efficient and environmental-friendly batteries and electrochemical  capacitors~\cite{armand2008building,kotz2000principles,luo2015overview}  as electrolytes and ionic liquids are essential to the underlying conduction processes and energy storage. 

Fluctuation--induced forces (FIFs) comprise a remarkable aspect of fluctuation effects and arise when external objects disturb a correlated medium by imposing specific boundary conditions on its fluctuation modes~\cite{kardar99friction}. 
If the correlations in the medium are long-ranged, they then give rise to FIFs that can persist between objects at large distances and exhibit a number of  universal features~\cite{gambassi2009review,gambassinature, casimir1948attraction,fisher1978wall}. 
Fluctuation forces have many applications in  nanosciences~\cite{RMP2010} and colloidal systems
~\cite{maciolekreview} and they are also prevalent in nonequilibrium and driven settings \cite{najafi2004soret,dean2010out,aminov2015neqfif,rohwer2017transfif,gross2018surface,gross2019dynamics,rohwer2018neqfif,activecasimir,deanbrownian2014, deannoneqtune2016}. 
%
For an electrolyte in thermal equilibrium, the spatial range of the FIF is limited by the Debye screening length, which is often of the order of a few nanometers~\cite{israelachvili,jancovici2004screening,lee2018casimir}.  
Since nonlocal correlations generically emerge from conserved dynamics~\cite{garrido90conservative,grinstein90conservation,hwa89dissipative}, it may not be surprising that nonequilibrium FIFs  in driven electrolytes become long-ranged~\cite{mahdisoltani2021long}.  %
Since the underlying cause for this FIF, namely anisotropy in the conserved dynamics, is rather generic, similar nonequilibrium forces might also be relevant to modeling electrokinetic and biological systems where there are many instances of ionic currents in confined spaces~\cite{bazant2009towards,holm2001electrostatic}.  
%

\begin{figure}[b]
		\centering
		\includegraphics[width=.55\linewidth]{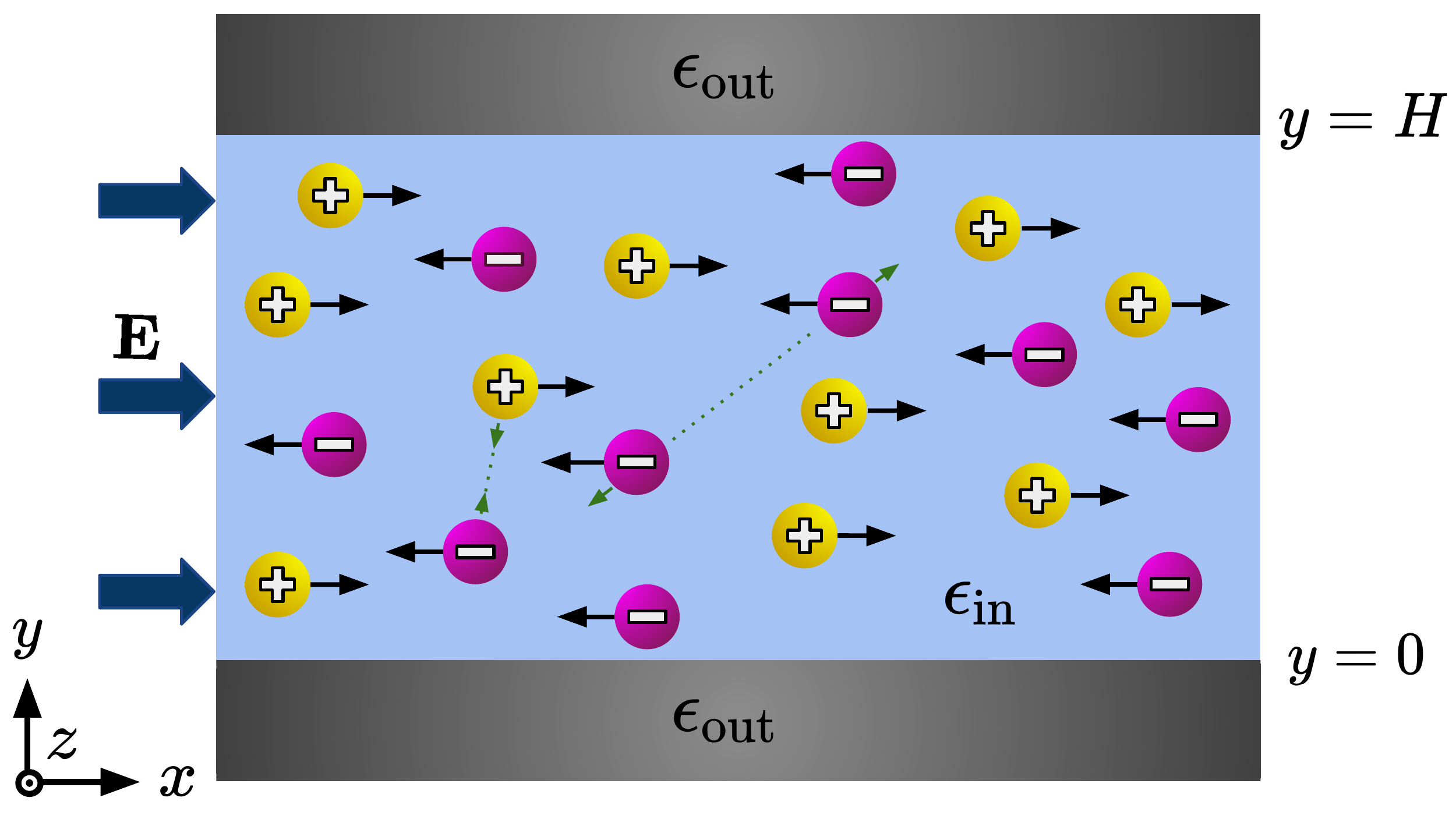}
	\caption{Schematics of a  driven electrolyte in flat geometry in $d=3$ dimensions. The external electric field $\bm{E} = E \hat{\bm{e}}_x$ is switched on at $t=0$ and drives the positive and negative charges in opposite directions (black arrows) from their previously equilibrium configuration; in addition, there are electrostatic interactions among the ions  (green arrows). The long-ranged correlation functions of this driven electrolyte are computed both in the bulk and in confinement, which in the latter case give rise to FIFs on the boundary plates.  
	Depicted here is the confined systems which is open along the $x$ and $z$ axes, and it is closed by uncharged plates with symmetric Neumann boundary conditions in the $y$ direction. 
	} \label{fig:schematic}
\end{figure}

In a previous work~\cite{mahdisoltani2021long}, we used the Dean--Kawasaki formalism~\cite{dean96langevin,kawasaki1994,ddftreview} along with scaling arguments to investigate the steady-state fluctuations in a simple strong electrolyte that is driven by a constant external electric field.  
It was shown that the anisotropy caused by the electric field leads to the so-called \textit{generically scale invariant} dynamics~\cite{tauber}, and the correlation functions in the steady-state become long-ranged and take power-law forms, despite the screening effects.  
These correlations then lead to nonequilibrium FIFs on the  confining boundaries, which in flat geometry and at the  steady state, depends on the plate separation $H$ as $H^{-d}$, where $d$ is the spatial dimensionality. The FIF in this case has unique non-monotonic changes as the  strength of the applied electric field is increased. 
In the present manuscript, we look into the transient  features of the FIF after the electric is switched on at $t=0$.  
To elucidate the fundamental concepts, we keep our focus on the case of a strong binary electrolyte which is initially at thermal equilibrium and is confined between two neutral parallel plates. The electrolyte is acted upon at time $t=0$ by a DC electric field parallel with the boundaries (see Fig.~\ref{fig:schematic}).
We investigate the  temporal variations of the FIF and  show that the FIF exhibits a diffusive behavior in time where its difference with the steady-state value decays as $t^{-d/2}$ in $d$ spatial dimensions (with a possible crossover in the value of the exponent, see Section~\ref{sec:stress}). 
These results cast some light on the emergent behavior in electrolytes out of thermal equilibrium and demonstrate that the dynamical and nonequilibrium effects in electrolytes can be manifestly different from their screened counterparts in equilibrium.  

The manuscript is organized as follows: 
in Section~\ref{sec:formalism}, we start from the Langevin equations that govern the dynamics of the ions and then  derive the stochastic density equations based on the Dean--Kawasaki approach. This is then accompanied by a   linearization and quasi-stationary approximation schemes. We also discuss the applicability of these approximations to settings where the electric field is a slowly varying function of time and, in addition, to solutions with different mobilities of charge species (which could be relevant for maintaining an unscreened electric field in the bulk). 
In Section~\ref{sec:corrfunc}, we use the linearized density equations to derive the charge and density correlation functions; first, this is done for a bulk solution and then for an electrolyte that is confined between two flat boundaries. 
In Section~\ref{sec:stress}, we compute the stress exerted by the electrolyte on the confining plates. To this end,  we use Maxwell stress and make use of the obtained correlation functions to obtain analytical expressions for the normal stress on the boundaries. The steady-state and transient parts of the stress amplitude are investigated using asymptotic expansions and numerical evaluations. 
Section~\ref{sec:conc} contains the summary and final concluding remarks. 
There are also four appendices containing the scaling analysis of the nonlinearities (Appendix~\ref{app:scaling}), the calculation of the two-point correlation functions of the linearized dynamics  at equal times  (Appendix~\ref{appendix:exact-equal-time-corrs}) and at different times (Appendix~\ref{appendix:exact-outof-time-corrs}) without using the quasi-stationary approximation, and a simplification of the transient part of the FIF  amplitude (Appendix~\ref{app:simpleAt}).

%
%

\section{Formalism}     \label{sec:formalism}
To analyze the dynamics of the electrolyte in the presence of the external electric field, in Section~\ref{subsec:Langevin} we start from the single particle Langevin equations that describe the motion of the individual ions, and then we recast the set of Langevin equations in terms of the time evolution of their instantaneous density using the Dean--Kawasaki approach~\cite{dean96langevin,kawasaki1994,ddftreview}. 
The density equations are then linearized in Section~\ref{subsec:linear} whose applicability on relaxing the assumptions of static electric field and equal ionic mobilities are discussed in Sections~\ref{subsec:slow-tdependence}~and~\ref{subsec:unequal-mobilities}.

\subsection{Stochastic density equations}  \label{subsec:Langevin}

The electrolyte we study here is composed of a symmetric collection of charged particles with charges $\pm Q$ that have equal mobilities
$ \mu_+ = \mu_- = \mu $. 
These particles undergo Brownian motion due to the microscopic collisions with the implicit solvent molecules (we will neglect the hydrodynamic effects of the solvent,  assuming a finite (screened) range of hydrodynamic interactions \cite{brady2006,ajdarinote,lowenlane,poncet2017universal,nagele}). 
In $d$ spatial dimensions, the overdamped trajectory $\bm{r}^\pm_a(t)$ of a cation or anion, labeled by $a$ ($a = 1,2,\ldots,N$) is governed by the Langevin equation
\begin{equation}       \label{eq:langevinparticle}
    \dot{\bm{r}}_a^\pm(t) = 
    \pm \mu Q \left[ -\nabla\phi\big(\bm{r}_a^\pm(t)\big) + \bm{E} \right] + 
    \sqrt { 2 D } \, \bm{\eta}_a^\pm (t),
\end{equation}    
where $\bm{E} = E \hat{\bm{e}}_x$ is the external electric field along the $x$ axis and $\bm{\eta}_a^\pm$ represent independent Gaussian white noises with correlations 
$\langle \eta_{a i}^\pm (t) \eta_{b j}^\pm (t') \rangle = 
\delta_{ab} \, \delta_{ij} \, \delta ( t - t' )$
and zero means.  
The fluctuation--dissipation relation connects the noise strength $D$ to the mobility $\mu$ through the Einstein relation $\mu = \beta D$, where $\beta = 1/(k_{\rm B} T)$ represents the inverse temperature of the system. 
The electrostatic potential field $\phi(\bm{r},t)$ is created by the ions in the system, and in the Gaussian  units it satisfies the Poisson equation
\begin{align}
    -\nabla^2 \phi(\bm{r},t) =\frac{S_d Q}{\epsilon_{\rm in} } \left( C^+ (\bm{r},t) - C^- (\bm{r},t) \right), 
\end{align}
where $S_d= 2 \pi^{d/2} / \Gamma(d/2)$ is the surface area of $d$-dimensional unit sphere, $\epsilon_{\rm in}$ is the permittivity of the electrolyte, and the instantaneous density of charge species is defined as 
\begin{align}
    C^\pm (\bm{r},t) = \sum_{a=1}^N \delta^d \left(\bm{r}-\bm{r}^\pm_a(t) \right).
\end{align}
%
%


Using the stochastic equations of Dean and Kawasaki,  
one then obtains the exact stochastic dynamics of $C^\pm$ as continuity equations, i.e. 
$\partial_t C^\pm + \nabla\cdot \bm{J}^\pm = 0 $, where the stochastic currents are given by
\begin{align}   \label{eq:Jpmfull}
\bm{J}^{\pm} (\bm{r},t) = - D \nabla C^\pm \pm C^\pm \mu  Q \left(  - \nabla\phi + \bm{E} \right) - \sqrt{2 D C^\pm} \, \bm{\zeta}^\pm (\bm{r},t).
\end{align}
Here, $\bm{\zeta}^\pm (\bm{r},t)$ are uncorrelated Gaussian noise fields characterized by zero averages and 
$\langle \zeta_i^\pm(\bm{r},t) \zeta_j^\pm(\bm{r}',t') \rangle =\delta_{ij} \delta^d(\bm{r}-\bm{r}') \delta(t-t')$.

We next shift our focus to the dynamics in terms of the number density $\mathcal{C} = C^+ + C^-$ and the charge density $\rho = C^+ - C^-$, for which the density equations can be recast as
\begin{align}   \label{eq:fullcontinuity}
    \partial_t \mathcal{C} + \nabla\cdot\bm{J}_c  = 0,
    \qquad \text{and} \qquad 
    \partial_t \rho + \nabla\cdot\bm{J}_\rho = 0,
\end{align}
where the density and charge currents are defined as  
$\bm{J}_c = \bm{J}^+ + \bm{J}^-$ 
and 
$\bm{J}_\rho = \bm{J}^+ - \bm{J}^-$ 
and they explicitly read
\begin{align}    \label{eq:Jcrhofull}   
    \bm{J}_c = - D\nabla \mathcal{C} + \rho \mu Q ( -\nabla \phi + \bm{E}) - \sqrt{2D \mathcal{C}} \bm{\zeta}_c, 
    \qquad \qquad
    \bm{J}_\rho = - D\nabla\rho + \mathcal{C} \mu Q(-\nabla\phi+\bm{E}) - \sqrt{2D \mathcal{C}}\bm{\zeta}_\rho. 
\end{align}
Note that the noise fields $\sqrt{2D \mathcal{C}} \bm{\zeta}_{c,\rho}$ are obtained from the addition and subtraction of the Gaussian noise fields $\sqrt{2D C^\pm}\bm{\zeta^\pm}$; consequently, these uncorrelated fields also have zero averages with correlations given by
$$\langle \zeta_{\rho i}(\bm{r},t) \zeta_{\rho j}(\bm{r}',t') \rangle =
\langle \zeta_{c i}(\bm{r},t) \zeta_{c j}(\bm{r}',t') \rangle =
\delta_{ij} \delta^d(\bm{r}-\bm{r}') \delta(t-t').$$ 
Substituting the density and charge currents $\bm{J}_c$ and $\bm{J}_\rho$ into the corresponding continuity equations, we find for the dynamics 
\begin{align}  
    \partial_t \mathcal{C}  =  \label{eq:fullC}
     D\nabla^2 \mathcal{C} + \mu Q \nabla\cdot (\rho\nabla\phi) - \mu Q E\partial_x\rho + \nabla\cdot\left( \sqrt{2D \mathcal{C}} \bm{\zeta}_c \right), \\
    \partial_t \rho = \label{eq:fullrho}
     D\nabla^2 \rho + \mu Q \nabla\cdot (\mathcal{C} \nabla\phi) - \mu Q E\partial_x \mathcal{C} + \nabla\cdot\left( \sqrt{2D \mathcal{C}} \bm{\zeta}_\rho \right),
\end{align}
where now the electric potential satisfies 
$-\nabla^2 \phi = (S_d Q/\epsilon_{\rm in}) \rho$. 
Note that the electric field introduces a source term $\propto \partial_x \rho$ to the dynamics of $\mathcal{C}$ and vice versa; this is the origin of the long-range behavior we will obtain later, since it leads to the modification of the effective density diffusion coefficient parallel to the electric field.

\subsection{Linearized density equations}   \label{subsec:linear}

As Equations~\eqref{eq:fullC} and \eqref{eq:fullrho} contain nonlinear terms and multiplicative stochasticity, they are in general difficult to analyze.  
It is thus customary 
to focus on the fluctuations of the densities fields about a uniform background baseline $C_0$  (i.e, $C^\pm = C_0 + \delta C^\pm$ with $|\delta C^\pm| \ll C_0$), and to only keep terms in leading order of the fluctuations~\cite{demery2014generalized,demery2016conductivity,zorkot2016power,zorkot2018nanopore,poncet2017universal}. 
A scaling analysis of the nonlinearities also shows that they become negligible in the macroscopic limit (see Appendix~\ref{app:scaling}). 
We define the density fluctuation field $c(\bm{r},t)$ and the charge fluctuation field (in units of $Q$) $\rho(\bm{r},t)$ as
\begin{align}
    c (\bm{r},t) = \delta C^+  + \delta C^- = C^+ + C^- - 2 C_0, 
    \qquad\qquad 
    \rho (\bm{r},t) = \delta C^+ - \delta C^- = C^+ - C^-. \label{eq:crhodef} 
\end{align}
Expanding Eqs.~\eqref{eq:fullC}~and~\eqref{eq:fullrho} and keeping the linear terms, one arrives at the following linear stochastic equations for the dynamics of $c$ and $\rho$: 
\begin{align}   
    \partial_t c &= 
    D\nabla^2 c -\mu Q E \partial_x \rho + \sqrt{4 D C_0} \eta_c, \label{eq:linc}\\
    \partial_t \rho &= 
    D\nabla^2 \rho  - \mu Q E \partial_x c - D\kappa^2 \rho + \sqrt{4 D C_0} \eta_\rho. \label{eq:linrho}
\end{align}
In these equations, we have defined the Debye length $\kappa^{-1}$ through
\begin{align}   \label{eq:kappadef}
    \kappa^2 = \frac{2 S_d C_0 Q^2}{\epsilon_{\rm in} k_B T},
\end{align}
and the linearized noise correlations now read 
$\langle \eta_\rho (\bm{r},t) \eta_\rho (\bm{r}',t') \rangle = \langle \eta_c (\bm{r},t) \eta_c (\bm{r}',t') \rangle = 
-\nabla^2 \delta^d (\bm{r}-\bm{r}')  \delta (t-t') $,
while 
$\eta_\rho$ and $\eta_c$ have zero averages and are uncorrelated.
Note that in the absence of the electric field ($E=0$), the linearized dynamics of the density fluctuations, Eq.~\eqref{eq:linc}, describes a normal diffusion process. 
On the other hand, with the electric field set to zero, the charge dynamics of Eq.~\eqref{eq:linrho} is a relaxational process as a result of the linear term $\propto D\kappa^2 \rho$ on the r.h.s, and it shows the exponential screening effects  beyond the Debye length $\kappa^{-1}$ and the Debye relaxation time $(D\kappa^2)^{-1}$.    

The linearized density and charge equations~\eqref{eq:linc}~and~\eqref{eq:linrho} will be used in Section~\ref{sec:corrfunc} to compute the correlation functions. This linear set of equations allows for the full characterization of the correlation functions in the Fourier space (see Appendices~\ref{appendix:exact-equal-time-corrs}~and~\ref{appendix:exact-outof-time-corrs}).  
However, since we are interested in the long-distance asymptotic limit, a convenient simplification can already be made by noting that for time and length scales beyond those set by the Debye screening processes, the charge fluctuation has a quasi-stationary solution given by 
\begin{align} \label{eq:rho(c)}
    \rho \approx - \kappa^{-2} \beta Q E \partial_x c,  
\end{align}
which is obtained by neglecting the temporal and spatial derivatives of the charge density 
\footnote{Note that the noise is also discarded at this level since it will only have short range contributions to the correlation, see Ref.~\cite{mahdisoltani2021long}}.   
Since the typical Debye screening length is of the order of $\kappa^{-1} \sim 1 - 10 \,\rm{nm}$ \cite{israelachvili}, such approximation is justified for studying the dynamics of the electrolyte beyond the screening scale. Accordingly, the corresponding FIFs that will be calculated below can in principle be realized in settings where the boundary separations are larger than the screening scale of the electrolyte which, for instance, may be the case for wet ion channels such as mechanosensitive channels~\cite{Martinac2004}, synthetic nanopores~\cite{Siwy2002}, and force measurement setups with large inter-plate separations~\cite{perez2019surface,stoneAC}.  

Equation~\eqref{eq:rho(c)} shows that the charge density which persists beyond the relaxation time $\sim 1/(D\kappa^2)$ is proportional to the gradient of the number density along the direction of the electric field.
This is in principle due to the bias introduced to the dynamics of the charged particles by the external field, which competes with the electrostatic forces that tend to relax any excess charge in the bulk of the electrolyte.  
Upon substituting Eq.~\eqref{eq:rho(c)} back into the number density dynamics Eq.~\eqref{eq:linc}, we arrive at  an \textit{anisotropic} diffusion equation that reads
\begin{equation}       \label{eq:anisodiffusion}
    \partial_t c = 
    D \left(\calE^2 \, \partial_x^2 + \nabla^2 \right) c
    + \sqrt{4 D C_0} \, \eta_c. 
\end{equation}
where the dimensionless electric field $\calE$ is defined as 
\begin{align}   \label{eq:calEdef}
    \calE = \frac{\beta Q E}{\kappa} = 
      \left[\frac{\epsilon_{\rm in} E^2 / (2 S_d)}{ k_{\rm B} T C_0}\right]^{1/2}. 
\end{align}
%
Equation~\eqref{eq:anisodiffusion} is the central result of this section and one of the main points of this work as it also underlies the long-range FIF that will be calculated later. 
Note that the Einstein relation $\mu = \beta D$ that holds between the mobility and the noise strength in the single particle Langevin dynamics  \eqref{eq:langevinparticle}, is no longer valid at the macroscopic level of Eq.~\eqref{eq:anisodiffusion}. 
This introduces a mismatch between the noisy fluctuations and the dissipative forces in the conserved dynamics of the density fluctuations, as a result of which the dynamics of $c$ represents a realization of generic scale invariance \cite{tauber,grinstein90conservation, garrido90conservative}. 
%

Finally, it is worth noting that $\calE$ can be regarded as the ratio of the energy density of the electric field ($\frac{\epsilon_{\rm in} E^2}{2 S_d}$) to the thermal energy density ($C_0 k_B T$) of the electrolyte, which is also the ratio between the typical values of the corresponding stress components.   
Alternatively, $\calE$ could be interpreted as the average distance traversed along the electric field by a charged particle during the charge relaxation process ($\text{velocity} \times \text{time} \sim (\mu Q E) \times (D\kappa^2)^{-1}$), divided by the Debye length ($\kappa^{-1}$) which determines the spatial extent of the counterion cloud around a cation or anion. 
As such, $\calE$ qualitatively encodes the extent of the deformation of the counterion atmospheres from their equilibrium (symmetric) forms as a result of the applied external field~\cite{onsagerlong1932}.

\subsection{Slowly varying electric fields} \label{subsec:slow-tdependence}

In the preceding analysis, we have assumed the electric field is DC, i.e., it is constant in time. However, maintaining a constant electric field in the bulk of an electrolyte is difficult in experimental setups, as  ions will quickly accumulate on the electrodes with opposite charges and screen their electric field to a short-ranged residual field acting only on the boundary layers. 
This naturally raises the question of whether the analysis that led to derivation of Eq.~\eqref{eq:anisodiffusion} can be extended to cases where the driving field is time-dependent? 
To answer this, we note that the main simplifying approximation that allowed us to obtain the anisotropic diffusion process, Eq.~\eqref{eq:anisodiffusion}, is the quasi-stationary relation between charge and density fluctuations  Eq.~\eqref{eq:rho(c)}.  
In physical terms, this approximation assumes the charge relaxation processes take place over time-scales that are much faster when compared to other dynamical scales. 
Such an approximation is therefore applicable to time-dependent electric fields as far as the variations in the field strength over the charge relaxation time is negligible; in that case, the ionic atmosphere surrounding each ion will have enough time to rearrange itself to the moving position of the central ion before the applied field changes considerably~\cite{wrightelectrolyte}. 
As the frequency set by the relaxation processes is often of order of 
few nano-seconds, the condition of slowly changing electric field is, in fact, often met in setups with oscillatory electric fields. 
In such cases, the anisotropic diffusion of the density fluctuation field $c(\bm{r},t)$ is therefore given by 
\begin{align}
    \partial_t c = D \left(\calE^2(t) \, \partial_x^2 + \nabla^2 \right) c + \sqrt{4D C_0} \, \eta_c,
\end{align}
with $\calE(t) = \beta Q E(t)/\kappa$. This equations shows that with a slowly varying electric field, the diffusion coefficient for density fluctuations along the direction of the field becomes a time-dependent factor. It is finally worth noting that as far as the time-scale for the variations of this diffusion coefficient is slower with respect to the charge relaxation processes, it can, in principle, be faster than other (diffusive) time-scales in the system.

\subsection{Unequal mobilities of cations and anions} 
\label{subsec:unequal-mobilities}

Another simplifying assumption in deriving Eq.~\eqref{eq:anisodiffusion} was the equal mobilities and diffusivities for the cations and anions in the electrolyte solution.  
However, often the cations and anions in an electrolyte have different ionic radii and as such their mobilities and diffusivities are not the same~\cite{israelachvili}. 
A crucial observation is that in oscillatory electric fields, such differences in the ionic mobilities lead to a mismatch in the spatial range of the motion of the ions within the solution which, in turn, gives rise to an average electric field in the bulk of the solution~\cite{amrei2018oscillating}.
In fact, this observation suggests a way of maintaining unscreened fields in the bulk of the electrolyte.   
Having this in mind, we now show that once an electric field is set in the bulk, a difference in the mobilities $\mu^\pm$ 
of the charge species would not change the macroscopic dynamics described by Eq.~\eqref{eq:anisodiffusion}. 

Let us consider the mobilities and diffusivities of the cations $(+)$ and anions $(-)$ to be given by 
\begin{align}
    D^\pm = D \pm \delta D, 
    \qquad \text{and} \qquad 
    \mu^\pm = \mu \pm \delta \mu,
\end{align}
and assume they are related through the Einstein relation 
$\mu^\pm = \beta D^\pm$ (note that $\delta D$ and $\delta \mu$ may be positive or negative and are not necessarily small). 
Following the same steps outlined in Section~\ref{subsec:Langevin}, and using the definition of the number density $\mathcal{C}=C^+ + C^-$ and the charge density  $\rho=C^+ - C^-$,  we obtain the modified version of the full number density and charge currents as 
\begin{align}
    \bm{J}_c = -D\nabla \mathcal{C} - \delta D \nabla\rho + \left[\rho \mu + \mathcal{C} \delta\mu\right] Q (-\nabla\phi+\bm{E}) 
     - \sqrt{2 \left[D \mathcal{C} + \delta D \rho\right] } \bm{\zeta}_c, \label{eq:Jcgeneralized}
    \\
    \bm{J}_\rho = - D\nabla\rho - \delta D \nabla \mathcal{C} + \left[\mathcal{C} \mu + \rho \delta \mu\right] Q (-\nabla\phi+\bm{E})  
     - \sqrt{2 \left[D \mathcal{C} + \delta D \rho\right] } \bm{\zeta}_\rho. \label{eq:Jrhogeneralized} 
\end{align}
A similar linearization procedure as in Section~\ref{subsec:linear} yields the dynamics of the fluctuations in the number and charge densities according to
\begin{align}
    \partial_t c &= D\nabla^2 c + \delta D \nabla^2 \rho - \delta D \kappa^2 \rho       
     - \mu Q E \partial_x \rho - \delta\mu Q E \partial_x c  - \sqrt{4DC_0}\eta_c, \label{eq:lincgeneral}
     \\
    \partial_t \rho &= D\nabla^2\rho + \delta D \nabla^2 c - D\kappa^2 \rho      
     -\mu Q E \partial_x c - \delta \mu Q E \partial_x \rho - \sqrt{4 D C_0} \eta_\rho. \label{eq:linrhogeneral}
\end{align}
It could be seen that the quasi-stationary approximation of Eq.~\eqref{eq:rho(c)} still holds for Eq.~\eqref{eq:linrhogeneral} since the added terms $\propto \delta D$ and $\propto \delta \mu$ with higher number of derivatives are not relevant at macroscopic scales (this can be seen more rigorously through a scaling analysis similar to Appendix~\ref{app:scaling}). 
Therefore to obtain the dynamics of the density fluctuations $c$, we substitute Eq.~\eqref{eq:rho(c)} into the modified density dynamics given by Eq.~\eqref{eq:lincgeneral}; 
noting that $\delta\mu / \delta D = \mu / D = \beta$, we recover the same anisotropic diffusion equation~\eqref{eq:anisodiffusion} derived for the case with equal mobilities and diffusivities 
(with the only modification being that the diffusion coefficient should now be replaced by $D= (D^+ + D^-)/2$). 

We conclude that as long as an electric field is set in the bulk of the electrolyte, the difference in the diffusivities of the cations and anions may only affect the microscopic dynamics of the particles, but it does not change the effective macroscopic equations governing the long-distance dynamics of the density and the charge fluctuations.

\section{Correlation Functions}  \label{sec:corrfunc}
We now turn to calculating the correlations of the density fluctuation field $c(\bm{r},t)$ and the charge fluctuation field  $\rho (\bm{r},t)$ after a quench in the electric field. 
In Section~\ref{subsec:bulkcorr}, we first compute the correlation functions of the density fluctuations in the bulk, using the appropriate Fourier representation of Eq.~\eqref{eq:anisodiffusion}, and then make use of Eq.~\eqref{eq:rho(c)} to obtain the bulk charge correlations. 
We then compute the correlations in the presence of neutral confining boundaries parallel with the electric field in Section~\ref{subsec:confinedcorr}. 
Our focus will be on the case that a constant (DC)  electric field is switched on at $t=0$, before which the electrolyte is assumed to have been at thermal equilibrium. 

\subsection{Bulk correlation functions} \label{subsec:bulkcorr}

Let us first derive the bulk correlations in the absence of any confining boundaries. The translational symmetry makes it convenient to work with the spatial Fourier transform of Eq.~\eqref{eq:anisodiffusion} which reads 
\begin{align}   \label{eq:Fourier-anisotropic}
    \partial_t c (\bm{k},t) = - D \left(\calE^2 k_x^2 + \bm{k}^2 \right) c (\bm{k},t) + \sqrt{4 D C_0} \eta_c (\bm{k},t).  
\end{align}
We use the Fourier convention 
$f(\bm{k}) = \int \dif^d \bm{r} \, e^{-i\bm{k}\cdot\bm{r}} f(\bm{r})$ with 
$\bm{r} = (y, \bm{s}) \in \mathbb{R}^d$ 
where $\bm{s}=(s_1=x, s_2, \ldots, s_{d-1}) \in \mathbb{R}^{d-1}$, 
and $\bm{k} = \bm{k}_s + k_y \hat{\bm{e}}_y$ with 
$\bm{k}_s=(k_{s_1}=k_x, k_{s_2},\ldots, k_{s_{d-1}})$.  
The noise correlation in the Fourier space reads $ \langle \eta_c (\bm{k},t) \eta_c(\bm{k}',t') \rangle = (2\pi)^d \bm{k}^2 \delta^d (\bm{k}+\bm{k}') \delta(t-t')$. 
Using an integration factor, Eq.~\ref{eq:Fourier-anisotropic} can be solved as 
\begin{align}   \label{eq:cFouriersln}
    c(\bm{k},t) = c(\bm{k},0) \, e^{-D(\calE^2 k_x^2 + \bm{k}^2) t} + 
    \sqrt{4 D C_0} \int_0^t \dif\tau \, 
    \eta_c (\bm{k},\tau) \,
    e^{-D(\calE^2 k_x^2 + \bm{k}^2) (t-\tau)}. 
\end{align}
Note that for a time-dependent electric field, this expression should be modified to
\begin{align}   \label{eq:csol-tdependentE}
    c(\bm{k},t) = 
    c(\bm{k},0) \, e^{-D k_x^2 \int_{0}^t \calE^2(u) \, \dif u } \,
    e^{-D \bm{k}^2 t} +
    \sqrt{4 D C_0} \int_0^t \dif\tau \, 
    \eta_c (\bm{k},\tau) \,
    e^{-D k_x^2 \int_{\tau}^t \calE^2(u) \, \dif u } \,
    e^{-D \bm{k}^2 (t-\tau)}.
\end{align}
In other words, the expression for the density fluctuations when the electric field is time-dependent can be obtained from that of the static electric field by making the substitution 
$\calE^2 \to (\int_{t_1}^{t_2} \calE^2(u) \, \dif u)/(t_2-t_1)$. For instance, for a periodic drivin field $\calE(t) = \calE_0 \cos(\Omega t)$, this substitution reads $\calE^2 \to (\calE_0^2/2)\left(1+ \sin(2\Omega t)/(2\Omega t) \right) $, which at long time scales simply reads $\calE^2 \to \calE_0^2/2$. 
Keeping in mind that this substitution is also be applicable to the results that will follow, from here on we will solely focus on the case of static electric fields.

Equipped with Eq.~\eqref{eq:cFouriersln}, it is now  straightforward to obtain the density correlations as
\begin{equation}    \label{eq:ccbulk-1IC}
\begin{split}
    \langle c(\bm{k},t) c(\bm{k}',t') \rangle_{\rm bulk} &=  
     c(\bm{k},0) c(\bm{k}',0) e^{-D(\calE^2 k_x^2 +\bm{k}^2)t } e^{-D ( \calE^2 {k'_x}^2 + {\bm{k}'}^2 t' )}  \\
    &+ (2\pi)^d \delta^d (\bm{k}+\bm{k}') \, 2C_0 
    \left[ e^{-D(\calE^2 k_x^2 +\bm{k}^2)(t'-t)} - e^{-D(\calE^2 k_x^2 +\bm{k}^2)(t'+t)} \right]\\
    &+ (2\pi)^d \delta^d (\bm{k}+\bm{k}') \frac{-2 C_0 \calE^2 k_x^2}{\calE^2 k_x^2 + \bm{k}^2}
    \left[ e^{-D(\calE^2 k_x^2 +\bm{k}^2)(t'-t)} - e^{-D(\calE^2 k_x^2 +\bm{k}^2)(t'+t)} \right]
\end{split}    
\end{equation}
where the averaging is performed over the noise realizations, and we have assumed $t'\geq t$ without loss of generality. 
The first line of Eq.~\eqref{eq:ccbulk-1IC} represents the decay of the initial conditions; the second line shows the diffusive propagation of the density fluctuations between two different times, as well as the establishment of the steady-state correlations; and the third line encodes both the steady-state and the transient correlations of the nonequilibrium fluctuations which vanish for $\calE=0$. 
Hereafter, we restrict the calculation to electric field quenches applied to electrolyte solutions which are initially in thermal equilibrium at $t=0$; performing an ensemble averaging over the thermal initial configurations (denoted by $\langle \ldots \rangle_{\rm IC}$) implies
\begin{align}
    \langle c(\bm{k},0) c(\bm{k}',0) \rangle_{\rm IC} = (2\pi)^d \delta^d(\bm{k}+\bm{k}') \, 2C_0. 
\end{align}
A similar averaging on Eq.~\eqref{eq:ccbulk-1IC} for $t=t'$ gives the density correlation functions at equal times according to 
\begin{align}   \label{eq:c2bulk-Fourier}
    \langle c(\bm{k},t) c(\bm{k}',t) \rangle_{\rm bulk, IC} &= 
    (2\pi)^d \delta^d (\bm{k}+\bm{k}') \Bigg[ 2C_0 
    \underbrace{ - \frac{2C_0 \calE^2 k_x^2}{\calE^2 k_x^2 + \bm{k}^2} \left(1-e^{-2D(\calE^2 k_x^2 + \bm{k}^2)t} \right)}_{\equiv c^{(2)}_{\mathrm{bulk}}(\bm{k},t)} \Bigg]
\end{align}
which also defines the distinct part of the bulk correlation function, $c^{(2)}_{\mathrm{bulk}}(\bm{k},t)$~\cite{hansen}. Note that the local part of the correlation function ($\propto C_0 \delta^d(\bm{k}+\bm{k}')$) represents the screened equilibrium correlations in the long distance limit (see Appendix~\ref{appendix:exact-equal-time-corrs} for the full expression without taking this limit). 
Transforming $c^{(2)}_{\rm bulk}(\bm{k},t)$ to the real space yields
\begin{equation}   \label{eq:c2bulk-real}
\begin{split}
    c^{(2)}_{\rm bulk} (\bm{r},t) = 
    &-\frac{2C_0 \calE^2 (1-d  \tilde{x}^2/\tilde{r}^2)}{ S_d(\calE^2+1)^{3/2} \tilde{r}^d}
    +  \int 
    \dif^d \tilde{\bm{r}}' \, 
    \frac{2C_0 \calE^2 (1-d  {\tilde{x}}^{\prime 2}/\tilde{r}^{\prime 2})}{ S_d(\calE^2+1)^2 \,  \tilde{r}^{\prime d}} \,
   \frac{\exp\left(-\frac{(\tilde{\bm{r}}-\tilde{\bm{r}}')^2 }{8 D t}\right)}{  (8\pi D t)^{d/2}  }, 
\end{split}    
\end{equation}
where $\tilde{\bm{r}}$ is obtained from $\bm{r}$ by substituting $x \to \tilde{x} = x / \sqrt{\calE^2+1}$. 
In Eq.~\eqref{eq:c2bulk-real}, the first term on the r.h.s gives the nonequilibirum steady-state correlation function which vanishes without the external electric field, and in $d$ spatial dimensions it decays as $\sim r^{-d}$ with distance; moreover, this term is manifestly anisotropic with a dipolar character~\cite{tauber}. 
The second term on the r.h.s, on the other hand, represents the transient effects with their long-time decay governed by the power-law tail $\sim t^{-d/2}$. 

Next, we turn to the charge fluctuation correlation functions. 
Incorporating the quasi-stationary charge profile of  Eq.~\eqref{eq:rho(c)}, one obtain the bulk charge correlations upon taking derivatives of Eq.~\eqref{eq:c2bulk-Fourier} as 
\begin{align}   \label{eq:rho2bulk-Fourier}
    \langle \rho(\bm{k},t) \rho(\bm{k}',t) \rangle_{\rm bulk, IC} &= (2\pi)^d \delta^d (\bm{k}+\bm{k}') \Bigg[ 2C_0 + \frac{2C_0 \calE^2 k_x^2}{\kappa^2}
     \underbrace{- \frac{2 C_0 \calE^4 k_x^4}{\kappa^2(\calE^2 k_x^2 + \bm{k}^2)} 
     \left( 1- e^{-2 D(\calE^2 k_x^2 + \bm{k}^2)t } \right)}_{\equiv \rho^{(2)}_{\rm bulk}(\bm{k},t)} \Bigg].  
\end{align}
Note that we have not included the second term in the brackets in $\rho^{(2)}$ since it originates from the density self-correlations in Eq.~\eqref{eq:c2bulk-Fourier}; the exact expressions provided in Appendix~\ref{appendix:exact-equal-time-corrs} reveal that this term represents the asymmetry in the screened correlation functions caused by the external field and is another local contribution to the charge density ($\propto \partial_x^2 \delta^d(\bm{r})$ in real-space). 
Such local terms will not contribute to the long distance behavior of the fluctuation-induced forces (Section~\ref{sec:stress}), and therefore will be discarded from the subsequent calculations. 
The real-space form of the non-local charge correlations $\rho^{(2)}_{\rm bulk}$ can also be obtained from Eq.~\eqref{eq:c2bulk-real} by taking derivatives with respect to the $x$ coordinate (i.e., along the electric field) in accordance with Eq.~\eqref{eq:rho(c)}; it is therefore seen that such charge fluctuations are also long-range correlated in the electrolyte solution.

Equations~\eqref{eq:c2bulk-Fourier}~and~\eqref{eq:rho2bulk-Fourier} give both the transient as well as the steady-state correlation functions, within the length and time scales where the approximation Eq.~\eqref{eq:rho(c)} holds. 
In Appendices~\ref{appendix:exact-equal-time-corrs}~and~\ref{appendix:exact-outof-time-corrs}, without making use of this approximation, we derive the density and charge two-point correlation at steady-state, for fluctuation field considered at equal times as well as at different times. 
One can go further and obtain the full correlations in the transient regimes by solving the stochastic dynamics of  Eqs.~\eqref{eq:linc}~and~\eqref{eq:linrho}, which in the matrix form read
\begin{align}
    \partial_t 
    \begin{pmatrix} c(\bm{k},t) \\ \rho(\bm{k},t) \end{pmatrix} =
    -\underbrace{
    \begin{pmatrix} D\bm{k}^2 & i\mu Q E k_x \\ 
    i \mu Q E k_x & D(\bm{k}^2+\kappa^2) \end{pmatrix} 
    }_{\equiv \mathcal{M}}
    \begin{pmatrix} c(\bm{k},t) \\ \rho(\bm{k},t) \end{pmatrix}
    + 
    \begin{pmatrix} \sqrt{4 D C_0}\, \eta_c(\bm{k},t) \\ \sqrt{4 D C_0} \, \eta_\rho(\bm{k},t) \end{pmatrix}. 
\end{align}
The formal solution of this matrix equation are given by 
\begin{align}
    \begin{pmatrix} c (\bm{k},t) \\ \rho (\bm{k},t) \end{pmatrix} =
    e^{-t \mathcal{M}} 
    \begin{pmatrix} c (\bm{k},0) \\ \rho (\bm{k},0) \end{pmatrix} +
    \int_0^t \dif s \, e^{ -(t-s)\mathcal{M} }
    \begin{pmatrix} \sqrt{4 D C_0}\, \eta_c (\bm{r},t) \\ \sqrt{4 D C_0} \, \eta_\rho (\bm{r},t) \end{pmatrix},
\end{align}
which then allows for computing the full correlation functions. These expressions are rather cumbersome and will not be evaluated here. Nevertheless, it is clear that the eigenvalues of $\mathcal{M}$ control the approach of the correlation functions toward their steady-state form. These eigenvalues read 
\begin{align}   \label{eq:eigenvals}
    \lambda_\pm = D(\bm{k}^2 + \frac{\kappa^2}{2}) \pm \frac{1}{2} \sqrt{ D^2 \kappa^4 - 4 \calE^2 D^2 \kappa^2 k_x^2}. 
\end{align}
The relative values of $\kappa$, $\calE$, and $k_x$ determine whether the eigenvalues have an imaginary part.
Two different dynamical behaviors are inferred from these eigenvalues: 
for $ k_x < \kappa/(2\calE) $, both eigenvalues remain real, and the full solution is the superposition of a (fast) decaying term with relaxation time $1/(D\kappa^2)$, and a soft diffusive mode. 
For $ k_x > \kappa /(2\calE)$, on the other hand, the eigenvalues also acquire an imaginary part; in this case, both contributions to the full solution are damped oscillatory with relaxation time $1/(D\kappa^2)$. 
(A similar behavior also appears in two-point correlation functions 
evaluated at different times, while 
the equal-time correlations at steady state  remain the same in both regimes, see Appendix~\ref{appendix:exact-outof-time-corrs}.)
These results imply that Eq.~\eqref{eq:anisodiffusion} captures the diffusive dynamics of the electrolyte at length scales beyond $2\calE/\kappa$, while for smaller length scales the dynamics is relaxational. 
This therefore introduces an additional scale for the dynamics of the solution should be taken into account together with the fact that the approximate charge profile given by Eq.~\eqref{eq:rho(c)}is  already restricted to scales beyond $\kappa^{-1}$.

\subsection{Correlation functions in flat confinement}   \label{subsec:confinedcorr}

We now turn to computing the density and charge correlation functions of the driven electrolyte in the presence of flat boundaries located at $y=0$ and $y=H$ (Fig.~\ref{fig:schematic}). %
The boundaries are assumed to be impenetrable (i.e., they are blocking electrodes), and therefore they impose no-flux Neumann boundary conditions on Eq.~\eqref{eq:anisodiffusion}, namely $\partial_y c |_{y=0} = \partial_y c|_{y=H} = 0$. 
One can construct the corresponding solutions that  satisfy these boundary conditions by making use of the Neumann eigenmodes $\cos(p_n y)$ with $p_n = \frac{n\pi}{H}$ through 
\footnote{A more involved approach is to use the method of images together with the real-space diffusive Green's functions, and then make use of the Poisson summation formula to represent the solution in terms of the Neumann modes \cite{barton1989}.}
\begin{align}
    c (\bm{r},t) =
     \sideset{}{'}\sum_{n=0}^\infty   \,\frac{2}{H} \cos(p_n y) \int \frac{d^{d-1}\bm{k}_s}{(2\pi)^{d-1}} \, 
     e^{i \bm{k}_s\cdot\bm{s}}  \, c_n (\bm{k}_s,t), 
\end{align}
where $\bm{s}=(s_1=x, s_2, \ldots,s_{d-1}) \in \mathbb{R}^{d-1}$ is the position along the boundary surfaces, 
$\bm{k}_s = (k_{s_1}=k_x, k_{s_2},\ldots,k_{s_{d-1}})$ is the corresponding momentum vector (with $k_s = |\bm{k}_s|$), 
and $\sum'_n$ indicates a summation where the $n=0$ term takes an additional factor of $1/2$. 
Performing the similar transformation on the noise term $\eta_c$, the correlation functions between the density modes $c_n (\bm{k}_s,t)$ are then obtained, which after simplification read
\begin{equation}
\begin{split}
    \langle c_n (\bm{k}_s,t) c_{n'} (\bm{k}'_s,t') \rangle &=   \, 
     (2\pi)^{d-1} \delta^{d-1}(\bm{k}_s+\bm{k}'_s) \, \delta_{n,n'} (1+\delta_{n,0})
     \\
     &\quad \times \frac{H C_0 (p_n^2 + k_s^2)}{\calE^2 k_x^2 + p_n^2 + k_s^2} 
     \left[ e^{-D(\calE^2 k_x^2 + p_n^2 +k_s^2)(t'-t)} - e^{-D(\calE^2 k_x^2 + p_n^2 +\bm{k}_s^2)(t'+t)} \right]. \nonumber
\end{split}
\end{equation}
This expression could be used in order to obtain the following partial Fourier transformation of the density correlation functions along the $\bm{s}$ coordinates:
\begin{equation}
\begin{split}
    &\langle c(y;\bm{k}_s;t) c(y';\bm{k}'_s;t') \rangle =
      \, (2\pi)^{d-1} \delta^{d-1}(\bm{k}_s+\bm{k}'_s)
     \\
    &\qquad\quad  \times \frac{4C_0}{H}\sideset{}{'}\sum_{n=0}^\infty  \cos(p_n y) \cos(p_n y') \, 
    \frac{ p_n^2 + k_s^2 }{\calE^2 k_x^2 + p_n^2 + k_s^2}
    \left[ e^{-D(\calE^2 k_x^2 + p_n^2 +k_s^2)(t'-t)} - e^{-D(\calE^2 k_x^2 + p_n^2 +k_s^2)(t'+t)} \right].
\end{split}    
\end{equation}
Narrowing down to correlations of density fluctuations at equal times, similar to the case of bulk correlations we define
%
    $\langle c(y;\bm{k}_s;t) c(y';\bm{k}'_s;t) \rangle = 
    (2\pi)^{d-1} \delta^{d-2}(\bm{k}_s + \bm{k}'_s)
     \left[ 2C_0 \delta(y-y') + c^{(2)} (y,y';\bm{k}_s;t) \right],$
%
and get 
\begin{equation}    \label{eq:c2confined}
\begin{split}
    c^{(2)} (y,y';\bm{k}_s;t) = 
     &- \frac{4 C_0}{H} \sideset{}{'}\sum_{n=0}^\infty  \cos(p_n y) \cos(p_n y') \, e^{-2 D (\calE^2 k_x^2 + p_n^2+k_s^2)t }  \\
     &\qquad -\frac{4 C_0 \calE^2 k_x^2 }{H}
    \sideset{}{'}\sum_{n=0}^\infty 
    \frac{\cos(p_n y) \cos(p_n y')}{\calE^2 k_x^2 + p_n^2 + k_s^2}  \left[1 - e^{-2D(\calE^2 k_x^2 +p_n^2 +k_s^2)t} \right].
\end{split}    
\end{equation}
Here, the first line is an equilibrium-like diffusive correlation which vanishes at long times, and the second line represents the nonequilibrium transients that in the steady-state give rise to a long-ranged term. 

Upon taking derivatives of Eq.~\eqref{eq:c2confined}, we get for the charge correlation functions at equal times
\begin{equation}    \label{eq:rho2confined}
\begin{split}
    \rho^{(2)} (y,y';\bm{k}_s;t) = 
    &\frac{4 C_0 \calE^2 k_x^2}{\kappa^2 H} \sideset{}{'}\sum_{n=0}^\infty  \cos(p_n y) \cos(p_n y') \left[1 - e^{-2 D (\calE^2 k_x^2 + p_n^2+k_s^2)t}\right]  \\
    & -\frac{4 C_0 \calE^4 k_x^4 }{\kappa^2 H}
    \sideset{}{'}\sum_{n=0}^\infty 
    \frac{\cos(p_n y) \cos(p_n y')}{\calE^2 k_x^2 + p_n^2 + k_s^2} \left[1 - e^{-2D(\calE^2 k_x^2 +p_n^2 +k_s^2)t} \right].
\end{split}    
\end{equation}
Note that the first line hear is due to the short-ranged screened correlations (since 
$\sideset{}{'}\sum\limits_{n=0}^\infty (2/H) \cos(p_n y) \cos(p_n y') = \delta(y-y')$), and the second line is the long-ranged nonequilibrium contribution. 

It is worth mentioning that the correlation functions for the two half-spaces out of the confined space (i.e., for $y<0$ and $y>H$) are obtained from Eqs.~\eqref{eq:c2confined}~and~\eqref{eq:rho2confined} by making the substitution 
$\frac{1}{H} \sideset{}{'} \sum_n g(p_n) \to \int_{-\infty}^\infty \frac{\dif p}{2\pi} g(p)$ 
where $p_n = \frac{n \pi}{H}$ and $g(p_n)$ stand for the summand in Eqs.~\eqref{eq:c2confined} and Eq.~\eqref{eq:rho2confined}. These bulk correlations will be used to compute the stress exerted from the electrolyte outside the confinement, which are then subtracted from the internal stress to obtain the net total stress on the boundaries.

\section{Stress tensor} \label{sec:stress}

In this section, we turn to calculating the stress exerted by the driven electrolyte on confining parallel boundaries.  
Since the system under consideration is out of thermal equilibrium due to the driving electric field, the stress or pressure cannot be calculated from thermodynamics. Based on the mechanical definition of stress, one can instead obtain stress formulas that also work in nonequilibrium conditions~\cite{irving1950hydro,kruger2018neqstress}; such procedure for electric forces yields the well-known (electrostatic) Maxwell stress~\cite{jackson,woodson1968electromechanical} which also  circumvents the ambiguity often faced in deriving the stress tensor from body forces (due to the freedom in inverting the divergence operator). 
In Section~\ref{sec:generalMax}, we derive a simplified formula for the noise-averaged Maxwell stress exerted by a general charge distribution confined between two parallel plates, assuming the charge distribution is invariant under translations along the boundaries.  
In Section~\ref{subsec:confinedstress} we implement the charge correlations given by Eq.~\eqref{eq:rho2confined} into this formulation to obtain the FIF, whose steady-state and transient parts are analyzed in Sections~\ref{subsec:steadystress}~and~\ref{subsec:transientstress}. 

\subsection{General expression for Maxwell stress in plane parallel geometry} \label{sec:generalMax}

For an electrostatic potential field $\phi$, the noise-averaged Maxwell stress tensor in $d$ spatial dimensions reads
\begin{align}   \label{eq:maxwellstress}
    \langle \sigma_{ij} \rangle= 
\frac{\epsilon_{\rm in}}{2 S_d} \Big( 2 \langle \nabla_i \phi \, \nabla_j \phi \rangle
- \delta_{ij} \langle (\nabla\phi)^2 \rangle \Big),
\end{align}
where the electric potential satisfies the Poisson equation
$-\nabla^2 \phi = S_d \rho Q/ \epsilon_{\rm in}$. With free boundary conditions, the solution to this Poisson equation is given by
$   \phi_{\rm free} (\bm{r},t) = 
    \frac{ Q}{\epsilon_{\rm in} (d-2)} \int \dif^{d} \bm{r}' \, \frac{\rho(\bm{r}',t) }{|\bm{r}-\bm{r}'|^{d-2}} $
%
(for $d=2$, the potential is given by a logarithmic Coulomb form, but the electric field relation and the Maxwell stress formula remain unchanged.) 
To proceed with Eq.~\eqref{eq:maxwellstress}, the correlation function of the electric potential $\phi (\bm{r},t)$ in confinement are needed. 
We thus first construct the solutions of the Poisson equation taking into account the electrostatic boundary conditions. Assuming the boundaries located at $y=0$~and~$H$ do not carry free charges, the electrostatic boundary conditions read 
\begin{equation}   \label{eq:electroBC}
\begin{split}
\epsilon_{\rm in} \partial_y \phi \big\rvert_{0^+} &= \epsilon_{\rm out} \partial_y \phi_{\rm out}\big\rvert_{0^-},
\qquad\qquad
\nabla_{\bm{s}} \phi \big\rvert_{0^+} = \nabla_{\bm{s}} \phi \big\rvert_{0^-}
\\
\epsilon_{\rm in} \partial_y \phi \big\rvert_{ H^-} &= \epsilon_{\rm out} \partial_y \phi_{\rm out}\big\rvert_{ H^+}, 
\qquad \qquad 
\nabla_{\bm{s}} \phi \big\rvert_{ H^-} = \nabla_{\bm{s}} \phi \big\rvert_{ H^+},
\end{split}
\end{equation}
%
where $\nabla_{\bm{s}} = \sum\limits_{i=1}^{d-1} \partial_{s_i} \hat{\bm{e}}_{s_i}$. 
The potential $\phi$ that satisfies these boundary conditions can be obtained using the method of electrostatic image charges~\cite{jackson}. 
We group the (infinite number of) image charges into two sets: image charges in group $\mathbb{I}_L$ are obtained from first imaging the source charge distribution in the left boundary ($y=0$), and then in the right boundary, and so on. For this group, the location of the image charges are
\begin{align}
    \mathbb{I}_L=
    &  \left\lbrace y_n =  (-1)^{n} \left( 2 \floor*{ \frac{n}{2} } H + y_{\rm source} \right); \quad 
    \bm{s}_n = \bm{s}_{\rm source} \quad 
    \bigg| n \in \mathbb{N} \right\rbrace. 
\end{align}
On the other hand, image charges in group $\mathbb{I}_R$ are obtained by first imaging the source charge distribution in the right boundary ($y=H$), and then in the left boundary and so on; in this case, the image locations are  given by 
\begin{align}
    \mathbb{I}_R=
    &  \left\lbrace y_n =  (-1)^{n+1} \left( 2 \ceil*{ \frac{n}{2} } H - y_{\rm source} \right); \quad \bm{s}_n = \bm{s}_{\rm source} \quad  
    \bigg| n \in \mathbb{N} \right\rbrace.   
\end{align}
(Note that $\floor*{\ldots}$ and $\ceil*{\ldots}$ are the floor and the ceiling functions, respectively·) 
For both groups, the $n$th image has electric charge $Q_n = \lambda^n  Q_{\rm src}$ where we have defined the dielectric contrast $\lambda$ between the electrolyte solution and the boundary material as ($\epsilon_{\rm in}$ and $\epsilon_{\rm out}$ are the permittivities of the solvent and the boundaries, respectively) 
\begin{align}
    \lambda = \frac{ \epsilon_{\rm in} - \epsilon_{\rm out}}{\epsilon_{\rm in} + \epsilon_{\rm out}},
\end{align}
which determines the electric charge ratio between successive images. The electric potential $\phi$ that satisfies the boundary conditions~\eqref{eq:electroBC} is then calculated as 
\begin{equation}    \label{eq:phiimage}
\begin{split}
    &\phi (y; \bm{k}_s; t) = 
     \frac{ Q}{\epsilon_{\rm in} (d-2)}
    \int_0^H \dif y' \, \rho ( y'; \bm{k}_s ;t) \frac{e^{-k_s|y-y'|}}{k_s}
    + \sum_{\mathbb{I}_L,\mathbb{I}_R}
    \frac{ Q_n}{\epsilon_{\rm in} (d-2)}
 \int_0^H \dif y' \, \rho ( y'; \bm{k}_s; t) \frac{e^{-k_s|y-y'_n|}}{k_s}, 
\end{split} 
\end{equation}
where the first term is the electric potential created directly by the source charge distribution, and we have made use of the translation invariance of the system  along the direction parallel to the surfaces (i.e., the $\bm{s}$ coordinates) by taking the Fourier transform of $\phi_{\rm free}$ with momentum $\bm{k}_s \in \mathbb{R}^{d-1}$ along them. 
The noise-averaged normal component of the Maxwell stress corresponding to this electric potential is then calculated as
\begin{align}
    \langle \sigma_{yy} ( \bm{r} , t  ) \rangle &=  
    \frac{Q^2 S_d }{8 \epsilon_{\rm in}} 
    \int_0^H \dif y' 
    \int_0^H \dif y''
    \int \frac{\dif^{d-1}\bm{k}_s}{(2\pi)^{d-1}} \,
    \rho^{(2)} (y',y'';\bm{k}_s ;t) \nonumber\\
    &\qquad \times 
    \Bigg[ 
    \sum_{n', n'' \in \mathbb{I}_L,\mathbb{I}_R} \lambda^{n'+n''}
     e^{-k_s |y-y'_{n'}|}  \,
     e^{-k_s |y-y''_{n''}|} \,
     \lbrace \mathrm{sgn} (y-y'_{n'}) \,
     \mathrm{sgn} (y-y''_{n''})-1 \rbrace
      \Bigg]. 
\end{align}
%

To calculate the (excess) stress at the location of the boundaries, we implement  Eq.~\eqref{eq:phiimage} and perform the summation over the image charges, noting that
\begin{align}
    \sum_{n=1}^\infty
    (\pm \lambda)^n \, e^{-b \floor*{\frac{n}{2}} }
    =\frac{\lambda ( \lambda \pm e^b )}{e^b - \lambda^2}, \qquad \text{and} \qquad 
    \sum_{n=1}^\infty
    (\pm \lambda)^n \, e^{-b \ceil*{\frac{n}{2}} }
    =\frac{ \lambda ( \lambda \pm 1)}{e^b - \lambda^2}.  
\end{align}
One can readily show that the stress at the location of the two plates are equal, and after some algebraic manipulation it can be written as
\begin{equation}        \label{eq:sigmayygen}
\begin{split} 
    \langle \sigma_{yy} (t) \rangle = 
     \frac{-\lambda Q^2 S_d}{2 \epsilon_{\rm in}}
    &\int_0^H \dif y' 
    \int_0^H \dif y''
    \int \frac{\dif^{d-1}\bm{k}_s}{(2\pi)^{d-1}} 
    \rho^{(2)} (y',y'';\bm{k}_s;t) \\
    &\times \left[ e^{-k_s y'} \frac{e^{2k_s H} }{e^{2k_s H} - \lambda^2} + e^{k_s y'} \frac{\lambda}{e^{2k_s H}-\lambda^2}\right] 
    \left[ e^{-k_s y''} \frac{e^{2k_s H} }{e^{2k_s H} - \lambda^2} + e^{k_s y''} \frac{\lambda}{e^{2k_s H}-\lambda^2}\right].  
\end{split}    
\end{equation}
Note that this expression holds for a generic charge distribution which is invariant with respect to translations along the $\bm{s}$ coordinates (i.e., parallel to the surfaces) and has the charge correlation function $\rho^{(2)} (y',y'';\bm{k}_s;t)$.  
%

\subsection{Stress exerted by the confined driven electrolyte } \label{subsec:confinedstress}

We now substitute the charge correlation function  Eq.~\eqref{eq:rho2confined} into the stress formula in Eq.~\eqref{eq:sigmayygen} and perform the integrations over $y'$ and $y''$ to obtain the stress for the specific settings at hand (see Fig.~\ref{fig:schematic}). 
Note that the resulting expressions contain formally divergent terms that will be removed upon subtracting  the bulk stress (which is necessary to obtain the net force acting on the boundaries). 
The net force per unit area of the right boundary is then obtained as
\begin{align}   \label{eq:F/S}
    \frac{F(t)}{S} = - \frac{k_B T}{H^d} \calE^4 \calA ( \calE,\lambda;t)  ,
\end{align}
with the dimensionless amplitudes $\calA$ defined as
\begin{equation}   \label{eq:calA-full}
\begin{split}
     \calA (\calE,\lambda;t) 
     =  \lambda \int \dif^{d-1} \bm{\nu}_s \,
     \Bigg\lbrace  
     &\sideset{}{'}\sum_{n=0}^\infty 
     \calR_n ( \lambda,\nu_s) \, 
     g (\calE,n,\bm{\nu}_s)
      \left[1 - e^{-2\pi^2 \tau (n^2 + \calE^2 \nu_x^2  +\nu_s^2)} \right] 
      \\ 
      &\qquad\qquad  
     - \int_0^\infty \dif n \, g (\calE,n,\bm{\nu}_s)  
     \left[1 - e^{-2\pi^2 \tau (n^2 + \calE^2 \nu_x^2  +\nu_s^2)} \right] \Bigg\rbrace. 
\end{split}     
\end{equation}
Here, $\tau = D t /H^2$ and we have also defined 
\begin{align}       \label{eq:calRcalY-def}
    \calR_n (\lambda, \nu_s) 
     =\begin{cases}
     &\left(\dfrac{e^{\pi\nu_s} - 1}{e^{\pi\nu_s} - \lambda} \right)^2 \equiv \calY_- (\lambda,\nu_s) \qquad n \,\,\, \text{even}, \\[.5cm]
     &\left(\dfrac{e^{\pi\nu_s} + 1}{e^{\pi\nu_s} + \lambda} \right)^2 \equiv \calY_+ (\lambda,\nu_s) \qquad n \,\,\, \text{odd},
     \end{cases}
\end{align}
 where $\lambda \calR_n$ represents the electrostatic response from the image charges (note that the amplitude vanishes for $\lambda=0$ where there is no such response). Moreover, we used the definition  
\begin{align}
    g (\calE,n,\bm{\nu}_s ) &= 
    \frac{  2^{1-d} \nu_x^4 \nu_s^2  }{(n^2+ \calE^2\nu_x^2 +  {\nu}_s^2)(n^2 + \nu_s^2 )^2} \, . 
\end{align}
where $\bm{\nu}_s=(\nu_{s_1}=\nu_x,\nu_{s_2},\ldots,\nu_{s_{d-1}}) \in \mathbb{R}^{d-1}$ is a dimensionless vector and $\nu_s = |\bm{\nu}_s|$. 

To analyze the time-dependent stress amplitude given by Eq.~\eqref{eq:calA-full}, it is convenient to separate its steady-state and transient parts as
\begin{align}
    \calA ( \calE, \lambda; t) = \calAs (\calE, \lambda) + \calAt (\calE,\lambda;t) 
\end{align}
where $\calAs$ denotes the long-time steady-state amplitude and is given by the general expression
\begin{align}       \label{eq:calAs-def}
    \calAs (\calE, \lambda) = 
    \lambda \int \dif^{d-1} \bm{\nu}_s \,
     \Bigg\lbrace  
     &\sideset{}{'}\sum_{n=0}^\infty 
     \calR_n ( \lambda,\nu_s) \, 
     g (\calE,n,\bm{\nu}_s)
     - \int_0^\infty \dif n \, g (\calE,n,\bm{\nu}_s)   \Bigg\rbrace,
\end{align}
and $\calAt$ is the transient part of the amplitude which reads
\begin{align}       \label{eq:calAt-def}
     \calAt (\calE,\lambda;t) 
     =  -\lambda \int \dif^{d-1} \bm{\nu}_s \,
     \Bigg\lbrace  
     &\sideset{}{'}\sum_{n=0}^\infty 
     \calR_n ( \lambda,\nu_s) \, 
     g (\calE,n,\bm{\nu}_s) \,
       e^{-2\pi^2 \tau (n^2 + \calE^2 \nu_x^2  +\nu_s^2)}  
     - \int_0^\infty \dif n \, g (\calE,n,\bm{\nu}_s)  \,
      e^{-2\pi^2 \tau (n^2 + \calE^2 \nu_x^2  +\nu_s^2)}  \Bigg\rbrace. 
\end{align}  
The transient part $\calAt$ determines the initial behavior of the force amplitude after the quench, but it  vanishes at long times. 
As will be shown in the following section, for $\tau \gg 1$, the temporal decay of $\calAt$ takes a power-law form (with possible crossover regimes) which corresponds to the long-time tails of diffusion processes.

\subsection{Steady-state stress amplitude}  \label{subsec:steadystress}

The steady-state properties of the stress and the amplitude were investigated in Ref.~\cite{mahdisoltani2021long}. Here, we present a brief account of the results.
Performing the summations involved in Eq.~\eqref{eq:calAs-def} and after some simplifications we arrive at the following expression for $\calA$ in $d$ spatial dimensions
\begin{equation}    \label{eq:calAs-fullexpand}
\begin{split}
    \calAs (\calE,\lambda) = \frac{\lambda}{2^{d-1}} &\int \dif^{d-1} \bm{\nu}_s \, \, 
    \frac{\pi \nu_s}{4\calE^4}
    \Bigg\lbrace  
     \Big(1- \frac{\calE^2 \cos^2\theta}{2} \Big) 
     \left[2 - \calY_-(\lambda,\nu_s) \,
    \coth(\frac{\pi\nu_s}{2})
    - \calY_+(\lambda,\nu_s) \,  
    \tanh(\frac{\pi\nu_s}{2}) \right]
    \\
    & \qquad\qquad - \frac{ 2 - \calY_-(\lambda,\nu_s)  \coth( \frac{\pi\nu_s}{2} \sqrt{\calE^2\cos^2\theta +1})  - \calY_+(\lambda,\nu_s)  \tanh(\frac{\pi\nu_s}{2}\sqrt{\calE^2 \cos^2\theta +1}) }{ \calE^4 \sqrt{\calE^2 \cos^2 \theta+1}} 
    \\
    &\qquad\qquad +\frac{1}{4} \pi \nu_s \calE^2 \cos^2\theta
    \Big[  \calY_-(\lambda,\nu_s) \, \mathrm{csch}^2 (\frac{\pi\nu_s}{2}) 
    - \calY_+(\lambda,\nu_s) \, \mathrm{sech}^2 (\frac{\pi\nu_s}{2}) \Big] 
    \Bigg\rbrace,
\end{split}    
\end{equation}
where $\nu_s \cos\theta = \nu_x$ and $\calY_\pm (\lambda,\nu_s)$ are defined in Eq.~\eqref{eq:calRcalY-def}. 
For $d=3$, the some integrations in Eq.~\eqref{eq:calAs-fullexpand} can be carried out, yielding  
\begin{equation}   \label{eq:As3dfull}
\begin{split}
    &\calAs (\calE,\lambda) = \frac{\lambda \zeta(3)}{16 \pi} \frac{\calE^2 + 2}{ \calE^4 (\calE^2+1)^{3/2}} 
    +\frac{\calE^2 - 4}{32\pi \calE^4} \Big[ (\lambda - \frac{1}{\lambda}) \, \Li_2(\lambda^2) +\frac{1}{2} (\lambda+\frac{1}{\lambda}) \, \Li_3(\lambda^2) \Big] + \frac{3 \,  \Li_3(\lambda^2)}{32\pi\calE^2} 
    \\
    &+ \frac{\lambda \pi}{16} \int_0^{2\pi} \dif\theta \int_0^\infty \nu_s^2 \dif \nu_s \frac{ \left[\calY_-(\lambda,\nu_s) -1 \right] \coth( \frac{\pi\nu_s}{2} \sqrt{\calE^2\cos^2\theta +1})  +  \left[\calY_+ (\lambda,\nu_s) -1\right] \tanh(\frac{\pi\nu_s}{2}\sqrt{\calE^2 \cos^2\theta +1})}{ \calE^4 \sqrt{\calE^2 \cos^2 \theta+1}}, 
\end{split}    
\end{equation}
where ${\Li}_n(z) = \sum_{k=1}^\infty \dfrac{z^k}{k^n}$ is the polylogarithm function. 
This expression can be evaluated by numerical methods,  and we also report simple asymptotic expressions for its limiting cases in Table.~\ref{tab:As-asympt} (for more details on the derivation of the asymptotic formulas and also similar analysis for the $d=2$ case, see the supplemental material of Ref.~\cite{mahdisoltani2021long}).

%
\begin{table}[t]
    \centering
    \begin{ruledtabular}
         \begin{tabular}{c  c  c}
          &   $\calE \ll 1$ 
          & $\calE \gg 1$     \\ \hline
         {$\lambda \!\ll \! 1$} \rule{0pt}{5ex}   
         & ${ \dfrac{9\left( 4 \zeta(5) + 1 \right)}{512 \pi} \lambda^2 \!+\!  \dfrac{9\left( 4 \zeta(5)-1 \right)}{1024 \pi } \lambda}$ 
         & $\dfrac{ \lambda (6\lambda-1) }{64 \pi}\calE^{-2}  $\\[.3cm]
         {$\lambda \!=\! 1$} 
         & $\dfrac{9 \, \zeta (3) }{64 \pi}$ 
         & $\dfrac{\zeta (3)  }{ 8 \pi } \calE^{-2} $ \\[.3cm]
          $\lambda \!=\! -1$ 
          & $\dfrac{9 \zeta (3) }{256 \pi}$ 
          & $\dfrac{\zeta (3)  }{16 \pi }\calE^{-2}$ \\
        \end{tabular}
        \end{ruledtabular}
        \caption{Asymptotic expressions for the  dimensionless steady-state stress amplitude $\calAs (\calE,\lambda)$ in $d=3$, obtained from expanding Eq.~\eqref{eq:As3dfull} 
        (corrections are $\mathcal{O}(\calE^2)$ for $\calE \ll 1$,  and $\mathcal{O}(\calE^{-4})$ for $\calE \gg 1$). 
        \label{tab:As-asympt}}
    \end{table} 

\subsection{Transient stress amplitude} \label{subsec:transientstress}

We now focus on the transient part of the stress amplitude as per Eq.~\eqref{eq:calAt-def}. First, we note that the initial rate of change of $\calAt$ is given by
\begin{align}      
    \frac{\partial \calAt}{\partial \tau} \Big\rvert_{\tau= 0 }
    &=
    2\pi^2\lambda \int \dif^{d-1}\bm{\nu}_s \,  
    \Bigg\lbrace
    \sideset{}{'} \sum_{n=0}^\infty \calR_n(\lambda,\nu_s) \frac{2^{1-d} \nu_s^6 \cos^4\theta}{(n^2+\nu_s^2)^2} - 
    \int_0^\infty \dif n \, \frac{2^{1-d} \nu_s^6 \cos^4\theta}{(n^2+\nu_s^2)^2}
    \Bigg\rbrace 
    \\
    &= \frac{ \pi^3 \lambda}{2^{d+1}}
    \int \dif^{d-1} \bm{\nu}_s \, 
    (\nu_s^3 \cos^4 \theta) 
    \Bigg\lbrace 
        \frac{\pi \nu_s}{2} 
        \Big[ \calY_-(\lambda,\nu_s) \, \csch^2 (\frac{\pi\nu_s}{2}) - \calY_+ (\lambda,\nu_s) \, \sech^2 (\frac{\pi\nu_s}{2}) \Big]  \label{eq:dtAt-def}
    \\
    &\hspace{5cm} - \Big[2 - \calY_- (\lambda,\nu_s) \coth(\frac{\pi\nu_s}{2}) - \calY_+(\lambda,\nu_s) \tanh(\frac{\pi\nu_s}{2}) \Big] 
    \Bigg\rbrace \nonumber
\end{align}
where we have used $\nu_s \cos\theta = \nu_x$. Note that this initial rate is independent of the electric field, and it is a function of the dielectric contrast $\lambda$ only. 
For $d=3$, carrying out the integrals in Eq.~\eqref{eq:dtAt-def} yields
\begin{align}   \label{eq:dtAtd3}
    \frac{\partial \calAt}{\partial \tau} \Big\rvert_{\tau=0} = 
    \frac{9}{64 \pi} \Bigg[ (\lambda-\frac{1}{\lambda}) \, \Li_4 (\lambda^2)
    + \Big( 5 + \frac{1}{2} (\lambda + \frac{1}{\lambda})\Big) \, \Li_5 (\lambda^2)  \Bigg],
\end{align}
which can be approximated for small dielectric contrasts to second order in $\lambda$ as
\begin{align}
    \frac{\partial \calAt}{\partial \tau} \Big\rvert_{\tau=0} = \frac{9\lambda ( -1 + 10 \lambda)}{128 \pi}  + \mathcal{O} (\lambda^3).
\end{align}
Since the total stress at $t=0$ vanishes, this shows that for $ 0 \lesssim \lambda \lesssim 0.1$ the stress amplitude initially decreases and becomes negative (i.e., the force between the plates is initially repulsive). Having calculated the initial rate of $\calAt$ and using the asymptotic expansions of $\calAs$ given in Table~\ref{tab:As-asympt} which determine the long-term behavior of the stress, one can see there are a number of different dynamical behaviors that the FIF exhibits
\begin{itemize}
    \item for $\lambda \lesssim -0.31$ and $\lambda \gtrsim 0.17$, both the initial slope $\partial_\tau \calA \big\rvert_{\tau=0}$ and the stready-state amplitude $\calAs$ are positive. 
    \item for $-0.31 \lesssim \lambda <0$, the initial slope of $\calAt$ is positive, and $\calAs$ is negative for weak electric fields while it becomes positive for strong fields.
    \item for $0<\lambda\lesssim 0.1$, the initial slope is negative, and $\calAs$ is positive for weak fields and negative for strong fields. 
    \item for $0.1 \lesssim \lambda \lesssim 0.17$, the initial slope is positive, and $\calAs$ is positive for small electric fields while it is negative for strong fields. 
\end{itemize}

To investigate how $\calAt$ decays at long times, we define $\tilde{\nu}_s = \sqrt{\tau} \nu_s$ upon which  Eq.~\eqref{eq:calAt-def} reads
\begin{equation}    \label{eq:Atscaled}
\begin{split}
    \calAt = \frac{-\lambda}{(\sqrt{4\tau})^{d-1}} \int \dif^{d-1} \tilde{\nu}_s \, &\Bigg\lbrace
    \sideset{}{'}\sum_{n=0}^\infty \left( \frac{e^{\pi \tilde{\nu}_s/\sqrt{\tau}} \pm 1}{e^{\pi \tilde{\nu}_s/\sqrt{\tau}} \pm \lambda}\right)^2 \, \frac{\tilde{\nu}_x^4 {\tilde{\nu}_s}^2 \, e^{-2\pi^2 (n^2 \tau + \calE^2 \tilde{\nu}_x^2 + \tilde{\nu}_s^2)} }{\left(n^2 \tau + \calE^2 \tilde{\nu}_x^2 + \tilde{\nu}_s^2 \right)(n^2 \tau +\tilde{\nu}_s^2)^2} 
    \\
    &\qquad - \int_0^\infty \frac{\dif \tilde{n}}{\sqrt{\tau}} \, 
    \frac{\tilde{\nu}_x^4 {\tilde{\nu}_s}^2 \, e^{-2\pi^2 (\tilde{n}^2  + \calE^2 \tilde{\nu}_x^2 + \tilde{\nu}_s^2)} }{\left(\tilde{n}^2 + \calE^2 \tilde{\nu}_x^2 + \tilde{\nu}_s^2 \right)(\tilde{n}^2 +\tilde{\nu}_s^2)^2} 
    \Bigg\rbrace.
\end{split}
\end{equation}
(In the second line we have defined $\tilde{n}=\sqrt{\tau} n$ for the integration variable.) 
Since the exponential factors suppress the integrands for large values of $\tilde{\nu}_s$, the final outcome of the integration is effectively determined by the behavior of the integrand for $\tilde{\nu}_s \sim \mathcal{O}(1)$. 
In addition, for $\tau \to \infty$, to leading order the summation is given by the $n=0$ term. 
We thus only keep this contribution and also expand the exponential factor $e^{\pi\tilde{\nu}_s/\sqrt{\tau}} $ to obtain
\begin{equation}    \label{eq:calAtscaleexpanded}
\begin{split}
    \lim_{\tau \to \infty} \calAt = \frac{-\lambda}{(\sqrt{4\tau})^{d-1}} \int \dif^{d-1} \tilde{\nu}_s \, &\Bigg\lbrace
    \frac{1}{2} \left( \frac{{\pi \tilde{\nu}_s/\sqrt{\tau}} }{1 + \pi \tilde{\nu}_s/\sqrt{\tau} - \lambda}\right)^2 \, \frac{\cos^4\theta  }{1+\calE^2 \cos^2 \theta } \, e^{-2\pi^2 \tilde{\nu}_s^2 ( 1 + \calE^2 \cos^2 \theta )}
    \\
    &\qquad - \frac{1}{\sqrt{\tau}}\int_0^\infty \dif \tilde{n} \, 
    \frac{ {\tilde{\nu}_s}^6 \cos^4\theta \, e^{-2\pi^2 \left(\tilde{n}^2 + \tilde{\nu}_s^2 (1+\calE^2 \cos^2\theta) \right)} }{\left(\tilde{n}^2 + \tilde{\nu}_s^2 (1+\calE^2 \cos^2\theta) \right)(\tilde{n}^2 +\tilde{\nu}_s^2)^2} 
    \Bigg\rbrace 
    + \mathcal{O}(\tau^{-1}). 
\end{split}
\end{equation}
From the second line, it becomes evident that the transient force amplitude due to the bulk electrolyte outside the boundaries decays as $\sim \tau^{-d/2}$; this is, in fact,  the usual power-law tail of the diffusion process in $d$ spatial dimensions. 
On the other hand, the temporal decay of the transient stress coming from the electrolyte confined between the plates exhibits two different regimes: 
for $\tau \lesssim \pi^2/(1-\lambda)^2$ the decay is governed by the power-law form $\sim \tau^{-(d-1)/2}$, 
while for $\tau \gtrsim \pi^2/(1-\lambda)^2$ it follows the form $\sim \tau^{-(d+1)/2}$. Since these expressions are obtained for $\tau \gg 1$, the first of these two regimes is only accessible when $1-\lambda \ll 1$. 
(However, note that $\calAt$ may also exhibit a sign change at times comparable to the crossover time which in principle can make it difficult to observe the crossover between the two regimes.) 

\begin{figure*}[t]
   \centering
  \includegraphics[width=0.99\textwidth,keepaspectratio]{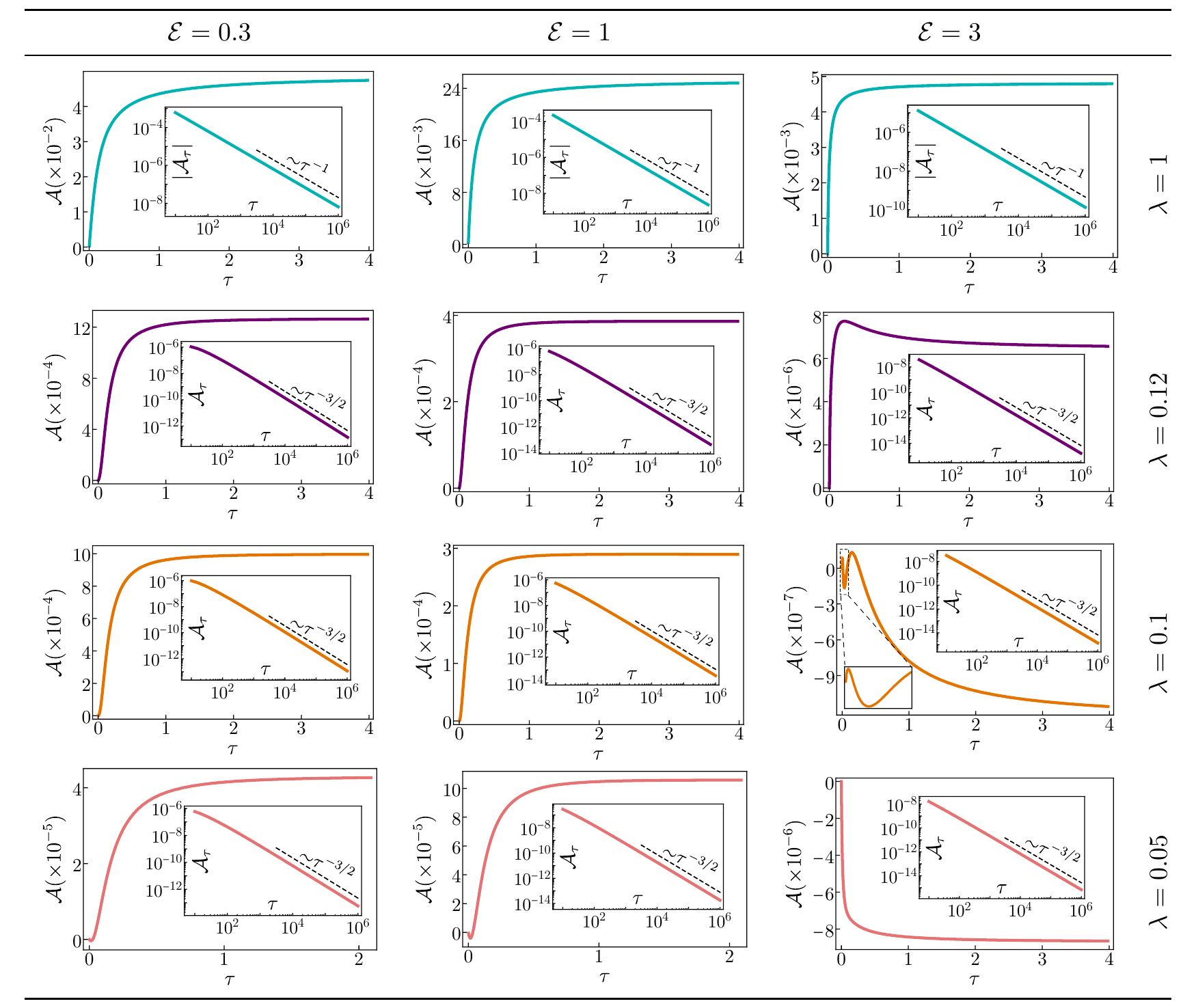}
    \caption{The full FIF amplitude $\calA = \calAs + \calAt$ as a function of the reduced time $\tau = \frac{D t}{H^2}$ for positive values of the dielectric contrast $\lambda$ and different values of the electric field strength $\calE$.  
    In each case, the inset shows the  transient part of the force amplitude,  $\calA_\tau$ (or its absolute value if  $\calA_\tau$ is negative),  as a function of $\tau$ in logarithmic scales. For $\lambda = 1$, the transient amplitude $\calAt$ decays to zero as $\tau^{-1}$; for other values of $\lambda$, the decay is governed by $\tau^{-3/2}$.  
    The long-time limit of the full amplitude $\calA$ corresponds to the FIF at the steady state and can change sign for $-0.17 \lesssim \lambda \lesssim 0.31$~\cite{mahdisoltani2021long}. Note that for $\lambda=0.12$, the sign change is not observable for $\calE=3$ since it requires very large values of the electric field strength.  
	} \label{fig:Aplots-positivelambda}
\end{figure*}

\begin{figure*}[t]
   \centering
  \includegraphics[width=0.99\textwidth,keepaspectratio]{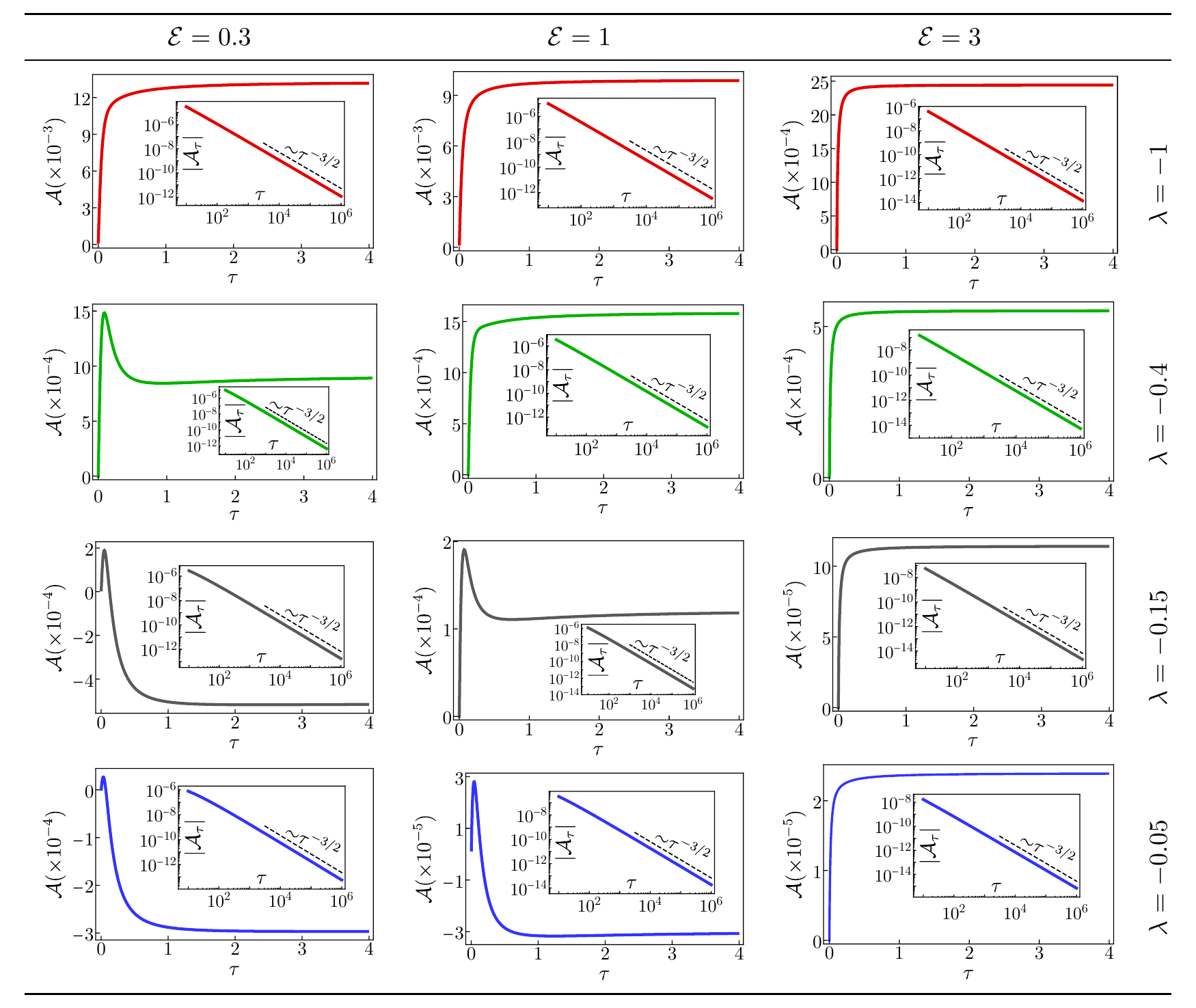}
     \caption{The full FIF amplitude $\calA = \calAs + \calAt$ as a function of the reduced time $\tau = \frac{D t}{H^2}$ for negative values of the dielectric contrast $\lambda$ and different values of the electric field strength $\calE$.  
    In each case, the inset shows the  transient part of the force amplitude,  $\calA_\tau$,  as a function of $\tau$ in logarithmic scales. In these cases, $\calAt$ is negative, and its magnitude decays to zero as $\tau^{-3/2}$.  
	} \label{fig:Aplots-negativelambda}
\end{figure*}

The above analysis shows that at the longest time scales, the decay of the FIF amplitude is governed by the diffusive tails of the bulk electrolyte outside the boundaries (and not those of the confined electrolyte between the plates). It is worth mentioning that in Eq.~\eqref{eq:calAtscaleexpanded}, the sign of this asymptotic long-time behavior is determined by $\lambda$. A comparison with the (sign of the) steady-state amplitude $\calAs$ shows that for values of dielectric contrast and with weak applied electric fields $\calE \ll 1$, the total amplitude~$\calA$ overshoots $\calA_s$ at a finite time before approaching it at long times. This phenomenon also happens for $\lambda \gtrsim 0.17$ with strong applied fields. 
In Figs.~\ref{fig:Aplots-positivelambda}~and~\ref{fig:Aplots-negativelambda}, the temporal variations of the full FIF amplitude $\calA$ and its transient part $\calAt$ as obtained from the numerical evaluation of Eqs.~\eqref{eq:calAs-def}~and~\eqref{eq:calAt-def} (and the corresponding expressions in $d=3$) are shown. One can observe that initial variations, long-time decays, and the sign (and magnitude) of the amplitude at steady states agree with the analysis we have presented in this section. 


\section{Concluding remarks} \label{sec:conc}

In this work, we studied the correlation functions and the fluctuation effects in a strong electrolyte in the transient regime after an electric field quench that drives the solution out of thermal equilibrium state. The density and charge  fluctuations are generally long-range correlated both in this transient period as well as in the long-time nonequilibrium steady state, as a result of the generic scale invariance of the stochastic dynamics. 
Such fluctuations, in turn, give rise to novel long-ranged forces on the confining boundaries. We analyze these forces as a function of the time elapsed from the electric field quench, which together with the steady-state result of Ref.~\cite{mahdisoltani2021long} provide a complete  account of the forces at different time scales. 
We find that the FIF scales with the plate separation as $H^{-d}$ in $d$ spatial dimensions, and in general it has a diffusive character in its approach toward the steady-state form (see Eqs.~\eqref{eq:F/S}~and~\eqref{eq:calA-full}).  
This diffusive approach gives rise to power-law temporal decays of the transient part of the force at long times. 
Figs.~\ref{fig:Aplots-positivelambda}~and~\ref{fig:Aplots-negativelambda} show the variety of temporal variations of the total force amplitude $\calA$ depending on the dielectric contrast $\lambda$ and the dimensionless electric field $\calE$, as depicted in Figs.~\ref{fig:Aplots-positivelambda}~and~\ref{fig:Aplots-negativelambda} for $d=3$. 
While the initial slope of the force amplitude is solely determined by $\lambda$ through Eq.~\eqref{eq:dtAt-def} (which reduces to Eq.~\eqref{eq:dtAtd3} for $d=3$), the long-time trend of the force is also affected by the strength of the electric field and is given by Eq.~\eqref{eq:calAs-fullexpand} (Eq.~\eqref{eq:As3dfull} in $d=3$ dimensions). 
The strength and the direction (i.e., attraction or repulsion) of the steady-state force can be controlled by the strength of the applied electric field. On the other hand, the early time temporal variations of the force amplitude can be non-monotonic and in some cases, it can result in changes in the sign of the force. 
These rich features point towards unexplored methods of  force manipulation in practical applications, for example to control neutral colloidal particles that are immersed in an electrolyte solution. 
Moreover, they resemble some of the experimentally observed temporal patterns of force variation in surface measurements~\cite{perez2019surface, stoneAC}; this implies  that fluctuation effects which are generally discarded in mean-field models can indeed be relevant to understanding the force generation mechanisms in charged solutions out of equilibrium. 

Although to derive the correlation functions, which form the basis of the force calculation, we have assumed a static external electric field, in Section~\ref{subsec:slow-tdependence} we argued these results are also applicable when the electric field varies slowly over time. Since the time-scale of the charge relaxation is often of order of a few nanoseconds, this condition is, in fact, met in many experiments where oscillatory electric fields are used. The long-time description employed here can then be used by making minimal modifications as described  below Eq.~\eqref{eq:csol-tdependentE}. 
It has been shown that an oscillatory driving field accompanied by different mobility coefficients for cations and anions can give rise to steady electric fields within the electrolyte~\cite{amrei2018oscillating}; however, once an electric field driving the ions in opposite directions is set up in the electrolyte, the analysis in Section~\ref{subsec:unequal-mobilities} shows that the long-distance asymptotic limit of the fluctuations is governed equations similar to the case of equal mobilities of charge species.   

The present work uncovers a novel dynamical mechanism for generating long-ranged forces in driven charged fluids which  had remained obscure within mean-field approaches. 
Although we have focused on a very simple setup here (constant electric field, flat boundaries, symmetric binary electrolyte), these results can in principle be extended to more complex settings since the underlying notion of generic scale invariance in anisotropic  conserved dynamics pertains to a wide range of driven charged systems. 
It would be particularly relevant for experimental setups to investigate how oscillatory electric fields that are perpendicular to the confining boundaries would change the FIF studied here (note that static fields will lead to charge accumulation on the electrodes which eventually screen the applied field withing a short distance).


\appendix

\section{Scaling Analysis of the Nonlinear Terms}  \label{app:scaling}

In this section, we look into the scaling analysis of the full stochastic equations of motion, 
and show that the nonlinearities are irrelevant in the macroscopic limit for $d>2$.  

The full (not linearized) Dean--Kawasaki stochastic equations of motion for density and charge are given by Eqs.~\eqref{eq:fullcontinuity}~and~\eqref{eq:Jcrhofull}. 
First, we determine the Gaussian scaling exponents through the linearized dynamics, Eqs.~\eqref{eq:linc}~and~\eqref{eq:linrho}. 
We consider the scaling of these equations according to 
$\bm{r} \to b \bm{r}, t \to b^z t, \rho \to b^{\chi_\rho} \rho, c \to b^{\chi_c} c, $
which yield 
\begin{align}
    \partial_t c &= 
    \left( b^{-2 + z} \right) D\nabla^2 c 
    - \left( b^{-1+z+\chi_\rho-\chi_c} \right) \mu  Q E \partial_x\rho
    + \left(b^{-1 + z/2 - d/2 - \chi_c}\right) \sqrt{4 D C_0} \eta_c, \label{app_eq:linc}\\
    \partial_t \rho &= 
    \left(b^{-2+z} \right) D\nabla^2 \rho 
    - \left( b^{-1+z-\chi_\rho+\chi_c} \right) \mu  Q E\partial_x c
    - \left(b^z\right) D\kappa^2 \rho 
    + \left(b^{-1 + z/2 - d/2 - \chi_\rho}\right) \sqrt{4 D C_0} \eta_\rho. \label{app_eq:linrho}
\end{align}
From the first equation, the mean-field exponents are obtained as
\begin{align}
    z=2, \qquad \chi_c = 1 + \chi_\rho = -d/2.
\end{align}
Note that the equation of motion for $\rho$ is clearly not scale-invariant since on the r.h.s of Eq.~\eqref{app_eq:linrho}, the second term ($\propto \mu Q E \partial_x c$) and the  third term ($\propto D \kappa^2 \rho$) scale as $b^z$, and therefore they grow under rescaling. 
This reflects the relaxational nature of the charge dynamics that underlies the bulk electroneutrality, and which we have also made use of to obtain the quasi-stationary solution for $\rho$ (Eq.~\eqref{eq:rho(c)}). 
Using these mean-field exponents, we can then examine the scaling behavior of the nonlinear terms in the dynamics of density fluctuations $c$. Note that the nonlinear term in Eq.~\eqref{eq:fullcontinuity} is 
$\mu Q \nabla\cdot(\rho\nabla\phi)$
which 
scales as 
$b^{z-\chi_c+2\chi_\rho} = b^{-d/2}$.
Since the corresponding scaling exponent is always negative, this nonlinearity is irrelevant in the renormalization group (RG) sense in the vicinity of the Gaussian fixed point. 

One may also consider other nonlinear terms that could potentially be incorporated into the dynamics of $c$, e.g., those generated through a coarse-graining procedure. 
The most general nonlinearity is given by $a_{lmnp} \nabla^l \rho^m c^n (\nabla\phi)^p$, where only gradients of the electric potential $\phi$ are considered since a constant shift in the potential shall not make any observable differences in the dynamics. There are some restriction on the form of this nonlinearity (see Ref.~\cite{mahdisoltani2021chemotaxis} for a similar analysis in the context of chemotaxis): 
as the dynamics is in essence a conservation law represented through a continuity equation, we should have $l\geq 1$; 
interaction terms should (at least) contain one field and thus $m+n+p \geq 1$, and they should also be scalars (and remain unchanged under spatial inversion), hence $l+p$ must be an even number; finally, we only consider local interactions (in terms of the fields) and therefore $m,n,p \geq 0$.  
The scaling dimension of such a term in the Langevin equation of $c$ is then given by: $[a_{lmnp}] = 2-l-m -\frac{d}{2}(m+n+p-1)$. 
By analyzing this expression, one finds that for spatial  dimensions larger than $d=2$, all possible nonlinear terms are irrelevant, i.e., they have negative scaling exponents. For the case of $d=2$, there is one marginal nonlinearity, namely $\nabla \cdot ( c \nabla \phi)$. 

The scaling analysis outlined above shows that the nonlinear effects can be discarded in the vicinity of the Gaussian fixed point for $d>2$. 
It remains for future studies to  use renormalization group (RG) analysis to examine the marginal term in the two-dimensional case, as well as the existence of other non-Gaussian fixed points in the phase space for which the scaling exponents differ from those of the linearized theory.

\section{Equal-time correlation functions} \label{appendix:exact-equal-time-corrs}

In this appendix, we provide a detailed computation of the density and charge two-point correlation functions at equal times, directly from the linearized stochastic dynamics of Eqs.~\eqref{eq:linc}~and~\eqref{eq:linrho}, both in the bulk as well as in flat confinement. 

\subsection{Bulk correlations}

Using the Fourier convention 
$f(\bm{r},t) = \int \frac{\dif^d \bm{k}}{(2\pi)^d} \frac{\dif\omega}{2\pi} e^{-i\omega t+i\bm{k}\cdot\bm{r}} f(\bm{k},\omega)$, the linearized  equations~\eqref{eq:linc} and \eqref{eq:linrho} can be expressed as follows
\begin{align}
    & c (\bm{k},\omega) = \frac{
     - i \mu Q E k_x \rho (\bm{k},\omega) + \sqrt{4 D C_0} \eta_c (\bm{k},\omega)}{-i\omega + D\bm{k}^2}, 
     \qquad \qquad 
     \rho (\bm{k},\omega) = \frac{- i \mu Q E k_x c(\bm{k},\omega) + \sqrt{4 D C_0} \eta_\rho (\bm{k},\omega)}{-i\omega + D (\bm{k}^2 + \kappa^2 )}.
\end{align}
These coupled equations can directly be solved for $c(\bm{k},\omega)$ and $\rho(\bm{k},\omega)$ in terms of the noise fields $\eta_c$ and $\eta_\rho$, from which the correlation functions read
%
\begin{align}
    &\left\langle c(\bm{k},\omega) c(\bm{k}',\omega') \right\rangle_{\rm bulk} = (2\pi)^{d+1} \delta^d ( \bm{k}+\bm{k}') \delta(\omega+\omega') \label{app_eq:cckomega}\\
    &\quad \times \frac{ \left(4 D C_0 \bm{k}^2\right)  \left(\omega^2 + \left( D(\bm{k}^2+\kappa^2) \right)^2 + \mu^2 Q^2 E^2 k_x^2 \right) }{ \Big( (-i\omega+D\bm{k}^2)(-i\omega+D(\bm{k}^2+\kappa^2))+\mu^2Q^2E^2k_x^2 \Big)
    \Big( (i\omega+D\bm{k}^2)(i\omega+D(\bm{k}^2+\kappa^2))+\mu^2Q^2E^2k_x^2 \Big)}, \nonumber 
    \\
    &\left\langle \rho(\bm{k},\omega) \rho(\bm{k}',\omega') \right\rangle_{\rm bulk} = (2\pi)^{d+1} \delta^d ( \bm{k}+\bm{k}') \delta(\omega+\omega') \label{app_eq:rhorhokomega}\\
    &\quad \times \frac{ \left(4 D C_0 \bm{k}^2\right)  \left(\omega^2 + \left( D\bm{k}^2 \right)^2 + \mu^2 Q^2 E^2 k_x^2 \right) }{ \Big( (-i\omega+D\bm{k}^2)(-i\omega+D(\bm{k}^2+\kappa^2))+\mu^2Q^2E^2k_x^2 \Big)
    \Big( (i\omega+D\bm{k}^2)(i\omega+D(\bm{k}^2+\kappa^2))+\mu^2Q^2E^2k_x^2 \Big)}. \nonumber
\end{align}
To compute the equal-time correlation functions, we need to perform the frequency integrations. For the density correlations, these can be expressed in the following form
\begin{align}
    \langle c(\bm{k},t) c(\bm{k}',t) \rangle_{\rm bulk} = (2\pi)^d \delta(\bm{k}+\bm{k}') \left[ 4 D C_0 \bm{k}^2 \int \frac{\dif \omega}{2\pi} \frac{\omega^2 + \alpha}{(\omega-\omega_{u}^+)(\omega-\omega_{u}^-)(\omega-\omega_{l}^+)(\omega-\omega_{l}^-)} \right], 
\end{align}
where $\alpha = D^2 (\bm{k}^2 + \kappa^2)^2 + \calE^2 D^2 \kappa^2 k_x^2$ (recall that $\calE = \mu Q E / (D\kappa)$), and  $\omega^\pm_{u}$ and $\omega^\pm_{l}$  are defined as
\begin{equation}   \label{eq:omegaul-def}
\begin{split}
   \omega_{u}^\pm &= i \lambda_\pm  = i D (\bm{k}^2 + \frac{\kappa^2}{2})
   \pm i \frac{D \kappa^2}{2} \sqrt{ 1 -  \frac{4 \calE^2 k_x^2}{\kappa^2} } \equiv i A \pm i B,  \\
    \omega_{l}^\pm &=  -i \lambda_\pm = - i D (\bm{k}^2 + \frac{\kappa^2}{2})
   \pm i \frac{D \kappa^2}{2} \sqrt{ 1 -  \frac{4 \calE^2 k_x^2}{\kappa^2} } \equiv -i A \pm i B,
\end{split}
\end{equation}
which represent the frequency poles in the upper half and lower half of the complex frequency plane, respectively. 
The frequency integration can be carried out, with the final result reading:
\begin{align}   \label{app_eq:ccfull}
    \langle c(\bm{k}) c(\bm{k}') \rangle_{\rm bulk} \equiv  \langle c(\bm{k},t) c(\bm{k}',t) \rangle_{\rm bulk}
    &= (2\pi)^d \delta^d(\bm{k}+\bm{k}') 
    \left[ 2C_0 - \frac{ 2C_0 \calE^2 \kappa^4 k_x^2}{ \left( 2\bm{k}^2 + \kappa^2 \right) \left(\bm{k}^2(\bm{k}^2+\kappa^2) + \calE^2 \kappa^2 k_x^2 \right)} \right].
\end{align}
Here, the first term in the bracket is independent of the applied electric field and represents the local correlations in equilibrium, while the second part vanishes for $\calE=0$ and is the nonequilibrium part of the density correlations. 
In the limit of $\bm{k}/\kappa \ll 1$, the above bulk density correlation is approximated by 
\begin{align}
    \langle c(\bm{k}) c(\bm{k}') \rangle_{\rm bulk} \approx (2\pi)^d \delta^d (\bm{k}+\bm{k}') 
    \left[2C_0 -\frac{2 C_0 \calE^2 k_x^2}{\bm{k}^2 + \calE^2 k_x^2} \right],
\end{align}
recovering Eq.~\eqref{eq:c2bulk-real} which was obtained using the quasi-stationary approximation of  Eq.~\eqref{eq:rho(c)}. 

For the charge correlations, a similar calculation yields 
\begin{equation}
\begin{split}
    \langle \rho(\bm{k}) \rho(\bm{k}') \rangle_{\rm bulk} &\equiv  \langle \rho(\bm{k},t) \rho(\bm{k}',t) \rangle_{\rm bulk} \\
    &= (2\pi)^d \delta(\bm{k}+\bm{k}') \left[ 
    \frac{2 C_0 (\bm{k}^2+\calE^2 k_x^2)}{\bm{k}^2+\kappa^2} - 
    \frac{2 C_0 \calE^2 k_x^2 \left( \bm{k}^4(3\kappa^2 + 2\bm{k}^2) + \calE^2 \kappa^2 k_x^2 (2\bm{k}^2 + \kappa^2) \right)}{(\bm{k}^2+\kappa^2)(2\bm{k}^2+\kappa^2)(\bm{k}^2(\bm{k}^2+\kappa^2)+\calE^2 \kappa^2 k_x^2)} \right] \label{app_eq:rhorhofull}
\end{split}    
\end{equation}
Similar to the density correlation functions, the first term in the brackets represents the short-ranged equilibrium correlations (note that in real space this gives the sum of a delta function and a screened-Coulomb (Yukawa-type) term which decays exponentially with Debye screening length  $\kappa^{-1}$); the second term, on the other hand, is the nonequilibrium part due to the external field, and it vanishes for a non-driven electrolyte (i.e., when $\calE=0$). 
For $\bm{k}/\kappa \ll 1$, expanding the charge correlation function yields
\begin{align}
    \langle \rho(\bm{k}) \rho(\bm{k}') \rangle_{\rm bulk} \approx
    (2\pi)^d \delta^d(\bm{k}+\bm{k}') 
    \Bigg[ 
        \frac{2 C_0 \bm{k}^2}{\kappa^2} + \frac{2 C_0 \calE^2 k_x^2}{\kappa^2} - \frac{2 C_0 \calE^4 k_x^4}{\kappa^2 ( \bm{k}^2 + \calE^2 k_x^2)}     
    \Bigg].
\end{align}
Note that this expression agrees with the long-distance limit of  Eq.~\eqref{eq:rho2bulk-Fourier} at late times. 

\subsection{Correlations in confinement}

To construct the solutions to Eqs.~\eqref{eq:linc}~and~\eqref{eq:linrho} in the presence of the no-flux boundaries at $y=0$ and $y=H$, we first need to identify the boundary conditions of $c$ and $\rho$. 
To this end, we consider the linearized version of the density and charge currents of Eq.~\eqref{eq:Jcrhofull}, which read
\begin{align}   
    \bm{j}_c &= - D\nabla c + \rho \mu Q \bm{E} - \sqrt{4DC_0} \bm{\zeta}_c, \label{eq_apx:jclin}
    \\
    \bm{j}_\rho &= - D\nabla\rho 
    - 2C_0 \mu Q \nabla\phi + (2C_0 + c) \mu Q \bm{E} - \sqrt{4DC_0}\bm{\zeta}_\rho.  \label{eq_apx:jrholin} 
\end{align}
The Neumann boundary conditions then read  $\hat{\bm{e}}_y\cdot \bm{j}_{c,\rho}  |_{y=0,H} = 0$. 
In the present setup where $\bm{E}=E\hat{\bm{e}}_x$
is perpendicular to the plates, the (unique) solutions to the Langevin dynamics can be obtained self-consistently using the ansatz 
$C^\pm = \sum_n C^\pm_n(\bm{s}) \cos(p_n y) $; the Poisson equation then implies that both $c$ and $\rho$ are superpositions of Neuman eigenfunctions.
Therefore, the density and charge profiles can be written as cosine series with coefficients given by 
\begin{align}
    c_n(\bm{k}_s;t) &= 
    \int_0^H \dif y \int \dif^{d-1} \bm{s} \, 
    e^{-i \bm{k}_s\cdot\bm{s}} \, \cos(p_n y) \, c(\bm{r},t),  \\
    \rho_n(\bm{k}_s;t) &= 
    \int_0^H \dif y \int \dif^{d-1} \bm{s} \, 
    e^{-i \bm{k}_s\cdot\bm{s}} \, \cos(p_n y) \, \rho (\bm{r},t).
\end{align}
Following the same steps as in the bulk calculation, it is  seen that $\langle c_n(\bm{k}_s) c_{n'} (\bm{k}'_s)\rangle$ 
and 
$\langle \rho_n (\bm{k}_s) \rho_{n'}(\bm{k}'_s)\rangle$ can be obtained from the bulk correlations, Eqs.~\eqref{app_eq:ccfull}~and~\eqref{app_eq:rhorhofull}, by making the following substitutions 
\begin{align}
    \bm{k}^2 \to p_n^2 + \bm{k}_s^2, 
    \qquad \text{and} \qquad 
    (2\pi)^d  \delta^d(\bm{k}+\bm{k}') \to 
     (2\pi)^{d-1} \delta^{d-1}(\bm{k}_s + \bm{k}'_s)  \delta_{n,n'} (1+\delta_{n,0}) \frac{H}{2},
\end{align}
which yield
\begin{align}
    &\left\langle c_n(\bm{k}_s,\omega) c_{n'}(\bm{k}'_s,\omega') \right\rangle =
    (2\pi)^{d} \delta^{d-1} ( \bm{k}_s+\bm{k}'_s)\delta(\omega+\omega') \,  \delta_{n,n'}(1+\delta_{n,0}) 
    2 D C_0 H\label{app_eq:ccnomega}\\
    &\qquad\qquad\qquad\qquad \times \frac{  (\bm{k}_s^2 + p_n^2)  \left(\omega^2 + \left( D(\bm{k}_s^2+p_n^2+\kappa^2) \right)^2 + \mu^2 Q^2 E^2 k_x^2 \right) }{ \Big( (-i\omega+D(\bm{k}_s^2+p_n^2))(-i\omega+D(\bm{k}_s^2+p_n^2+\kappa^2))+\mu^2Q^2E^2k_x^2 \Big)
    \Big( \omega \to -\omega \Big)}, \nonumber 
    \\
    &\left\langle \rho_n(\bm{k}_s,\omega) \rho_{n'}(\bm{k}'_s,\omega') \right\rangle= 
    (2\pi)^{d} \delta^{d-1} ( \bm{k}_s+\bm{k}'_s)  \delta(\omega+\omega') \, \delta_{n,n'}(1+\delta_{n,0})\,
    2 D C_0 H \label{app_eq:rhorhonomega}\\
    &\qquad\qquad\qquad\qquad \times \frac{  (\bm{k}_s^2 + p_n^2)  \left(\omega^2 + \left( D(\bm{k}_s^2+p_n^2) \right)^2 + \mu^2 Q^2 E^2 k_x^2 \right) }{ \Big( (-i\omega+D(\bm{k}_s^2+p_n^2))(-i\omega+D(\bm{k}_s^2+p_n^2+\kappa^2))+\mu^2Q^2E^2k_x^2 \Big)
    \Big( \omega \to -\omega \Big)}, \nonumber
\end{align}
The equal-time correlation functions can then be obtained by integrating over the frequencies, and they read
\begin{align}
    \langle c_n(\bm{k}_s) c_{n'}(\bm{k}'_s) \rangle 
    &= 
    (2\pi)^{d-1} \delta^{d-1} (\bm{k}_s+\bm{k}'_s) \delta_{n,n'}(1+\delta_{n,0}) \, C_0 H  \\
    &\qquad\quad \times
    \left[ 1 - \frac{ \calE^2 \kappa^4 k_x^2}{ \left( 2(\bm{k}_s^2 + p_n^2) + \kappa^2 \right) \left((\bm{k}_s^2+p_n^2)(\bm{k}_s^2+p_n^2+\kappa^2) + \calE^2 \kappa^2 k_x^2 \right)} \right], \nonumber 
    \\
    \langle \rho_n(\bm{k}_s) \rho_{n'}(\bm{k}_s') \rangle
    &=  (2\pi)^{d-1} \delta^{d-1}(\bm{k}_s+\bm{k}'_s)\delta_{n,n'}(1+\delta_{n,0}) \, C_0 H \\
    &\times \left[ 
    \frac{ \bm{k}_s^2+p_n^2+\calE^2 k_x^2}{\bm{k}_s^2+p_n^2+\kappa^2} - 
    \frac{ \calE^2 k_x^2 \left( (\bm{k}_s^2+p_n^2)^2 (3\kappa^2 + 2(\bm{k}_s^2+p_n^2)) + \calE^2 \kappa^2 k_x^2 (2(\bm{k}_s^2+p_n^2) + \kappa^2) \right)}{(\bm{k}_s^2+p_n^2+\kappa^2)(2(\bm{k}_s^2+p_n^2)+\kappa^2)((\bm{k}_s^2+p_n^2)(\bm{k}_s^2+p_n^2+\kappa^2)+\calE^2 \kappa^2 k_x^2)} \right].  \nonumber
\end{align}
Finally, the $y$-dependent correlation functions are obtained from these expression via 
\begin{align}
    \langle c(\bm{k}_s,y) c(\bm{k}'_s,y') \rangle = 
    \frac{4}{H^2}
    \sideset{}{'}\sum_{n=0}^\infty 
    \sideset{}{'}\sum_{n'=0}^\infty 
    \cos(p_n y) \cos(p_{n'} y') 
    \langle c_n(\bm{k}_s) c_{n'}(\bm{k}'_s) \rangle,
\end{align}
with a similar expression for the charge correlations.

\section{Steady-state bulk correlation functions at different times}  \label{appendix:exact-outof-time-corrs}

In this appendix, we extend the computation of the equal-time correlations to obtain the steady-state dynamics factors, which from Eq.~\eqref{app_eq:cckomega} can be evaluated via 
\begin{align}
    \lim_{t_0 \to \infty} \langle c(\bm{k},t_0 + t) c(\bm{k}',t_0) \rangle = (2\pi)^d \delta^d(\bm{k}+\bm{k}') \left[ 4 D C_0 \bm{k}^2 \int \frac{\dif \omega}{2\pi} \frac{e^{-i\omega t} \left(\omega^2 + \alpha\right) }{(\omega-\omega_{u}^+)(\omega-\omega_{u}^-)(\omega-\omega_{l}^+)(\omega-\omega_{l}^-)} \right],
\end{align}
where $\omega^\pm_u$ and $\omega^\pm_l$ are defined in Eqs~\eqref{eq:omegaul-def}. 
After carrying out the frequency integration gives
\begin{align}
    \lim_{t_0 \to \infty} \langle c(\bm{k},t_0+t) c(\bm{k}',t_0) \rangle &= 
    (2\pi)^d \delta^d(\bm{k}+\bm{k}') \,  D C_0 \bm{k}^2 \, \frac{e^{-A t} }{2 A B} \left[ e^{B t} \frac{-(A-B)^2 + \alpha}{A-B} -e^{-Bt} \frac{-(A+B)^2 + \alpha}{A+B} \right].  
\end{align}

In order to analyze this expression, we first consider the case of $ k_x < \kappa /(2\calE)$ for which $A=D(\bm{k}^2+\kappa^2/2)$ and $B=( D \kappa^2/2) \sqrt{1-4\calE^2 k_x^2/\kappa^2} $. Substituting these into the expression for density correlations, we obtain
\begin{equation}    \label{eq:ccdynamic-smallkx}
\begin{split} 
    \lim_{t_0 \to \infty} \langle c(\bm{k},t_0+t) &c(\bm{k}',t_0) \rangle = (2\pi)^d \delta^d(\bm{k}+\bm{k}') \frac{ C_0 \bm{k}^2 \, e^{-tD(\bm{k}^2 + \kappa^2/2)}}{\bm{k}^2 (\bm{k}^2+\kappa^2)+ \calE^2\kappa^2 k_x^2 }\\
    &\quad \times \Bigg[ 
    \exp\left\lbrace \frac{t D\kappa^2}{2} \sqrt{1 - \frac{4\calE^2 k_x^2}{\kappa^2} } \right\rbrace
    \left( \bm{k}^2 + \kappa^2 + \frac{2\kappa^2 \calE^2 k_x^2}{2\bm{k}^2+\kappa^2} + \frac{\kappa(\bm{k}^2+\kappa^2)}{\sqrt{\kappa^2-4\calE^2 k_x^2}} \right) \\ 
    &\hspace{2.5cm}+  
    \exp\left\lbrace \frac{-t D\kappa^2}{2} \sqrt{1 - \frac{4\calE^2 k_x^2}{\kappa^2} } \right\rbrace
    \left( \bm{k}^2 + \kappa^2 + \frac{2\kappa^2 \calE^2 k_x^2}{2\bm{k}^2+\kappa^2} - \frac{\kappa(\bm{k}^2+\kappa^2)}{\sqrt{\kappa^2-4\calE^2 k_x^2}} \right)
    \Bigg]. 
\end{split}
\end{equation}
For $\bm{k}/\kappa \ll 1$, by performing a Taylor expansion  and setting the relaxation terms $\propto e^{-t D \kappa^2}$ to zero in the long time behavior we arrive at 
\begin{align}
    \lim_{t_0 \to \infty} \langle c(\bm{k},t_0+t) &c(\bm{k}',t_0) \rangle \approx (2\pi)^d \delta^d(\bm{k}+\bm{k}') \, 
    2 C_0 
    \left[ \frac{\bm{k}^2 e^{ -t D (\bm{k}^2 + \calE^2 k_x^2)}}{\bm{k}^2 + \calE^2 k_x^2} 
    + \mathcal{O}\left( \frac{ \bm{k}^2}{\kappa^2} \right) \right] ,
\end{align}
For $ k_x > \kappa/(2\calE)$, on the other hand,  we instead have $B= (D\kappa^2/2) \sqrt{4 \calE^2 k_x^2/\kappa^2 - 1}$. 
In this case, the density correlation function reads
\begin{align}   \label{eq:ccdynamic-largekx}
    &\lim_{t_0 \to \infty} \langle c(\bm{k},t_0+t) c(\bm{k}',t_0) \rangle = (2\pi)^d \delta^d(\bm{k}+\bm{k}') \frac{ 2 C_0 \bm{k}^2 \, e^{-tD(\bm{k}^2 + \frac{\kappa^2}{2})}}{\bm{k}^2 (\bm{k}^2+\kappa^2)+ \calE^2\kappa^2 k_x^2 } \\
    &\qquad \times \left[  \left( \bm{k}^2 + \kappa^2 + \frac{2\kappa^2 \calE^2 k_x^2}{2\bm{k}^2 + \kappa^2} \right)
    \cos( \frac{t D \kappa}{2} \sqrt{4\calE^2 k_x^2 - \kappa^2} )
    + \frac{\kappa(\bm{k}^2 + \kappa^2)}{\sqrt{4\calE^2 k_x^2 -\kappa^2}} \sin( \frac{t D \kappa}{2} \sqrt{4\calE^2 k_x^2 - \kappa^2} ) \right], \nonumber
\end{align}
which shows that for large momenta the density correlations decay exponentially in time due to the presence of the exponential factor $e^{-t D \kappa^2/2}$.
Finally, note that for $t \to 0$, both Eqs.~\eqref{eq:ccdynamic-smallkx}~and~\eqref{eq:ccdynamic-largekx} reproduce the equal-time correlations of  Eq.~\eqref{app_eq:ccfull}. 

A similar line of calculation can be carried out for the the out-of-time charge correlation functions. 
When $ k_x < \kappa/(2\calE)$, one obtains
\begin{equation}
\begin{split}
    \lim_{t_0 \to \infty} \left\langle \rho (\bm{k},t_0+t) \rho(\bm{k}',t_0) \right\rangle &= 
    (2\pi)^d \delta^d(\bm{k}+\bm{k}') 
     \, 
    \frac{ C_0 \bm{k}^2 e^{-tD(\bm{k}^2 + \kappa^2/2)}}{\bm{k}^2 (\bm{k}^2+\kappa^2)+ \calE^2\kappa^2 k_x^2 }\\
    &\qquad \times \Bigg[  \exp\left\lbrace \frac{t D\kappa^2}{2} \sqrt{1 - \frac{4\calE^2 k_x^2}{\kappa^2} } \right\rbrace 
    \left( \bm{k}^2  + \frac{2\kappa^2 \calE^2 k_x^2}{2\bm{k}^2+\kappa^2} - \frac{\kappa \bm{k}^2}{\sqrt{\kappa^2-4\calE^2 k_x^2}} \right) \\
    &\hspace{3cm} +  \exp\left\lbrace-\frac{t D\kappa^2}{2} \sqrt{1 - \frac{4\calE^2 k_x^2}{\kappa^2} } \right\rbrace \left( \bm{k}^2 + \frac{2\kappa^2 \calE^2 k_x^2}{2\bm{k}^2+\kappa^2} + \frac{\kappa^2\bm{k}^2}{\sqrt{\kappa^2-4\calE^2 k_x^2}} \right)
    \Bigg]. 
\end{split}    
\end{equation}
The macroscopic behavior of this correlation function is obtained by taking the hydrodynamics limit (i.e.,  $\bm{k}/\kappa \ll 1$); in this case, Taylor expansion gives
\begin{equation}
\begin{split}
    \lim_{t_0 \to \infty} \left\langle \rho (\bm{k},t_0+t) \rho(\bm{k}',t_0) \right\rangle \approx 
    (2\pi)^d \delta^d (\bm{k}+\bm{k}') \, 
    \frac{2 C_0 \bm{k}^2}{\kappa^2}
    &\Bigg[ \frac{ \calE^2 k_x^2 \exp\lbrace -t D (\bm{k}^2 + \calE^2 k_x^2) \rbrace}{ \bm{k}^2 + \calE^2 k_x^2}\\
    &\qquad\qquad +  \exp\lbrace-tD(\kappa^2 +\bm{k}^2 - \calE^2 k_x^2)\rbrace +\mathcal{O}\left( \frac{ \bm{k}^2}{\kappa^2} \right) \Bigg], 
\end{split}    
\end{equation}
which, for $t=0$, agrees with the same limit of Eq.~\eqref{app_eq:rhorhofull}. 


For $k_x > \kappa/(2\calE)$, the charge correlations are given by the following damped oscillatory expression:
\begin{align}
    &\lim_{t_0 \to \infty} \langle \rho (\bm{k},t_0+t) \rho (\bm{k}',t_0) \rangle = (2\pi)^d \delta^d(\bm{k}+\bm{k}') \frac{ 2 C_0 \bm{k}^2 \, e^{-tD(\bm{k}^2 + \frac{\kappa^2}{2})}}{\bm{k}^2 (\bm{k}^2+\kappa^2)+ \calE^2\kappa^2 k_x^2 } \\
    &\qquad \times \left[  \left( \bm{k}^2 + \frac{2\kappa^2 \calE^2 k_x^2}{2\bm{k}^2 + \kappa^2} \right)
    \cos( \frac{t D \kappa}{2} \sqrt{4\calE^2 k_x^2 - \kappa^2} )
    + \frac{\kappa \bm{k}^2 }{\sqrt{4\calE^2 k_x^2 -\kappa^2}} \sin( \frac{t D \kappa}{2} \sqrt{4\calE^2 k_x^2 - \kappa^2} ) \right], \nonumber
\end{align}

\section{Simplifying $\calAt$}  \label{app:simpleAt}

In this section we give the explicit result for the second term on the r.h.s. of Eq.~\eqref{eq:calAt-def} which can be used to facilitate  numerical computations: 
%
\begin{equation}
\begin{split}
    \int_0^\infty \dif n \, 
    g (\calE,n,\bm{\nu}_s) \, e^{-2\pi^2 \tau (n^2 + \calE^2 \nu_s^2 +\nu_s^2)} = \, 
    &\frac{\pi \nu_s }{2^d \calE^4 \sqrt{1+\calE^2 \cos^2\theta}} \, 
    \erfc \left( \nu_s \sqrt{2\pi^2 \tau (1+\calE^2\cos^2\theta)} \right) \\
    &-\frac{\pi\nu_s }{2^d \calE^4} \,
    e^{-2\pi^2 \tau \nu_s^2 \calE^2 \cos^2\theta}\, 
    \erfc(\nu_s \sqrt{2\pi^2\tau}) 
    \\
    &+ \frac{ \nu_s^2 \cos^2\theta}{2^d \calE^2} \, \sqrt{2\pi^3 \tau} \,
    e^{-2\pi^2 \tau \nu_s^2 ( 1+\calE^2 \cos^2\theta)} \\
    &- \pi \nu_s \cos^2\theta \left( \frac{4\pi^2 \nu_s^2 \tau-1}{2^{d+1} \calE^2}\right) \,
    e^{-2\pi^2 \tau \nu_s^2 \calE^2 \cos^2\theta} \,
    \erfc(\nu_s \sqrt{2\pi^2 \tau}),
\end{split}    
\end{equation}
where $\nu_x = \nu_s \cos\theta$, and $\erfc(z) = 1 - \frac{2}{\sqrt{\pi}} \int_0^z e^{-u^2} \, \dif u $ is the complementary error function.

\bibliography{Master_bib}

\begin{thebibliography}{79}%
\makeatletter
\providecommand \@ifxundefined [1]{%
 \@ifx{#1\undefined}
}%
\providecommand \@ifnum [1]{%
 \ifnum #1\expandafter \@firstoftwo
 \else \expandafter \@secondoftwo
 \fi
}%
\providecommand \@ifx [1]{%
 \ifx #1\expandafter \@firstoftwo
 \else \expandafter \@secondoftwo
 \fi
}%
\providecommand \natexlab [1]{#1}%
\providecommand \enquote  [1]{``#1''}%
\providecommand \bibnamefont  [1]{#1}%
\providecommand \bibfnamefont [1]{#1}%
\providecommand \citenamefont [1]{#1}%
\providecommand \href@noop [0]{\@secondoftwo}%
\providecommand \href [0]{\begingroup \@sanitize@url \@href}%
\providecommand \@href[1]{\@@startlink{#1}\@@href}%
\providecommand \@@href[1]{\endgroup#1\@@endlink}%
\providecommand \@sanitize@url [0]{\catcode `\\12\catcode `\$12\catcode
  `\&12\catcode `\#12\catcode `\^12\catcode `\_12\catcode `\%12\relax}%
\providecommand \@@startlink[1]{}%
\providecommand \@@endlink[0]{}%
\providecommand \url  [0]{\begingroup\@sanitize@url \@url }%
\providecommand \@url [1]{\endgroup\@href {#1}{\urlprefix }}%
\providecommand \urlprefix  [0]{URL }%
\providecommand \Eprint [0]{\href }%
\providecommand \doibase [0]{https://doi.org/}%
\providecommand \selectlanguage [0]{\@gobble}%
\providecommand \bibinfo  [0]{\@secondoftwo}%
\providecommand \bibfield  [0]{\@secondoftwo}%
\providecommand \translation [1]{[#1]}%
\providecommand \BibitemOpen [0]{}%
\providecommand \bibitemStop [0]{}%
\providecommand \bibitemNoStop [0]{.\EOS\space}%
\providecommand \EOS [0]{\spacefactor3000\relax}%
\providecommand \BibitemShut  [1]{\csname bibitem#1\endcsname}%
\let\auto@bib@innerbib\@empty
\bibitem [{\citenamefont {Israelachvili}(2011)}]{israelachvili}%
  \BibitemOpen
  \bibfield  {author} {\bibinfo {author} {\bibfnamefont {J.~N.}\ \bibnamefont
  {Israelachvili}},\ }\href@noop {} {\emph {\bibinfo {title} {Intermolecular
  and surface forces}}}\ (\bibinfo  {publisher} {Academic press},\ \bibinfo
  {year} {2011})\BibitemShut {NoStop}%
\bibitem [{\citenamefont {Oosawa}(1971)}]{oosawa1971}%
  \BibitemOpen
  \bibfield  {author} {\bibinfo {author} {\bibfnamefont {F.}~\bibnamefont
  {Oosawa}},\ }\href@noop {} {\emph {\bibinfo {title} {Polyelectrolytes}}}\
  (\bibinfo  {publisher} {Marcel Dekker: New York},\ \bibinfo {year}
  {1971})\BibitemShut {NoStop}%
\bibitem [{\citenamefont {Verwey}\ and\ \citenamefont
  {Overbeek}(1948)}]{verweyoverbeek1948}%
  \BibitemOpen
  \bibfield  {author} {\bibinfo {author} {\bibfnamefont {E.~J.~W.}\
  \bibnamefont {Verwey}}\ and\ \bibinfo {author} {\bibfnamefont {J.~T.~G.}\
  \bibnamefont {Overbeek}},\ }\href@noop {} {\emph {\bibinfo {title} {Theory of
  the Stability of Lyophobic Colloids}}}\ (\bibinfo  {publisher} {Elsevier,
  Amsterdam},\ \bibinfo {year} {1948})\BibitemShut {NoStop}%
\bibitem [{\citenamefont {Levin}(2002)}]{levin2002electrostatic}%
  \BibitemOpen
  \bibfield  {author} {\bibinfo {author} {\bibfnamefont {Y.}~\bibnamefont
  {Levin}},\ }\bibfield  {title} {\bibinfo {title} {Electrostatic correlations:
  from plasma to biology},\ }\href
  {https://iopscience.iop.org/article/10.1088/0034-4885/65/11/201/meta}
  {\bibfield  {journal} {\bibinfo  {journal} {Rep. Prog. Phys.}\ }\textbf
  {\bibinfo {volume} {65}},\ \bibinfo {pages} {1577} (\bibinfo {year}
  {2002})}\BibitemShut {NoStop}%
\bibitem [{\citenamefont {Kardar}\ and\ \citenamefont
  {Golestanian}(1999)}]{kardar99friction}%
  \BibitemOpen
  \bibfield  {author} {\bibinfo {author} {\bibfnamefont {M.}~\bibnamefont
  {Kardar}}\ and\ \bibinfo {author} {\bibfnamefont {R.}~\bibnamefont
  {Golestanian}},\ }\bibfield  {title} {\bibinfo {title} {The ``friction'' of
  vacuum, and other fluctuation-induced forces},\ }\href
  {https://doi.org/10.1103/RevModPhys.71.1233} {\bibfield  {journal} {\bibinfo
  {journal} {Rev. Mod. Phys.}\ }\textbf {\bibinfo {volume} {71}},\ \bibinfo
  {pages} {1233} (\bibinfo {year} {1999})}\BibitemShut {NoStop}%
\bibitem [{\citenamefont {Onsager}\ and\ \citenamefont
  {Fuoss}(1932)}]{onsagerlong1932}%
  \BibitemOpen
  \bibfield  {author} {\bibinfo {author} {\bibfnamefont {L.}~\bibnamefont
  {Onsager}}\ and\ \bibinfo {author} {\bibfnamefont {R.~M.}\ \bibnamefont
  {Fuoss}},\ }\bibfield  {title} {\bibinfo {title} {Irreversible processes in
  electrolytes. diffusion, conductance and viscous flow in arbitrary mixtures
  of strong electrolytes},\ }\href {https://doi.org/10.1021/j150341a001}
  {\bibfield  {journal} {\bibinfo  {journal} {J. Phys. Chem.}\ }\textbf
  {\bibinfo {volume} {36}},\ \bibinfo {pages} {2689} (\bibinfo {year}
  {1932})}\BibitemShut {NoStop}%
\bibitem [{\citenamefont {Kavokine}\ \emph {et~al.}(2020)\citenamefont
  {Kavokine}, \citenamefont {Netz},\ and\ \citenamefont
  {Bocquet}}]{nanofluidsreview}%
  \BibitemOpen
  \bibfield  {author} {\bibinfo {author} {\bibfnamefont {N.}~\bibnamefont
  {Kavokine}}, \bibinfo {author} {\bibfnamefont {R.~R.}\ \bibnamefont {Netz}},\
  and\ \bibinfo {author} {\bibfnamefont {L.}~\bibnamefont {Bocquet}},\
  }\bibfield  {title} {\bibinfo {title} {Fluids at the nanoscale: From
  continuum to subcontinuum transport},\ }\href
  {https://www.annualreviews.org/doi/abs/10.1146/annurev-fluid-071320-095958}
  {\bibfield  {journal} {\bibinfo  {journal} {Annu. Rev. Fluid Mech.}\ }\textbf
  {\bibinfo {volume} {53}} (\bibinfo {year} {2020})}\BibitemShut {NoStop}%
\bibitem [{\citenamefont {Bazant}\ \emph {et~al.}(2009)\citenamefont {Bazant},
  \citenamefont {Kilic}, \citenamefont {Storey},\ and\ \citenamefont
  {Ajdari}}]{bazant2009towards}%
  \BibitemOpen
  \bibfield  {author} {\bibinfo {author} {\bibfnamefont {M.~Z.}\ \bibnamefont
  {Bazant}}, \bibinfo {author} {\bibfnamefont {M.~S.}\ \bibnamefont {Kilic}},
  \bibinfo {author} {\bibfnamefont {B.~D.}\ \bibnamefont {Storey}},\ and\
  \bibinfo {author} {\bibfnamefont {A.}~\bibnamefont {Ajdari}},\ }\bibfield
  {title} {\bibinfo {title} {Towards an understanding of induced-charge
  electrokinetics at large applied voltages in concentrated solutions},\ }\href
  {https://www.sciencedirect.com/science/article/abs/pii/S000186860900092X}
  {\bibfield  {journal} {\bibinfo  {journal} {Adv. Colloid. Interface Sci.}\
  }\textbf {\bibinfo {volume} {152}},\ \bibinfo {pages} {48} (\bibinfo {year}
  {2009})}\BibitemShut {NoStop}%
\bibitem [{\citenamefont {Debye}\ and\ \citenamefont
  {H\"{u}ckel}(1923)}]{debye1923}%
  \BibitemOpen
  \bibfield  {author} {\bibinfo {author} {\bibfnamefont {P.}~\bibnamefont
  {Debye}}\ and\ \bibinfo {author} {\bibfnamefont {E.}~\bibnamefont
  {H\"{u}ckel}},\ }\bibfield  {title} {\bibinfo {title} {The theory of
  electrolytes i. the lowering of the freezing point and related occurrences},\
  }\href@noop {} {\bibfield  {journal} {\bibinfo  {journal} {Physikalische
  Zeitschrift}\ }\textbf {\bibinfo {volume} {24}},\ \bibinfo {pages} {185}
  (\bibinfo {year} {1923})}\BibitemShut {NoStop}%
\bibitem [{\citenamefont {Kosterlitz}\ and\ \citenamefont
  {Thouless}(1973)}]{kosterlitz}%
  \BibitemOpen
  \bibfield  {author} {\bibinfo {author} {\bibfnamefont {J.~M.}\ \bibnamefont
  {Kosterlitz}}\ and\ \bibinfo {author} {\bibfnamefont {D.~J.}\ \bibnamefont
  {Thouless}},\ }\bibfield  {title} {\bibinfo {title} {Ordering, metastability
  and phase transitions in two-dimensional systems},\ }\href
  {https://iopscience.iop.org/article/10.1088/0022-3719/6/7/010/meta?casa_token=1X7FEXVN-xAAAAAA:jDbzZaNAsXenZE9cFcHJl6Gi_sFFSchxvnb8CA0mRpYwOvSxRCMAsUIdR8V0JA3Jm3Z4GbQBqA}
  {\bibfield  {journal} {\bibinfo  {journal} {J. Phys. Part C Solid}\ }\textbf
  {\bibinfo {volume} {6}},\ \bibinfo {pages} {1181} (\bibinfo {year}
  {1973})}\BibitemShut {NoStop}%
\bibitem [{\citenamefont {Alexander}\ \emph {et~al.}(1984)\citenamefont
  {Alexander}, \citenamefont {Chaikin}, \citenamefont {Grant}, \citenamefont
  {Morales}, \citenamefont {Pincus},\ and\ \citenamefont
  {Hone}}]{alexander1984charge}%
  \BibitemOpen
  \bibfield  {author} {\bibinfo {author} {\bibfnamefont {S.}~\bibnamefont
  {Alexander}}, \bibinfo {author} {\bibfnamefont {P.}~\bibnamefont {Chaikin}},
  \bibinfo {author} {\bibfnamefont {P.}~\bibnamefont {Grant}}, \bibinfo
  {author} {\bibfnamefont {G.}~\bibnamefont {Morales}}, \bibinfo {author}
  {\bibfnamefont {P.}~\bibnamefont {Pincus}},\ and\ \bibinfo {author}
  {\bibfnamefont {D.}~\bibnamefont {Hone}},\ }\bibfield  {title} {\bibinfo
  {title} {Charge renormalization, osmotic pressure, and bulk modulus of
  colloidal crystals: Theory},\ }\href
  {https://aip.scitation.org/doi/abs/10.1063/1.446600?casa_token=yrB9PV-KoXYAAAAA:5vY9QdPJrG1m-_ryiH00QzkxKPfmWwWd-82AvOkqSJBxb8NswUtWJIbRoOo0wN4Kn9jnSPtqMgT6}
  {\bibfield  {journal} {\bibinfo  {journal} {J. Chem. Phys.}\ }\textbf
  {\bibinfo {volume} {80}},\ \bibinfo {pages} {5776} (\bibinfo {year}
  {1984})}\BibitemShut {NoStop}%
\bibitem [{\citenamefont {Grosberg}\ \emph {et~al.}(2002)\citenamefont
  {Grosberg}, \citenamefont {Nguyen},\ and\ \citenamefont
  {Shklovskii}}]{grosberg2002colloquium}%
  \BibitemOpen
  \bibfield  {author} {\bibinfo {author} {\bibfnamefont {A.~Y.}\ \bibnamefont
  {Grosberg}}, \bibinfo {author} {\bibfnamefont {T.}~\bibnamefont {Nguyen}},\
  and\ \bibinfo {author} {\bibfnamefont {B.}~\bibnamefont {Shklovskii}},\
  }\bibfield  {title} {\bibinfo {title} {Colloquium: the physics of charge
  inversion in chemical and biological systems},\ }\href
  {https://journals.aps.org/rmp/abstract/10.1103/RevModPhys.74.329} {\bibfield
  {journal} {\bibinfo  {journal} {Rev. Mod. Phys.}\ }\textbf {\bibinfo {volume}
  {74}},\ \bibinfo {pages} {329} (\bibinfo {year} {2002})}\BibitemShut
  {NoStop}%
\bibitem [{\citenamefont {Wong}\ and\ \citenamefont
  {Pollack}(2010)}]{wong2010electrostatics}%
  \BibitemOpen
  \bibfield  {author} {\bibinfo {author} {\bibfnamefont {G.~C.}\ \bibnamefont
  {Wong}}\ and\ \bibinfo {author} {\bibfnamefont {L.}~\bibnamefont {Pollack}},\
  }\bibfield  {title} {\bibinfo {title} {Electrostatics of strongly charged
  biological polymers: ion-mediated interactions and self-organization in
  nucleic acids and proteins},\ }\href
  {https://www.annualreviews.org/doi/full/10.1146/annurev.physchem.58.032806.104436}
  {\bibfield  {journal} {\bibinfo  {journal} {Annu. Rev. Phys. Chem.}\ }\textbf
  {\bibinfo {volume} {61}},\ \bibinfo {pages} {171} (\bibinfo {year}
  {2010})}\BibitemShut {NoStop}%
\bibitem [{\citenamefont {Manning}(1978)}]{manning1978molecular}%
  \BibitemOpen
  \bibfield  {author} {\bibinfo {author} {\bibfnamefont {G.~S.}\ \bibnamefont
  {Manning}},\ }\bibfield  {title} {\bibinfo {title} {The molecular theory of
  polyelectrolyte solutions with applications to the electrostatic properties
  of polynucleotides},\ }\href
  {https://www.cambridge.org/core/journals/quarterly-reviews-of-biophysics/article/molecular-theory-of-polyelectrolyte-solutions-with-applications-to-the-electrostatic-properties-of-polynucleotides/80B53BB944EEF69F16B84D0BAA1FB837}
  {\bibfield  {journal} {\bibinfo  {journal} {Q. Rev. Biophys.}\ }\textbf
  {\bibinfo {volume} {11}},\ \bibinfo {pages} {179} (\bibinfo {year}
  {1978})}\BibitemShut {NoStop}%
\bibitem [{\citenamefont {Zribi}\ \emph {et~al.}(2006)\citenamefont {Zribi},
  \citenamefont {Kyung}, \citenamefont {Golestanian}, \citenamefont
  {Liverpool},\ and\ \citenamefont {Wong}}]{zribi2006condensation}%
  \BibitemOpen
  \bibfield  {author} {\bibinfo {author} {\bibfnamefont {O.~V.}\ \bibnamefont
  {Zribi}}, \bibinfo {author} {\bibfnamefont {H.}~\bibnamefont {Kyung}},
  \bibinfo {author} {\bibfnamefont {R.}~\bibnamefont {Golestanian}}, \bibinfo
  {author} {\bibfnamefont {T.~B.}\ \bibnamefont {Liverpool}},\ and\ \bibinfo
  {author} {\bibfnamefont {G.~C.}\ \bibnamefont {Wong}},\ }\bibfield  {title}
  {\bibinfo {title} {Condensation of {DNA-actin} polyelectrolyte mixtures
  driven by ions of different valences},\ }\href
  {https://journals.aps.org/pre/abstract/10.1103/PhysRevE.73.031911} {\bibfield
   {journal} {\bibinfo  {journal} {Phys. Rev. E}\ }\textbf {\bibinfo {volume}
  {73}},\ \bibinfo {pages} {031911} (\bibinfo {year} {2006})}\BibitemShut
  {NoStop}%
\bibitem [{\citenamefont {Golestanian}\ and\ \citenamefont
  {Liverpool}(2002)}]{golestanian2002conformational}%
  \BibitemOpen
  \bibfield  {author} {\bibinfo {author} {\bibfnamefont {R.}~\bibnamefont
  {Golestanian}}\ and\ \bibinfo {author} {\bibfnamefont {T.~B.}\ \bibnamefont
  {Liverpool}},\ }\bibfield  {title} {\bibinfo {title} {Conformational
  instability of rodlike polyelectrolytes due to counterion fluctuations},\
  }\href {https://journals.aps.org/pre/abstract/10.1103/PhysRevE.66.051802}
  {\bibfield  {journal} {\bibinfo  {journal} {Phys. Rev. E}\ }\textbf {\bibinfo
  {volume} {66}},\ \bibinfo {pages} {051802} (\bibinfo {year}
  {2002})}\BibitemShut {NoStop}%
\bibitem [{\citenamefont {Golestanian}(2000)}]{golestanian2000dynamics}%
  \BibitemOpen
  \bibfield  {author} {\bibinfo {author} {\bibfnamefont {R.}~\bibnamefont
  {Golestanian}},\ }\bibfield  {title} {\bibinfo {title} {Dynamics of
  counterion condensation},\ }\href
  {https://iopscience.iop.org/article/10.1209/epl/i2000-00402-x/meta?casa_token=XBSlzkmhUp0AAAAA:_acmTVwvetG00pZJLk02tT-1B2hpnmWfLe0zLcGmAwkRB0gusgcyn7TYePY0zXVmEhJda-QQ1g}
  {\bibfield  {journal} {\bibinfo  {journal} {EPL}\ }\textbf {\bibinfo {volume}
  {52}},\ \bibinfo {pages} {47} (\bibinfo {year} {2000})}\BibitemShut {NoStop}%
\bibitem [{\citenamefont {Netz}(2003)}]{netz2003electrofriction}%
  \BibitemOpen
  \bibfield  {author} {\bibinfo {author} {\bibfnamefont {R.}~\bibnamefont
  {Netz}},\ }\bibfield  {title} {\bibinfo {title} {Electrofriction and dynamic
  stern layers at planar charged surfaces},\ }\href
  {https://journals.aps.org/prl/abstract/10.1103/PhysRevLett.91.138101}
  {\bibfield  {journal} {\bibinfo  {journal} {Phys. Rev. Lett.}\ }\textbf
  {\bibinfo {volume} {91}},\ \bibinfo {pages} {138101} (\bibinfo {year}
  {2003})}\BibitemShut {NoStop}%
\bibitem [{\citenamefont {Boroudjerdi}\ \emph {et~al.}(2005)\citenamefont
  {Boroudjerdi}, \citenamefont {Kim}, \citenamefont {Naji}, \citenamefont
  {Netz}, \citenamefont {Schlagberger},\ and\ \citenamefont
  {Serr}}]{boroudjerdi2005statics}%
  \BibitemOpen
  \bibfield  {author} {\bibinfo {author} {\bibfnamefont {H.}~\bibnamefont
  {Boroudjerdi}}, \bibinfo {author} {\bibfnamefont {Y.-W.}\ \bibnamefont
  {Kim}}, \bibinfo {author} {\bibfnamefont {A.}~\bibnamefont {Naji}}, \bibinfo
  {author} {\bibfnamefont {R.~R.}\ \bibnamefont {Netz}}, \bibinfo {author}
  {\bibfnamefont {X.}~\bibnamefont {Schlagberger}},\ and\ \bibinfo {author}
  {\bibfnamefont {A.}~\bibnamefont {Serr}},\ }\bibfield  {title} {\bibinfo
  {title} {Statics and dynamics of strongly charged soft matter},\ }\href
  {https://www.sciencedirect.com/science/article/pii/S0370157305002991?casa_token=chkY-f3IS00AAAAA:CokTorIPox9S2gxUup28FmYt2wfMAurmRUeCuh2p_w99-G6nEX_fjRGDqGlBxEKC0hmvh8uslw}
  {\bibfield  {journal} {\bibinfo  {journal} {Phys. Rep.}\ }\textbf {\bibinfo
  {volume} {416}},\ \bibinfo {pages} {129} (\bibinfo {year}
  {2005})}\BibitemShut {NoStop}%
\bibitem [{\citenamefont {Wright}(2007)}]{wrightelectrolyte}%
  \BibitemOpen
  \bibfield  {author} {\bibinfo {author} {\bibfnamefont {M.~R.}\ \bibnamefont
  {Wright}},\ }\href@noop {} {\emph {\bibinfo {title} {An introduction to
  aqueous electrolyte solutions}}}\ (\bibinfo  {publisher} {John Wiley \&
  Sons},\ \bibinfo {year} {2007})\BibitemShut {NoStop}%
\bibitem [{\citenamefont {Zorkot}\ \emph {et~al.}(2016)\citenamefont {Zorkot},
  \citenamefont {Golestanian},\ and\ \citenamefont
  {Bonthuis}}]{zorkot2016power}%
  \BibitemOpen
  \bibfield  {author} {\bibinfo {author} {\bibfnamefont {M.}~\bibnamefont
  {Zorkot}}, \bibinfo {author} {\bibfnamefont {R.}~\bibnamefont
  {Golestanian}},\ and\ \bibinfo {author} {\bibfnamefont {D.~J.}\ \bibnamefont
  {Bonthuis}},\ }\bibfield  {title} {\bibinfo {title} {The power spectrum of
  ionic nanopore currents: the role of ion correlations},\ }\href
  {https://pubs.acs.org/doi/abs/10.1021/acs.nanolett.5b04372} {\bibfield
  {journal} {\bibinfo  {journal} {Nano Lett.}\ }\textbf {\bibinfo {volume}
  {16}},\ \bibinfo {pages} {2205} (\bibinfo {year} {2016})}\BibitemShut
  {NoStop}%
\bibitem [{\citenamefont {Zorkot}\ and\ \citenamefont
  {Golestanian}(2018)}]{zorkot2018nanopore}%
  \BibitemOpen
  \bibfield  {author} {\bibinfo {author} {\bibfnamefont {M.}~\bibnamefont
  {Zorkot}}\ and\ \bibinfo {author} {\bibfnamefont {R.}~\bibnamefont
  {Golestanian}},\ }\bibfield  {title} {\bibinfo {title} {Current fluctuations
  across a nano-pore},\ }\href
  {https://iopscience.iop.org/article/10.1088/1361-648X/aab016/meta} {\bibfield
   {journal} {\bibinfo  {journal} {J. Phys. Condens. Matter}\ }\textbf
  {\bibinfo {volume} {30}},\ \bibinfo {pages} {134001} (\bibinfo {year}
  {2018})}\BibitemShut {NoStop}%
\bibitem [{\citenamefont {D{\'e}mery}\ and\ \citenamefont
  {Dean}(2016)}]{demery2016conductivity}%
  \BibitemOpen
  \bibfield  {author} {\bibinfo {author} {\bibfnamefont {V.}~\bibnamefont
  {D{\'e}mery}}\ and\ \bibinfo {author} {\bibfnamefont {D.~S.}\ \bibnamefont
  {Dean}},\ }\bibfield  {title} {\bibinfo {title} {The conductivity of strong
  electrolytes from stochastic density functional theory},\ }\href
  {https://iopscience.iop.org/article/10.1088/1742-5468/2016/02/023106/meta}
  {\bibfield  {journal} {\bibinfo  {journal} {J. Stat. Mech.: Theory Exp.}\
  }\textbf {\bibinfo {volume} {2016}}\bibinfo  {number} { (2)},\ \bibinfo
  {pages} {023106}}\BibitemShut {NoStop}%
\bibitem [{\citenamefont {Perkin}(2012)}]{Perkin2012}%
  \BibitemOpen
\bibfield  {number} {  }\bibfield  {author} {\bibinfo {author} {\bibfnamefont
  {S.}~\bibnamefont {Perkin}},\ }\bibfield  {title} {\bibinfo {title} {Ionic
  liquids in confined geometries},\ }\href {https://doi.org/10.1039/c2cp23814d}
  {\bibfield  {journal} {\bibinfo  {journal} {Phys. Chem. Chem. Phys.}\
  }\textbf {\bibinfo {volume} {14}},\ \bibinfo {pages} {5052} (\bibinfo {year}
  {2012})}\BibitemShut {NoStop}%
\bibitem [{\citenamefont {Gebbie}\ \emph {et~al.}(2013)\citenamefont {Gebbie},
  \citenamefont {Valtiner}, \citenamefont {Banquy}, \citenamefont {Fox},
  \citenamefont {Henderson},\ and\ \citenamefont
  {Israelachvili}}]{gebbie2013dilute}%
  \BibitemOpen
  \bibfield  {author} {\bibinfo {author} {\bibfnamefont {M.~A.}\ \bibnamefont
  {Gebbie}}, \bibinfo {author} {\bibfnamefont {M.}~\bibnamefont {Valtiner}},
  \bibinfo {author} {\bibfnamefont {X.}~\bibnamefont {Banquy}}, \bibinfo
  {author} {\bibfnamefont {E.~T.}\ \bibnamefont {Fox}}, \bibinfo {author}
  {\bibfnamefont {W.~A.}\ \bibnamefont {Henderson}},\ and\ \bibinfo {author}
  {\bibfnamefont {J.~N.}\ \bibnamefont {Israelachvili}},\ }\bibfield  {title}
  {\bibinfo {title} {Ionic liquids behave as dilute electrolyte solutions},\
  }\href {https://www.pnas.org/content/pnas/110/24/9674.full.pdf} {\bibfield
  {journal} {\bibinfo  {journal} {Proc. Natl. Acad. Sci. U.S.A.}\ }\textbf
  {\bibinfo {volume} {110}},\ \bibinfo {pages} {9674} (\bibinfo {year}
  {2013})}\BibitemShut {NoStop}%
\bibitem [{\citenamefont {Smith}\ \emph {et~al.}(2016)\citenamefont {Smith},
  \citenamefont {Lee},\ and\ \citenamefont {Perkin}}]{smith2016electrostatic}%
  \BibitemOpen
  \bibfield  {author} {\bibinfo {author} {\bibfnamefont {A.~M.}\ \bibnamefont
  {Smith}}, \bibinfo {author} {\bibfnamefont {A.~A.}\ \bibnamefont {Lee}},\
  and\ \bibinfo {author} {\bibfnamefont {S.}~\bibnamefont {Perkin}},\
  }\bibfield  {title} {\bibinfo {title} {The electrostatic screening length in
  concentrated electrolytes increases with concentration},\ }\href
  {https://pubs.acs.org/doi/abs/10.1021/acs.jpclett.6b00867} {\bibfield
  {journal} {\bibinfo  {journal} {J. Phys. Chem. Lett.}\ }\textbf {\bibinfo
  {volume} {7}},\ \bibinfo {pages} {2157} (\bibinfo {year} {2016})}\BibitemShut
  {NoStop}%
\bibitem [{\citenamefont {Gebbie}\ \emph {et~al.}(2017)\citenamefont {Gebbie},
  \citenamefont {Smith}, \citenamefont {Dobbs}, \citenamefont {Warr},
  \citenamefont {Banquy}, \citenamefont {Valtiner}, \citenamefont {Rutland},
  \citenamefont {Israelachvili}, \citenamefont {Perkin}, \citenamefont {Atkin}
  \emph {et~al.}}]{gebbie2017long}%
  \BibitemOpen
  \bibfield  {author} {\bibinfo {author} {\bibfnamefont {M.~A.}\ \bibnamefont
  {Gebbie}}, \bibinfo {author} {\bibfnamefont {A.~M.}\ \bibnamefont {Smith}},
  \bibinfo {author} {\bibfnamefont {H.~A.}\ \bibnamefont {Dobbs}}, \bibinfo
  {author} {\bibfnamefont {G.~G.}\ \bibnamefont {Warr}}, \bibinfo {author}
  {\bibfnamefont {X.}~\bibnamefont {Banquy}}, \bibinfo {author} {\bibfnamefont
  {M.}~\bibnamefont {Valtiner}}, \bibinfo {author} {\bibfnamefont {M.~W.}\
  \bibnamefont {Rutland}}, \bibinfo {author} {\bibfnamefont {J.~N.}\
  \bibnamefont {Israelachvili}}, \bibinfo {author} {\bibfnamefont
  {S.}~\bibnamefont {Perkin}}, \bibinfo {author} {\bibfnamefont
  {R.}~\bibnamefont {Atkin}}, \emph {et~al.},\ }\bibfield  {title} {\bibinfo
  {title} {Long range electrostatic forces in ionic liquids},\ }\href
  {https://pubs.rsc.org/am/content/articlehtml/2017/cc/c6cc08820a} {\bibfield
  {journal} {\bibinfo  {journal} {Chem. Commun.}\ }\textbf {\bibinfo {volume}
  {53}},\ \bibinfo {pages} {1214} (\bibinfo {year} {2017})}\BibitemShut
  {NoStop}%
\bibitem [{\citenamefont {Perez-Martinez}\ \emph {et~al.}(2017)\citenamefont
  {Perez-Martinez}, \citenamefont {Smith}, \citenamefont {Perkin} \emph
  {et~al.}}]{perez2017scaling}%
  \BibitemOpen
  \bibfield  {author} {\bibinfo {author} {\bibfnamefont {C.~S.}\ \bibnamefont
  {Perez-Martinez}}, \bibinfo {author} {\bibfnamefont {A.~M.}\ \bibnamefont
  {Smith}}, \bibinfo {author} {\bibfnamefont {S.}~\bibnamefont {Perkin}}, \emph
  {et~al.},\ }\bibfield  {title} {\bibinfo {title} {Scaling analysis of the
  screening length in concentrated electrolytes},\ }\href
  {https://journals.aps.org/prl/abstract/10.1103/PhysRevLett.119.026002}
  {\bibfield  {journal} {\bibinfo  {journal} {Phys. Rev. Lett.}\ }\textbf
  {\bibinfo {volume} {119}},\ \bibinfo {pages} {026002} (\bibinfo {year}
  {2017})}\BibitemShut {NoStop}%
\bibitem [{\citenamefont {Perez-Martinez}\ and\ \citenamefont
  {Perkin}(2019)}]{perez2019surface}%
  \BibitemOpen
  \bibfield  {author} {\bibinfo {author} {\bibfnamefont {C.~S.}\ \bibnamefont
  {Perez-Martinez}}\ and\ \bibinfo {author} {\bibfnamefont {S.}~\bibnamefont
  {Perkin}},\ }\bibfield  {title} {\bibinfo {title} {Surface forces generated
  by the action of electric fields across liquid films},\ }\href
  {https://pubs.rsc.org/en/content/articlelanding/2019/sm/c9sm00143c#!divAbstract}
  {\bibfield  {journal} {\bibinfo  {journal} {Soft Matter}\ }\textbf {\bibinfo
  {volume} {15}},\ \bibinfo {pages} {4255} (\bibinfo {year}
  {2019})}\BibitemShut {NoStop}%
\bibitem [{\citenamefont {Feng}\ \emph {et~al.}(2019)\citenamefont {Feng},
  \citenamefont {Chen}, \citenamefont {Bi}, \citenamefont {Goodwin},
  \citenamefont {Postnikov}, \citenamefont {Brilliantov}, \citenamefont
  {Urbakh},\ and\ \citenamefont {Kornyshev}}]{kornyshev}%
  \BibitemOpen
  \bibfield  {author} {\bibinfo {author} {\bibfnamefont {G.}~\bibnamefont
  {Feng}}, \bibinfo {author} {\bibfnamefont {M.}~\bibnamefont {Chen}}, \bibinfo
  {author} {\bibfnamefont {S.}~\bibnamefont {Bi}}, \bibinfo {author}
  {\bibfnamefont {Z.~A.~H.}\ \bibnamefont {Goodwin}}, \bibinfo {author}
  {\bibfnamefont {E.~B.}\ \bibnamefont {Postnikov}}, \bibinfo {author}
  {\bibfnamefont {N.}~\bibnamefont {Brilliantov}}, \bibinfo {author}
  {\bibfnamefont {M.}~\bibnamefont {Urbakh}},\ and\ \bibinfo {author}
  {\bibfnamefont {A.~A.}\ \bibnamefont {Kornyshev}},\ }\bibfield  {title}
  {\bibinfo {title} {Free and bound states of ions in ionic liquids,
  conductivity, and underscreening paradox},\ }\href
  {https://doi.org/10.1103/PhysRevX.9.021024} {\bibfield  {journal} {\bibinfo
  {journal} {Phys. Rev. X}\ }\textbf {\bibinfo {volume} {9}},\ \bibinfo {pages}
  {021024} (\bibinfo {year} {2019})}\BibitemShut {NoStop}%
\bibitem [{\citenamefont {Richter}\ \emph {et~al.}(2020)\citenamefont
  {Richter}, \citenamefont {\ifmmode~\dot{Z}\else \.{Z}\fi{}uk}, \citenamefont
  {Szymczak}, \citenamefont {Paczesny}, \citenamefont {Bk{a}k}, \citenamefont
  {Szymborski}, \citenamefont {Garstecki}, \citenamefont {Stone}, \citenamefont
  {Ho\l{}yst},\ and\ \citenamefont {Drummond}}]{stoneAC}%
  \BibitemOpen
  \bibfield  {author} {\bibinfo {author} {\bibfnamefont {L.}~\bibnamefont
  {Richter}}, \bibinfo {author} {\bibfnamefont {P.~J.}\ \bibnamefont
  {\ifmmode~\dot{Z}\else \.{Z}\fi{}uk}}, \bibinfo {author} {\bibfnamefont
  {P.}~\bibnamefont {Szymczak}}, \bibinfo {author} {\bibfnamefont
  {J.}~\bibnamefont {Paczesny}}, \bibinfo {author} {\bibfnamefont {K.~M.}\
  \bibnamefont {Bk{a}k}}, \bibinfo {author} {\bibfnamefont {T.}~\bibnamefont
  {Szymborski}}, \bibinfo {author} {\bibfnamefont {P.}~\bibnamefont
  {Garstecki}}, \bibinfo {author} {\bibfnamefont {H.~A.}\ \bibnamefont
  {Stone}}, \bibinfo {author} {\bibfnamefont {R.}~\bibnamefont {Ho\l{}yst}},\
  and\ \bibinfo {author} {\bibfnamefont {C.}~\bibnamefont {Drummond}},\
  }\bibfield  {title} {\bibinfo {title} {Ions in an ac electric field: Strong
  long-range repulsion between oppositely charged surfaces},\ }\href
  {https://doi.org/10.1103/PhysRevLett.125.056001} {\bibfield  {journal}
  {\bibinfo  {journal} {Phys. Rev. Lett.}\ }\textbf {\bibinfo {volume} {125}},\
  \bibinfo {pages} {056001} (\bibinfo {year} {2020})}\BibitemShut {NoStop}%
\bibitem [{\citenamefont {Armand}\ and\ \citenamefont
  {Tarascon}(2008)}]{armand2008building}%
  \BibitemOpen
  \bibfield  {author} {\bibinfo {author} {\bibfnamefont {M.}~\bibnamefont
  {Armand}}\ and\ \bibinfo {author} {\bibfnamefont {J.-M.}\ \bibnamefont
  {Tarascon}},\ }\bibfield  {title} {\bibinfo {title} {Building better
  batteries},\ }\href {https://www.nature.com/articles/451652a} {\bibfield
  {journal} {\bibinfo  {journal} {Nature}\ }\textbf {\bibinfo {volume} {451}},\
  \bibinfo {pages} {652} (\bibinfo {year} {2008})}\BibitemShut {NoStop}%
\bibitem [{\citenamefont {K{\"o}tz}\ and\ \citenamefont
  {Carlen}(2000)}]{kotz2000principles}%
  \BibitemOpen
  \bibfield  {author} {\bibinfo {author} {\bibfnamefont {R.}~\bibnamefont
  {K{\"o}tz}}\ and\ \bibinfo {author} {\bibfnamefont {M.}~\bibnamefont
  {Carlen}},\ }\bibfield  {title} {\bibinfo {title} {Principles and
  applications of electrochemical capacitors},\ }\href
  {https://doi.org/10.1016/S0013-4686(00)00354-6} {\bibfield  {journal}
  {\bibinfo  {journal} {Electrochim. Acta}\ }\textbf {\bibinfo {volume} {45}},\
  \bibinfo {pages} {2483} (\bibinfo {year} {2000})}\BibitemShut {NoStop}%
\bibitem [{\citenamefont {Luo}\ \emph {et~al.}(2015)\citenamefont {Luo},
  \citenamefont {Wang}, \citenamefont {Dooner},\ and\ \citenamefont
  {Clarke}}]{luo2015overview}%
  \BibitemOpen
  \bibfield  {author} {\bibinfo {author} {\bibfnamefont {X.}~\bibnamefont
  {Luo}}, \bibinfo {author} {\bibfnamefont {J.}~\bibnamefont {Wang}}, \bibinfo
  {author} {\bibfnamefont {M.}~\bibnamefont {Dooner}},\ and\ \bibinfo {author}
  {\bibfnamefont {J.}~\bibnamefont {Clarke}},\ }\bibfield  {title} {\bibinfo
  {title} {Overview of current development in electrical energy storage
  technologies and the application potential in power system operation},\
  }\href {https://www.sciencedirect.com/science/article/pii/S0306261914010290}
  {\bibfield  {journal} {\bibinfo  {journal} {Appl. energy}\ }\textbf {\bibinfo
  {volume} {137}},\ \bibinfo {pages} {511} (\bibinfo {year}
  {2015})}\BibitemShut {NoStop}%
\bibitem [{\citenamefont {Gambassi}(2009)}]{gambassi2009review}%
  \BibitemOpen
  \bibfield  {author} {\bibinfo {author} {\bibfnamefont {A.}~\bibnamefont
  {Gambassi}},\ }\bibfield  {title} {\bibinfo {title} {{The Casimir effect:
  From quantum to critical fluctuations}},\ }\href
  {https://doi.org/10.1088/1742-6596/161/1/012037} {\bibfield  {journal}
  {\bibinfo  {journal} {J. Phys.: Conf. Ser.}\ }\textbf {\bibinfo {volume}
  {161}},\ \bibinfo {pages} {012037} (\bibinfo {year} {2009})}\BibitemShut
  {NoStop}%
\bibitem [{\citenamefont {Hertlein}\ \emph {et~al.}(2008)\citenamefont
  {Hertlein}, \citenamefont {Helden}, \citenamefont {Gambassi}, \citenamefont
  {Dietrich},\ and\ \citenamefont {Bechinger}}]{gambassinature}%
  \BibitemOpen
  \bibfield  {author} {\bibinfo {author} {\bibfnamefont {C.}~\bibnamefont
  {Hertlein}}, \bibinfo {author} {\bibfnamefont {L.}~\bibnamefont {Helden}},
  \bibinfo {author} {\bibfnamefont {A.}~\bibnamefont {Gambassi}}, \bibinfo
  {author} {\bibfnamefont {S.}~\bibnamefont {Dietrich}},\ and\ \bibinfo
  {author} {\bibfnamefont {C.}~\bibnamefont {Bechinger}},\ }\bibfield  {title}
  {\bibinfo {title} {{Direct measurement of critical Casimir forces}},\ }\href
  {https://www.nature.com/articles/nature06443} {\bibfield  {journal} {\bibinfo
   {journal} {Nature}\ }\textbf {\bibinfo {volume} {451}},\ \bibinfo {pages}
  {172} (\bibinfo {year} {2008})}\BibitemShut {NoStop}%
\bibitem [{\citenamefont {Casimir}(1948)}]{casimir1948attraction}%
  \BibitemOpen
  \bibfield  {author} {\bibinfo {author} {\bibfnamefont {H.~B.}\ \bibnamefont
  {Casimir}},\ }\bibfield  {title} {\bibinfo {title} {On the attraction between
  two perfectly conducting plates},\ }\href
  {https://inspirehep.net/files/5b067228edaae87746b4aab970cbc6f2} {\bibfield
  {journal} {\bibinfo  {journal} {Proc. Kon. Ned. Akad. Wet.}\ }\textbf
  {\bibinfo {volume} {51}},\ \bibinfo {pages} {793} (\bibinfo {year}
  {1948})}\BibitemShut {NoStop}%
\bibitem [{\citenamefont {Fisher}\ and\ \citenamefont
  {Gennes}(1978)}]{fisher1978wall}%
  \BibitemOpen
  \bibfield  {author} {\bibinfo {author} {\bibfnamefont {M.~E.}\ \bibnamefont
  {Fisher}}\ and\ \bibinfo {author} {\bibfnamefont {P.}~\bibnamefont
  {Gennes}},\ }\bibfield  {title} {\bibinfo {title} {Wall phenomena in a
  critical binary mixture},\ }\href@noop {} {\bibfield  {journal} {\bibinfo
  {journal} {C. R. Acad. Sc. Paris B}\ }\textbf {\bibinfo {volume} {287}},\
  \bibinfo {pages} {207} (\bibinfo {year} {1978})}\BibitemShut {NoStop}%
\bibitem [{\citenamefont {French}\ \emph {et~al.}(2010)\citenamefont {French},
  \citenamefont {Parsegian}, \citenamefont {Podgornik}, \citenamefont {Rajter},
  \citenamefont {Jagota}, \citenamefont {Luo}, \citenamefont {Asthagiri},
  \citenamefont {Chaudhury}, \citenamefont {Chiang}, \citenamefont {Granick},
  \citenamefont {Kalinin}, \citenamefont {Kardar}, \citenamefont {Kjellander},
  \citenamefont {Langreth}, \citenamefont {Lewis}, \citenamefont {Lustig},
  \citenamefont {Wesolowski}, \citenamefont {Wettlaufer}, \citenamefont
  {Ching}, \citenamefont {Finnis}, \citenamefont {Houlihan}, \citenamefont {von
  Lilienfeld}, \citenamefont {van Oss},\ and\ \citenamefont {Zemb}}]{RMP2010}%
  \BibitemOpen
  \bibfield  {author} {\bibinfo {author} {\bibfnamefont {R.~H.}\ \bibnamefont
  {French}}, \bibinfo {author} {\bibfnamefont {V.~A.}\ \bibnamefont
  {Parsegian}}, \bibinfo {author} {\bibfnamefont {R.}~\bibnamefont
  {Podgornik}}, \bibinfo {author} {\bibfnamefont {R.~F.}\ \bibnamefont
  {Rajter}}, \bibinfo {author} {\bibfnamefont {A.}~\bibnamefont {Jagota}},
  \bibinfo {author} {\bibfnamefont {J.}~\bibnamefont {Luo}}, \bibinfo {author}
  {\bibfnamefont {D.}~\bibnamefont {Asthagiri}}, \bibinfo {author}
  {\bibfnamefont {M.~K.}\ \bibnamefont {Chaudhury}}, \bibinfo {author}
  {\bibfnamefont {Y.-m.}\ \bibnamefont {Chiang}}, \bibinfo {author}
  {\bibfnamefont {S.}~\bibnamefont {Granick}}, \bibinfo {author} {\bibfnamefont
  {S.}~\bibnamefont {Kalinin}}, \bibinfo {author} {\bibfnamefont
  {M.}~\bibnamefont {Kardar}}, \bibinfo {author} {\bibfnamefont
  {R.}~\bibnamefont {Kjellander}}, \bibinfo {author} {\bibfnamefont {D.~C.}\
  \bibnamefont {Langreth}}, \bibinfo {author} {\bibfnamefont {J.}~\bibnamefont
  {Lewis}}, \bibinfo {author} {\bibfnamefont {S.}~\bibnamefont {Lustig}},
  \bibinfo {author} {\bibfnamefont {D.}~\bibnamefont {Wesolowski}}, \bibinfo
  {author} {\bibfnamefont {J.~S.}\ \bibnamefont {Wettlaufer}}, \bibinfo
  {author} {\bibfnamefont {W.-Y.}\ \bibnamefont {Ching}}, \bibinfo {author}
  {\bibfnamefont {M.}~\bibnamefont {Finnis}}, \bibinfo {author} {\bibfnamefont
  {F.}~\bibnamefont {Houlihan}}, \bibinfo {author} {\bibfnamefont {O.~A.}\
  \bibnamefont {von Lilienfeld}}, \bibinfo {author} {\bibfnamefont {C.~J.}\
  \bibnamefont {van Oss}},\ and\ \bibinfo {author} {\bibfnamefont
  {T.}~\bibnamefont {Zemb}},\ }\bibfield  {title} {\bibinfo {title} {Long range
  interactions in nanoscale science},\ }\href
  {https://doi.org/10.1103/RevModPhys.82.1887} {\bibfield  {journal} {\bibinfo
  {journal} {Rev. Mod. Phys.}\ }\textbf {\bibinfo {volume} {82}},\ \bibinfo
  {pages} {1887} (\bibinfo {year} {2010})}\BibitemShut {NoStop}%
\bibitem [{\citenamefont {Macio{\l}ek}\ and\ \citenamefont
  {Dietrich}(2018)}]{maciolekreview}%
  \BibitemOpen
  \bibfield  {author} {\bibinfo {author} {\bibfnamefont {A.}~\bibnamefont
  {Macio{\l}ek}}\ and\ \bibinfo {author} {\bibfnamefont {S.}~\bibnamefont
  {Dietrich}},\ }\bibfield  {title} {\bibinfo {title} {{Collective behavior of
  colloids due to critical Casimir interactions}},\ }\href
  {https://journals.aps.org/rmp/abstract/10.1103/RevModPhys.90.045001}
  {\bibfield  {journal} {\bibinfo  {journal} {Rev. Mod. Phys.}\ }\textbf
  {\bibinfo {volume} {90}},\ \bibinfo {pages} {045001} (\bibinfo {year}
  {2018})}\BibitemShut {NoStop}%
\bibitem [{\citenamefont {Najafi}\ and\ \citenamefont
  {Golestanian}(2004)}]{najafi2004soret}%
  \BibitemOpen
  \bibfield  {author} {\bibinfo {author} {\bibfnamefont {A.}~\bibnamefont
  {Najafi}}\ and\ \bibinfo {author} {\bibfnamefont {R.}~\bibnamefont
  {Golestanian}},\ }\bibfield  {title} {\bibinfo {title} {{Forces induced by
  nonequilibrium fluctuations: The Soret-Casimir effect}},\ }\href
  {https://doi.org/10.1209/epl/i2004-10275-5} {\bibfield  {journal} {\bibinfo
  {journal} {EPL}\ }\textbf {\bibinfo {volume} {68}},\ \bibinfo {pages} {776}
  (\bibinfo {year} {2004})}\BibitemShut {NoStop}%
\bibitem [{\citenamefont {Dean}\ and\ \citenamefont
  {Gopinathan}(2010)}]{dean2010out}%
  \BibitemOpen
  \bibfield  {author} {\bibinfo {author} {\bibfnamefont {D.~S.}\ \bibnamefont
  {Dean}}\ and\ \bibinfo {author} {\bibfnamefont {A.}~\bibnamefont
  {Gopinathan}},\ }\bibfield  {title} {\bibinfo {title} {Out-of-equilibrium
  behavior of casimir-type fluctuation-induced forces for free classical
  fields},\ }\href
  {https://journals.aps.org/pre/abstract/10.1103/PhysRevE.81.041126} {\bibfield
   {journal} {\bibinfo  {journal} {Phys. Rev. E}\ }\textbf {\bibinfo {volume}
  {81}},\ \bibinfo {pages} {041126} (\bibinfo {year} {2010})}\BibitemShut
  {NoStop}%
\bibitem [{\citenamefont {Aminov}\ \emph {et~al.}(2015)\citenamefont {Aminov},
  \citenamefont {Kafri},\ and\ \citenamefont {Kardar}}]{aminov2015neqfif}%
  \BibitemOpen
  \bibfield  {author} {\bibinfo {author} {\bibfnamefont {A.}~\bibnamefont
  {Aminov}}, \bibinfo {author} {\bibfnamefont {Y.}~\bibnamefont {Kafri}},\ and\
  \bibinfo {author} {\bibfnamefont {M.}~\bibnamefont {Kardar}},\ }\bibfield
  {title} {\bibinfo {title} {Fluctuation-induced forces in nonequilibrium
  diffusive dynamics},\ }\href
  {https://journals.aps.org/prl/abstract/10.1103/PhysRevLett.114.230602}
  {\bibfield  {journal} {\bibinfo  {journal} {Phys. Rev. Lett.}\ }\textbf
  {\bibinfo {volume} {114}},\ \bibinfo {pages} {230602} (\bibinfo {year}
  {2015})}\BibitemShut {NoStop}%
\bibitem [{\citenamefont {Rohwer}\ \emph {et~al.}(2017)\citenamefont {Rohwer},
  \citenamefont {Kardar},\ and\ \citenamefont
  {Kr{\"u}ger}}]{rohwer2017transfif}%
  \BibitemOpen
  \bibfield  {author} {\bibinfo {author} {\bibfnamefont {C.~M.}\ \bibnamefont
  {Rohwer}}, \bibinfo {author} {\bibfnamefont {M.}~\bibnamefont {Kardar}},\
  and\ \bibinfo {author} {\bibfnamefont {M.}~\bibnamefont {Kr{\"u}ger}},\
  }\bibfield  {title} {\bibinfo {title} {Transient casimir forces from quenches
  in thermal and active matter},\ }\href
  {https://journals.aps.org/prl/abstract/10.1103/PhysRevLett.118.015702}
  {\bibfield  {journal} {\bibinfo  {journal} {Phys. Rev. Lett.}\ }\textbf
  {\bibinfo {volume} {118}},\ \bibinfo {pages} {015702} (\bibinfo {year}
  {2017})}\BibitemShut {NoStop}%
\bibitem [{\citenamefont {Gross}\ \emph {et~al.}(2018)\citenamefont {Gross},
  \citenamefont {Gambassi},\ and\ \citenamefont {Dietrich}}]{gross2018surface}%
  \BibitemOpen
  \bibfield  {author} {\bibinfo {author} {\bibfnamefont {M.}~\bibnamefont
  {Gross}}, \bibinfo {author} {\bibfnamefont {A.}~\bibnamefont {Gambassi}},\
  and\ \bibinfo {author} {\bibfnamefont {S.}~\bibnamefont {Dietrich}},\
  }\bibfield  {title} {\bibinfo {title} {{Surface-induced nonequilibrium
  dynamics and critical Casimir forces for model B in film geometry}},\ }\href
  {https://journals.aps.org/pre/abstract/10.1103/PhysRevE.98.032103} {\bibfield
   {journal} {\bibinfo  {journal} {Phys. Rev. E}\ }\textbf {\bibinfo {volume}
  {98}},\ \bibinfo {pages} {032103} (\bibinfo {year} {2018})}\BibitemShut
  {NoStop}%
\bibitem [{\citenamefont {Gross}\ \emph {et~al.}(2019)\citenamefont {Gross},
  \citenamefont {Rohwer},\ and\ \citenamefont {Dietrich}}]{gross2019dynamics}%
  \BibitemOpen
  \bibfield  {author} {\bibinfo {author} {\bibfnamefont {M.}~\bibnamefont
  {Gross}}, \bibinfo {author} {\bibfnamefont {C.~M.}\ \bibnamefont {Rohwer}},\
  and\ \bibinfo {author} {\bibfnamefont {S.}~\bibnamefont {Dietrich}},\
  }\bibfield  {title} {\bibinfo {title} {{Dynamics of the critical Casimir
  force for a conserved order parameter after a critical quench}},\ }\href
  {https://journals.aps.org/pre/abstract/10.1103/PhysRevE.100.012114}
  {\bibfield  {journal} {\bibinfo  {journal} {Phys. Rev. E}\ }\textbf {\bibinfo
  {volume} {100}},\ \bibinfo {pages} {012114} (\bibinfo {year}
  {2019})}\BibitemShut {NoStop}%
\bibitem [{\citenamefont {Rohwer}\ \emph {et~al.}(2018)\citenamefont {Rohwer},
  \citenamefont {Solon}, \citenamefont {Kardar},\ and\ \citenamefont
  {Kr\"uger}}]{rohwer2018neqfif}%
  \BibitemOpen
  \bibfield  {author} {\bibinfo {author} {\bibfnamefont {C.~M.}\ \bibnamefont
  {Rohwer}}, \bibinfo {author} {\bibfnamefont {A.}~\bibnamefont {Solon}},
  \bibinfo {author} {\bibfnamefont {M.}~\bibnamefont {Kardar}},\ and\ \bibinfo
  {author} {\bibfnamefont {M.}~\bibnamefont {Kr\"uger}},\ }\bibfield  {title}
  {\bibinfo {title} {Nonequilibrium forces following quenches in active and
  thermal matter},\ }\href {https://doi.org/10.1103/PhysRevE.97.032125}
  {\bibfield  {journal} {\bibinfo  {journal} {Phys. Rev. E}\ }\textbf {\bibinfo
  {volume} {97}},\ \bibinfo {pages} {032125} (\bibinfo {year}
  {2018})}\BibitemShut {NoStop}%
\bibitem [{\citenamefont {Ray}\ \emph {et~al.}(2014)\citenamefont {Ray},
  \citenamefont {Reichhardt},\ and\ \citenamefont
  {Reichhardt}}]{activecasimir}%
  \BibitemOpen
  \bibfield  {author} {\bibinfo {author} {\bibfnamefont {D.}~\bibnamefont
  {Ray}}, \bibinfo {author} {\bibfnamefont {C.}~\bibnamefont {Reichhardt}},\
  and\ \bibinfo {author} {\bibfnamefont {C.~J.~O.}\ \bibnamefont
  {Reichhardt}},\ }\bibfield  {title} {\bibinfo {title} {Casimir effect in
  active matter systems},\ }\href {https://doi.org/10.1103/PhysRevE.90.013019}
  {\bibfield  {journal} {\bibinfo  {journal} {Phys. Rev. E}\ }\textbf {\bibinfo
  {volume} {90}},\ \bibinfo {pages} {013019} (\bibinfo {year}
  {2014})}\BibitemShut {NoStop}%
\bibitem [{\citenamefont {Dean}\ and\ \citenamefont
  {Podgornik}(2014)}]{deanbrownian2014}%
  \BibitemOpen
  \bibfield  {author} {\bibinfo {author} {\bibfnamefont {D.~S.}\ \bibnamefont
  {Dean}}\ and\ \bibinfo {author} {\bibfnamefont {R.}~\bibnamefont
  {Podgornik}},\ }\bibfield  {title} {\bibinfo {title} {{Relaxation of the
  thermal Casimir force between net neutral plates containing Brownian
  charges}},\ }\href {https://doi.org/10.1103/PhysRevE.89.032117} {\bibfield
  {journal} {\bibinfo  {journal} {Phys. Rev. E}\ }\textbf {\bibinfo {volume}
  {89}},\ \bibinfo {pages} {032117} (\bibinfo {year} {2014})}\BibitemShut
  {NoStop}%
\bibitem [{\citenamefont {Dean}\ \emph {et~al.}(2016)\citenamefont {Dean},
  \citenamefont {Lu}, \citenamefont {Maggs},\ and\ \citenamefont
  {Podgornik}}]{deannoneqtune2016}%
  \BibitemOpen
  \bibfield  {author} {\bibinfo {author} {\bibfnamefont {D.~S.}\ \bibnamefont
  {Dean}}, \bibinfo {author} {\bibfnamefont {B.-S.}\ \bibnamefont {Lu}},
  \bibinfo {author} {\bibfnamefont {A.~C.}\ \bibnamefont {Maggs}},\ and\
  \bibinfo {author} {\bibfnamefont {R.}~\bibnamefont {Podgornik}},\ }\bibfield
  {title} {\bibinfo {title} {{Nonequilibrium Tuning of the Thermal Casimir
  Effect}},\ }\href {https://doi.org/10.1103/PhysRevLett.116.240602} {\bibfield
   {journal} {\bibinfo  {journal} {Phys. Rev. Lett.}\ }\textbf {\bibinfo
  {volume} {116}},\ \bibinfo {pages} {240602} (\bibinfo {year}
  {2016})}\BibitemShut {NoStop}%
\bibitem [{\citenamefont {Jancovici}\ and\ \citenamefont
  {{\v{S}}amaj}(2004)}]{jancovici2004screening}%
  \BibitemOpen
  \bibfield  {author} {\bibinfo {author} {\bibfnamefont {B.}~\bibnamefont
  {Jancovici}}\ and\ \bibinfo {author} {\bibfnamefont {L.}~\bibnamefont
  {{\v{S}}amaj}},\ }\bibfield  {title} {\bibinfo {title} {{Screening of
  classical Casimir forces by electrolytes in semi-infinite geometries}},\
  }\href
  {https://iopscience.iop.org/article/10.1088/1742-5468/2004/08/P08006/meta}
  {\bibfield  {journal} {\bibinfo  {journal} {J. Stat. Mech.: Theory Exp.}\
  }\textbf {\bibinfo {volume} {2004}}\bibinfo  {number} { (08)},\ \bibinfo
  {pages} {P08006}}\BibitemShut {NoStop}%
\bibitem [{\citenamefont {Lee}\ \emph {et~al.}(2018)\citenamefont {Lee},
  \citenamefont {Hansen}, \citenamefont {Bernard},\ and\ \citenamefont
  {Rotenberg}}]{lee2018casimir}%
  \BibitemOpen
\bibfield  {number} {  }\bibfield  {author} {\bibinfo {author} {\bibfnamefont
  {A.~A.}\ \bibnamefont {Lee}}, \bibinfo {author} {\bibfnamefont {J.-P.}\
  \bibnamefont {Hansen}}, \bibinfo {author} {\bibfnamefont {O.}~\bibnamefont
  {Bernard}},\ and\ \bibinfo {author} {\bibfnamefont {B.}~\bibnamefont
  {Rotenberg}},\ }\bibfield  {title} {\bibinfo {title} {Casimir force in dense
  confined electrolytes},\ }\href
  {https://www.tandfonline.com/doi/full/10.1080/00268976.2018.1478137?casa_token=WOxJR9kHcoYAAAAA%3A99DHHGBeVQeKfTUHs_pi0BHQFUtgKnI9f47-9xdYXV06mrVZq-vPU1Q-7FXVhySKDg_Aa6sWKyRox0Q}
  {\bibfield  {journal} {\bibinfo  {journal} {Mol. Phys.}\ }\textbf {\bibinfo
  {volume} {116}},\ \bibinfo {pages} {3147} (\bibinfo {year}
  {2018})}\BibitemShut {NoStop}%
\bibitem [{\citenamefont {Garrido}\ \emph {et~al.}(1990)\citenamefont
  {Garrido}, \citenamefont {Lebowitz}, \citenamefont {Maes},\ and\
  \citenamefont {Spohn}}]{garrido90conservative}%
  \BibitemOpen
  \bibfield  {author} {\bibinfo {author} {\bibfnamefont {P.~L.}\ \bibnamefont
  {Garrido}}, \bibinfo {author} {\bibfnamefont {J.~L.}\ \bibnamefont
  {Lebowitz}}, \bibinfo {author} {\bibfnamefont {C.}~\bibnamefont {Maes}},\
  and\ \bibinfo {author} {\bibfnamefont {H.}~\bibnamefont {Spohn}},\ }\bibfield
   {title} {\bibinfo {title} {Long-range correlations for conservative
  dynamics},\ }\href
  {https://journals.aps.org/pra/abstract/10.1103/PhysRevA.42.1954} {\bibfield
  {journal} {\bibinfo  {journal} {Phys. Rev. A}\ }\textbf {\bibinfo {volume}
  {42}},\ \bibinfo {pages} {1954} (\bibinfo {year} {1990})}\BibitemShut
  {NoStop}%
\bibitem [{\citenamefont {Grinstein}\ \emph {et~al.}(1990)\citenamefont
  {Grinstein}, \citenamefont {Lee},\ and\ \citenamefont
  {Sachdev}}]{grinstein90conservation}%
  \BibitemOpen
  \bibfield  {author} {\bibinfo {author} {\bibfnamefont {G.}~\bibnamefont
  {Grinstein}}, \bibinfo {author} {\bibfnamefont {D.-H.}\ \bibnamefont {Lee}},\
  and\ \bibinfo {author} {\bibfnamefont {S.}~\bibnamefont {Sachdev}},\
  }\bibfield  {title} {\bibinfo {title} {Conservation laws, anisotropy, and
  ‘‘self-organized criticality’’ in noisy nonequilibrium systems},\
  }\href {https://journals.aps.org/prl/abstract/10.1103/PhysRevLett.64.1927}
  {\bibfield  {journal} {\bibinfo  {journal} {Phys. Rev. Lett.}\ }\textbf
  {\bibinfo {volume} {64}},\ \bibinfo {pages} {1927} (\bibinfo {year}
  {1990})}\BibitemShut {NoStop}%
\bibitem [{\citenamefont {Hwa}\ and\ \citenamefont
  {Kardar}(1989)}]{hwa89dissipative}%
  \BibitemOpen
  \bibfield  {author} {\bibinfo {author} {\bibfnamefont {T.}~\bibnamefont
  {Hwa}}\ and\ \bibinfo {author} {\bibfnamefont {M.}~\bibnamefont {Kardar}},\
  }\bibfield  {title} {\bibinfo {title} {Dissipative transport in open systems:
  An investigation of self-organized criticality},\ }\href
  {https://journals.aps.org/prl/abstract/10.1103/PhysRevLett.62.1813}
  {\bibfield  {journal} {\bibinfo  {journal} {Phys. Rev. Lett.}\ }\textbf
  {\bibinfo {volume} {62}},\ \bibinfo {pages} {1813} (\bibinfo {year}
  {1989})}\BibitemShut {NoStop}%
\bibitem [{\citenamefont {Mahdisoltani}\ and\ \citenamefont
  {Golestanian}(2021)}]{mahdisoltani2021long}%
  \BibitemOpen
  \bibfield  {author} {\bibinfo {author} {\bibfnamefont {S.}~\bibnamefont
  {Mahdisoltani}}\ and\ \bibinfo {author} {\bibfnamefont {R.}~\bibnamefont
  {Golestanian}},\ }\bibfield  {title} {\bibinfo {title} {Long-range
  fluctuation-induced forces in driven electrolytes},\ }\href
  {https://journals.aps.org/prl/abstract/10.1103/PhysRevLett.126.158002}
  {\bibfield  {journal} {\bibinfo  {journal} {Phys. Rev. Lett.}\ }\textbf
  {\bibinfo {volume} {126}},\ \bibinfo {pages} {158002} (\bibinfo {year}
  {2021})}\BibitemShut {NoStop}%
\bibitem [{\citenamefont {Holm}\ \emph {et~al.}(2001)\citenamefont {Holm},
  \citenamefont {K{\'e}kicheff},\ and\ \citenamefont
  {Podgornik}}]{holm2001electrostatic}%
  \BibitemOpen
  \bibfield  {author} {\bibinfo {author} {\bibfnamefont {C.}~\bibnamefont
  {Holm}}, \bibinfo {author} {\bibfnamefont {P.}~\bibnamefont
  {K{\'e}kicheff}},\ and\ \bibinfo {author} {\bibfnamefont {R.}~\bibnamefont
  {Podgornik}},\ }\href@noop {} {\emph {\bibinfo {title} {Electrostatic effects
  in soft matter and biophysics}}},\ Vol.~\bibinfo {volume} {46}\ (\bibinfo
  {publisher} {Springer Science \& Business Media},\ \bibinfo {year}
  {2001})\BibitemShut {NoStop}%
\bibitem [{\citenamefont {Dean}(1996)}]{dean96langevin}%
  \BibitemOpen
  \bibfield  {author} {\bibinfo {author} {\bibfnamefont {D.~S.}\ \bibnamefont
  {Dean}},\ }\bibfield  {title} {\bibinfo {title} {{Langevin equation for the
  density of a system of interacting Langevin processes}},\ }\href
  {https://iopscience.iop.org/article/10.1088/0305-4470/29/24/001/meta}
  {\bibfield  {journal} {\bibinfo  {journal} {J. Phys. A}\ }\textbf {\bibinfo
  {volume} {29}},\ \bibinfo {pages} {L613} (\bibinfo {year}
  {1996})}\BibitemShut {NoStop}%
\bibitem [{\citenamefont {Kawasaki}(1994)}]{kawasaki1994}%
  \BibitemOpen
  \bibfield  {author} {\bibinfo {author} {\bibfnamefont {K.}~\bibnamefont
  {Kawasaki}},\ }\bibfield  {title} {\bibinfo {title} {Stochastic model of slow
  dynamics in supercooled liquids and dense colloidal suspensions},\ }\href
  {https://doi.org/10.1016/0378-4371(94)90533-9} {\bibfield  {journal}
  {\bibinfo  {journal} {Physica A}\ }\textbf {\bibinfo {volume} {208}},\
  \bibinfo {pages} {35} (\bibinfo {year} {1994})}\BibitemShut {NoStop}%
\bibitem [{\citenamefont {te~Vrugt}\ \emph {et~al.}(2020)\citenamefont
  {te~Vrugt}, \citenamefont {L{\"o}wen},\ and\ \citenamefont
  {Wittkowski}}]{ddftreview}%
  \BibitemOpen
  \bibfield  {author} {\bibinfo {author} {\bibfnamefont {M.}~\bibnamefont
  {te~Vrugt}}, \bibinfo {author} {\bibfnamefont {H.}~\bibnamefont
  {L{\"o}wen}},\ and\ \bibinfo {author} {\bibfnamefont {R.}~\bibnamefont
  {Wittkowski}},\ }\bibfield  {title} {\bibinfo {title} {Classical dynamical
  density functional theory: from fundamentals to applications},\ }\href
  {https://www.tandfonline.com/doi/full/10.1080/00018732.2020.1854965}
  {\bibfield  {journal} {\bibinfo  {journal} {Adv. Phys.}\ }\textbf {\bibinfo
  {volume} {69}},\ \bibinfo {pages} {121} (\bibinfo {year} {2020})}\BibitemShut
  {NoStop}%
\bibitem [{\citenamefont {T{\"a}uber}(2014)}]{tauber}%
  \BibitemOpen
  \bibfield  {author} {\bibinfo {author} {\bibfnamefont {U.~C.}\ \bibnamefont
  {T{\"a}uber}},\ }\href@noop {} {\emph {\bibinfo {title} {Critical dynamics: a
  field theory approach to equilibrium and non-equilibrium scaling behavior}}}\
  (\bibinfo  {publisher} {Cambridge University Press},\ \bibinfo {year}
  {2014})\BibitemShut {NoStop}%
\bibitem [{\citenamefont {Khair}\ and\ \citenamefont
  {Brady}(2006)}]{brady2006}%
  \BibitemOpen
  \bibfield  {author} {\bibinfo {author} {\bibfnamefont {A.~S.}\ \bibnamefont
  {Khair}}\ and\ \bibinfo {author} {\bibfnamefont {J.~F.}\ \bibnamefont
  {Brady}},\ }\bibfield  {title} {\bibinfo {title} {Single particle motion in
  colloidal dispersions: a simple model for active and nonlinear
  microrheology},\ }\href {https://doi.org/10.1017/S0022112006009608}
  {\bibfield  {journal} {\bibinfo  {journal} {J. Fluid Mech.}\ }\textbf
  {\bibinfo {volume} {557}},\ \bibinfo {pages} {73–117} (\bibinfo {year}
  {2006})}\BibitemShut {NoStop}%
\bibitem [{\citenamefont {Long}\ and\ \citenamefont
  {Ajdari}(2001)}]{ajdarinote}%
  \BibitemOpen
  \bibfield  {author} {\bibinfo {author} {\bibfnamefont {D.}~\bibnamefont
  {Long}}\ and\ \bibinfo {author} {\bibfnamefont {A.}~\bibnamefont {Ajdari}},\
  }\bibfield  {title} {\bibinfo {title} {A note on the screening of
  hydrodynamic interactions, in electrophoresis, and in porous media},\ }\href
  {https://link.springer.com/article/10.1007\%2Fs101890170139#citeas}
  {\bibfield  {journal} {\bibinfo  {journal} {Eur. Phys. J. E}\ }\textbf
  {\bibinfo {volume} {4}},\ \bibinfo {pages} {29} (\bibinfo {year}
  {2001})}\BibitemShut {NoStop}%
\bibitem [{\citenamefont {Dzubiella}\ \emph {et~al.}(2002)\citenamefont
  {Dzubiella}, \citenamefont {Hoffmann},\ and\ \citenamefont
  {L\"owen}}]{lowenlane}%
  \BibitemOpen
  \bibfield  {author} {\bibinfo {author} {\bibfnamefont {J.}~\bibnamefont
  {Dzubiella}}, \bibinfo {author} {\bibfnamefont {G.~P.}\ \bibnamefont
  {Hoffmann}},\ and\ \bibinfo {author} {\bibfnamefont {H.}~\bibnamefont
  {L\"owen}},\ }\bibfield  {title} {\bibinfo {title} {Lane formation in
  colloidal mixtures driven by an external field},\ }\href
  {https://doi.org/10.1103/PhysRevE.65.021402} {\bibfield  {journal} {\bibinfo
  {journal} {Phys. Rev. E}\ }\textbf {\bibinfo {volume} {65}},\ \bibinfo
  {pages} {021402} (\bibinfo {year} {2002})}\BibitemShut {NoStop}%
\bibitem [{\citenamefont {Poncet}\ \emph {et~al.}(2017)\citenamefont {Poncet},
  \citenamefont {B{\'e}nichou}, \citenamefont {D{\'e}mery},\ and\ \citenamefont
  {Oshanin}}]{poncet2017universal}%
  \BibitemOpen
  \bibfield  {author} {\bibinfo {author} {\bibfnamefont {A.}~\bibnamefont
  {Poncet}}, \bibinfo {author} {\bibfnamefont {O.}~\bibnamefont
  {B{\'e}nichou}}, \bibinfo {author} {\bibfnamefont {V.}~\bibnamefont
  {D{\'e}mery}},\ and\ \bibinfo {author} {\bibfnamefont {G.}~\bibnamefont
  {Oshanin}},\ }\bibfield  {title} {\bibinfo {title} {Universal long ranged
  correlations in driven binary mixtures},\ }\href
  {https://journals.aps.org/prl/abstract/10.1103/PhysRevLett.118.118002}
  {\bibfield  {journal} {\bibinfo  {journal} {Phys. Rev. Lett.}\ }\textbf
  {\bibinfo {volume} {118}},\ \bibinfo {pages} {118002} (\bibinfo {year}
  {2017})}\BibitemShut {NoStop}%
\bibitem [{\citenamefont {Nägele}(1996)}]{nagele}%
  \BibitemOpen
  \bibfield  {author} {\bibinfo {author} {\bibfnamefont {G.}~\bibnamefont
  {Nägele}},\ }\bibfield  {title} {\bibinfo {title} {On the dynamics and
  structure of charge-stabilized suspensions},\ }\href
  {https://doi.org/https://doi.org/10.1016/0370-1573(95)00078-X} {\bibfield
  {journal} {\bibinfo  {journal} {Phys. Rep}\ }\textbf {\bibinfo {volume}
  {272}},\ \bibinfo {pages} {215 } (\bibinfo {year} {1996})}\BibitemShut
  {NoStop}%
\bibitem [{\citenamefont {D{\'e}mery}\ \emph {et~al.}(2014)\citenamefont
  {D{\'e}mery}, \citenamefont {B{\'e}nichou},\ and\ \citenamefont
  {Jacquin}}]{demery2014generalized}%
  \BibitemOpen
  \bibfield  {author} {\bibinfo {author} {\bibfnamefont {V.}~\bibnamefont
  {D{\'e}mery}}, \bibinfo {author} {\bibfnamefont {O.}~\bibnamefont
  {B{\'e}nichou}},\ and\ \bibinfo {author} {\bibfnamefont {H.}~\bibnamefont
  {Jacquin}},\ }\bibfield  {title} {\bibinfo {title} {{Generalized Langevin
  equations for a driven tracer in dense soft colloids: construction and
  applications}},\ }\href
  {https://iopscience.iop.org/article/10.1088/1367-2630/16/5/053032/meta}
  {\bibfield  {journal} {\bibinfo  {journal} {New J. Phys.}\ }\textbf {\bibinfo
  {volume} {16}},\ \bibinfo {pages} {053032} (\bibinfo {year}
  {2014})}\BibitemShut {NoStop}%
\bibitem [{Note1()}]{Note1}%
  \BibitemOpen
  \bibinfo {note} {Note that the noise is also discarded at this level since it
  will only have short range contributions to the correlation, see Ref.~\cite
  {mahdisoltani2021long}}\BibitemShut {NoStop}%
\bibitem [{\citenamefont {Martinac}(2004)}]{Martinac2004}%
  \BibitemOpen
  \bibfield  {author} {\bibinfo {author} {\bibfnamefont {B.}~\bibnamefont
  {Martinac}},\ }\bibfield  {title} {\bibinfo {title} {Mechanosensitive ion
  channels: molecules of mechanotransduction},\ }\href
  {https://doi.org/10.1242/jcs.01232} {\bibfield  {journal} {\bibinfo
  {journal} {J. Cell Sci.}\ }\textbf {\bibinfo {volume} {117}},\ \bibinfo
  {pages} {2449} (\bibinfo {year} {2004})}\BibitemShut {NoStop}%
\bibitem [{\citenamefont {Siwy}\ and\ \citenamefont
  {Fuli\ifmmode~\acute{n}\else \'{n}\fi{}ski}(2002)}]{Siwy2002}%
  \BibitemOpen
  \bibfield  {author} {\bibinfo {author} {\bibfnamefont {Z.}~\bibnamefont
  {Siwy}}\ and\ \bibinfo {author} {\bibfnamefont {A.}~\bibnamefont
  {Fuli\ifmmode~\acute{n}\else \'{n}\fi{}ski}},\ }\bibfield  {title} {\bibinfo
  {title} {Fabrication of a synthetic nanopore ion pump},\ }\href
  {https://doi.org/10.1103/PhysRevLett.89.198103} {\bibfield  {journal}
  {\bibinfo  {journal} {Phys. Rev. Lett.}\ }\textbf {\bibinfo {volume} {89}},\
  \bibinfo {pages} {198103} (\bibinfo {year} {2002})}\BibitemShut {NoStop}%
\bibitem [{\citenamefont {Amrei}\ \emph {et~al.}(2018)\citenamefont {Amrei},
  \citenamefont {Bukosky}, \citenamefont {Rader}, \citenamefont {Ristenpart},\
  and\ \citenamefont {Miller}}]{amrei2018oscillating}%
  \BibitemOpen
  \bibfield  {author} {\bibinfo {author} {\bibfnamefont {S.~H.}\ \bibnamefont
  {Amrei}}, \bibinfo {author} {\bibfnamefont {S.~C.}\ \bibnamefont {Bukosky}},
  \bibinfo {author} {\bibfnamefont {S.~P.}\ \bibnamefont {Rader}}, \bibinfo
  {author} {\bibfnamefont {W.~D.}\ \bibnamefont {Ristenpart}},\ and\ \bibinfo
  {author} {\bibfnamefont {G.~H.}\ \bibnamefont {Miller}},\ }\bibfield  {title}
  {\bibinfo {title} {Oscillating electric fields in liquids create a long-range
  steady field},\ }\href
  {https://journals.aps.org/prl/abstract/10.1103/PhysRevLett.121.185504}
  {\bibfield  {journal} {\bibinfo  {journal} {Phys. Rev. Lett.}\ }\textbf
  {\bibinfo {volume} {121}},\ \bibinfo {pages} {185504} (\bibinfo {year}
  {2018})}\BibitemShut {NoStop}%
\bibitem [{\citenamefont {Hansen}\ and\ \citenamefont
  {McDonald}(1990)}]{hansen}%
  \BibitemOpen
  \bibfield  {author} {\bibinfo {author} {\bibfnamefont {J.-P.}\ \bibnamefont
  {Hansen}}\ and\ \bibinfo {author} {\bibfnamefont {I.~R.}\ \bibnamefont
  {McDonald}},\ }\href@noop {} {\emph {\bibinfo {title} {Theory of simple
  liquids}}}\ (\bibinfo  {publisher} {Elsevier},\ \bibinfo {year}
  {1990})\BibitemShut {NoStop}%
\bibitem [{Note2()}]{Note2}%
  \BibitemOpen
  \bibinfo {note} {A more involved approach is to use the method of images
  together with the real-space diffusive Green's functions, and then make use
  of the Poisson summation formula to represent the solution in terms of the
  Neumann modes \cite {barton1989}.}\BibitemShut {Stop}%
\bibitem [{\citenamefont {Irving}\ and\ \citenamefont
  {Kirkwood}(1950)}]{irving1950hydro}%
  \BibitemOpen
  \bibfield  {author} {\bibinfo {author} {\bibfnamefont {J.}~\bibnamefont
  {Irving}}\ and\ \bibinfo {author} {\bibfnamefont {J.~G.}\ \bibnamefont
  {Kirkwood}},\ }\bibfield  {title} {\bibinfo {title} {{The statistical
  mechanical theory of transport processes. IV. The equations of
  hydrodynamics}},\ }\href {https://aip.scitation.org/doi/10.1063/1.1747782}
  {\bibfield  {journal} {\bibinfo  {journal} {J. Chem. Phys.}\ }\textbf
  {\bibinfo {volume} {18}},\ \bibinfo {pages} {817} (\bibinfo {year}
  {1950})}\BibitemShut {NoStop}%
\bibitem [{\citenamefont {Kr{\"u}ger}\ \emph {et~al.}(2018)\citenamefont
  {Kr{\"u}ger}, \citenamefont {Solon}, \citenamefont {D{\'e}mery},
  \citenamefont {Rohwer},\ and\ \citenamefont {Dean}}]{kruger2018neqstress}%
  \BibitemOpen
  \bibfield  {author} {\bibinfo {author} {\bibfnamefont {M.}~\bibnamefont
  {Kr{\"u}ger}}, \bibinfo {author} {\bibfnamefont {A.}~\bibnamefont {Solon}},
  \bibinfo {author} {\bibfnamefont {V.}~\bibnamefont {D{\'e}mery}}, \bibinfo
  {author} {\bibfnamefont {C.~M.}\ \bibnamefont {Rohwer}},\ and\ \bibinfo
  {author} {\bibfnamefont {D.~S.}\ \bibnamefont {Dean}},\ }\bibfield  {title}
  {\bibinfo {title} {Stresses in non-equilibrium fluids: Exact formulation and
  coarse-grained theory},\ }\href
  {https://aip.scitation.org/doi/abs/10.1063/1.5019424} {\bibfield  {journal}
  {\bibinfo  {journal} {J. Chem. Phys.}\ }\textbf {\bibinfo {volume} {148}},\
  \bibinfo {pages} {084503} (\bibinfo {year} {2018})}\BibitemShut {NoStop}%
\bibitem [{\citenamefont {Jackson}(2007)}]{jackson}%
  \BibitemOpen
  \bibfield  {author} {\bibinfo {author} {\bibfnamefont {J.~D.}\ \bibnamefont
  {Jackson}},\ }\href@noop {} {\emph {\bibinfo {title} {Classical
  Electrodynamics}}}\ (\bibinfo  {publisher} {John Wiley \& Sons},\ \bibinfo
  {year} {2007})\BibitemShut {NoStop}%
\bibitem [{\citenamefont {Woodson}\ and\ \citenamefont
  {Melcher}(1968)}]{woodson1968electromechanical}%
  \BibitemOpen
  \bibfield  {author} {\bibinfo {author} {\bibfnamefont {H.~H.}\ \bibnamefont
  {Woodson}}\ and\ \bibinfo {author} {\bibfnamefont {J.~R.}\ \bibnamefont
  {Melcher}},\ }\href@noop {} {\emph {\bibinfo {title} {Electromechanical
  dynamics}}}\ (\bibinfo  {publisher} {Wiley},\ \bibinfo {year}
  {1968})\BibitemShut {NoStop}%
\bibitem [{\citenamefont {Mahdisoltani}\ \emph {et~al.}(2021)\citenamefont
  {Mahdisoltani}, \citenamefont {Zinati}, \citenamefont {Duclut}, \citenamefont
  {Gambassi},\ and\ \citenamefont {Golestanian}}]{mahdisoltani2021chemotaxis}%
  \BibitemOpen
  \bibfield  {author} {\bibinfo {author} {\bibfnamefont {S.}~\bibnamefont
  {Mahdisoltani}}, \bibinfo {author} {\bibfnamefont {R.~B.~A.}\ \bibnamefont
  {Zinati}}, \bibinfo {author} {\bibfnamefont {C.}~\bibnamefont {Duclut}},
  \bibinfo {author} {\bibfnamefont {A.}~\bibnamefont {Gambassi}},\ and\
  \bibinfo {author} {\bibfnamefont {R.}~\bibnamefont {Golestanian}},\
  }\bibfield  {title} {\bibinfo {title} {Nonequilibrium polarity-induced
  chemotaxis: Emergent galilean symmetry and exact scaling exponents},\ }\href
  {https://journals.aps.org/prresearch/abstract/10.1103/PhysRevResearch.3.013100}
  {\bibfield  {journal} {\bibinfo  {journal} {Phys. Rev. Research}\ }\textbf
  {\bibinfo {volume} {3}},\ \bibinfo {pages} {013100} (\bibinfo {year}
  {2021})}\BibitemShut {NoStop}%
\bibitem [{\citenamefont {Barton}\ and\ \citenamefont
  {Barton}(1989)}]{barton1989}%
  \BibitemOpen
  \bibfield  {author} {\bibinfo {author} {\bibfnamefont {G.}~\bibnamefont
  {Barton}}\ and\ \bibinfo {author} {\bibfnamefont {G.}~\bibnamefont
  {Barton}},\ }\href@noop {} {\emph {\bibinfo {title} {Elements of Green's
  functions and propagation: potentials, diffusion, and waves}}}\ (\bibinfo
  {publisher} {Oxford University Press},\ \bibinfo {year} {1989})\BibitemShut
  {NoStop}%
\end{thebibliography}%

\end{document}